\documentclass[fleqn,usenatbib]{mnras}

\usepackage{newtxtext,newtxmath}
\usepackage[T1]{fontenc}
\usepackage{ae,aecompl}
\usepackage{graphicx}
\usepackage{amsmath}
\usepackage{breqn}
\usepackage[normalem]{ulem}	% To enable strike-throughs with \sout{}
\usepackage{wasysym} % For the \astrosun symbol, etc
\usepackage{cleveref}
\usepackage{xcolor}

\crefname{section}{\S}{\S\S}

\newcommand{\mgii}{Mg\,\textsc{ii} }

\title[Galactic Conformity]{The Physical Origin of Galactic Conformity: From Theory to Observation}

\author[Ayromlou et al.]{Mohammadreza Ayromlou,$^{1,2}$\thanks{E-mail: ayromlou@uni-heidelberg.de}
Guinevere Kauffmann,$^{1}$
Abhijeet Anand,$^{1}$
Simon D. M. White$^{1}$
\\\\
$^{1}$Max Planck Institute for Astrophysics, Karl-Schwarzschild-Str. 1, 85741 Garching bei M{\"u}nchen, Germany\\
$^{2}$Universit{\"a}t Heidelberg, Zentrum f{\"u}r Astronomie, Institut f{\"u}r theoretische Astrophysik, Albert-Ueberle-Str. 2, 69120 Heidelberg, Germany
}

\date{}
\pubyear{2022}

\begin{document}
\label{firstpage}
\pagerange{\pageref{firstpage}--\pageref{lastpage}}
\maketitle

\begin{abstract}
We employ several galaxy formation models, in particular, L-GALAXIES, IllustrisTNG, and EAGLE, as well as observational samples from SDSS and DESI, to investigate galactic conformity, the observed large-scale correlation between the star-formation properties of central (primary) galaxies and those of their neighbours.
To analyse the models and observations uniformly, we introduce \textsc{CenSat}, a new algorithm to define whether a galaxy is a central or a satellite system based on an isolation criterion.
We find that the conformity signal is present, up to at least 5 Mpc from the centres of low- and intermediate-mass centrals in the latest version of L-GALAXIES (Ayromlou et al. 2021), IllustrisTNG, and EAGLE, as well as in SDSS and DESI observational samples. In comparison, the conformity signal is substantially weaker in an older version of L-GALAXIES (Henriques et al. 2020).
One of the main differences between this older model and the other models is its neglect of ram-pressure stripping of the gas reservoirs of galaxies except within the boundaries of massive cluster haloes. Our observational comparisons demonstrate that this difference significantly affects the observed large-scale conformity signal.
Furthermore, by examining the contribution of backsplash, fly-by, central, and satellite galaxies to the conformity signal, we show that much, but not all, of it arises from primary galaxies near massive systems.
Remaining tensions between the models and observations may be solved by modifying the physical prescriptions for how feedback processes affect the distribution and kinematics of gas and the environment around galaxies out to scales of several Megaparsecs.
\end{abstract}


%%%%%%%%%%%%%%%%% INTRODUCTION %%%%%%%%%%%%%%%%%%

\section{Introduction}
\label{sec: introduction}
Modern theories of galaxy formation and evolution are constructed within the framework of the Lambda cold dark matter ($\rm \Lambda CDM$) cosmology, which is calibrated to match observations of the cosmic microwave background \citep[e.g.][]{komatsu2011seven,planck2015_xiii}. According to the standard hierarchical theory of structure formation, gas (baryonic matter) falls into the potential wells of dark matter haloes, where it cools down and forms the  galaxies that can be observed today \citep[][]{white1978core,white1991galaxy}.

Observed galaxies come in different masses \citep{li2009distribution,d2015massive}, shapes \citep{Abraham2003morphology}, star formation activity levels \citep[][]{Gomez2003starFormation}, colours \citep[][]{Fukugita1995colors}, and ages \citep[][]{Terlevich2002galaxyAges}. An interesting aspect of galaxy properties is their connection with the  large scale structure of the Universe. One well-known example of such a connection is manifested by the dependence of the amplitude of the galaxy correlation function on large scales on the luminosities and colours of galaxies. Brighter and redder galaxies are shown to be more clustered than less bright and bluer ones \citep[e.g.][]{Zehavi2005correlationFunction}. Another recently discovered phenomenon, which is still an important puzzle and a matter of debate, is the large scale correlations between the star formation rates of neighbouring galaxies around isolated galaxies where star formation has shut down (often termed ``quenching''). This phenomenon is commonly known as galactic conformity and  is the topic of this research.

\cite{weinmann2006properties} first introduced the phrase "galactic conformity" to describe the strong correlations between the properties of satellite galaxies and their centrals in the SDSS data. The main result was an excess of the early-type fraction of satellites in the vicinity of early-type central galaxies compared to satellites around late-type centrals \footnote{In a group of galaxies within a dark matter halo, the galaxy that is considered as the central (primary) galaxy is the most massive or luminous galaxy in the group. The other galaxies in the group are labelled as satellites.}. This phenomenon investigated by \cite{weinmann2006properties} was "one-halo", or small scale, galactic conformity confined to galaxies within the virial radius of the halo. Later, \cite{kauffmann2013re} showed that galactic conformity in SDSS could extend to scales significantly larger than the halo virial radius, out to $\rm 4\, Mpc$. This finding is called large-scale, or two-halo, galactic conformity, and implies a correlation between the star formation rates of central galaxies and their neighbours (which can be  both centrals and satellites) out to scales well beyond the virial radius of the central. They suggested "pre-heating" as a possible physical origin for the observed conformity. In this scenario, gas is heated over large scales at high redshifts, influencing the cooling and star formation of distant galaxies, leading to the observed large-scale conformity signal. Furthermore, they analysed the \cite{guo2011dwarf} version of \textsc{L-Galaxies} for the conformity signal, but did not find a signal as significant as the one from the real data.

The small scale conformity  effect has been the subject of several observational and theoretical studies. \cite{Wang2012SatelliteAbundances} analysed the \cite{guo2011dwarf} version of \textsc{L-Galaxies} and argued that the correlation between the colours of central galaxies and their satellites is because red centrals reside in more massive and older  haloes in which satellite quenching has been more effective. 
This conclusions of this work are supported by an analysis of the distribution of pairwise velocities between satellite 
galaxies of $\sim 0.1L^{\star}$ brightness and their $\sim L^{\star}$ hosts in SDSS \citep{Phillips2014SatelliteQuenching},
where dynamical evidence was found  that the host haloes of passive central galaxies are more massive than the host haloes of actively star forming  central galaxies.

\cite{Hartley2015Conformity} found strong evidence of small scale galactic conformity at high redshifts ($z \lesssim 2)$ in the UKIRT Infrared Deep Sky Survey (UKIDSS; \citealt[][]{Lawrence2007Ukidss}). \cite{Kawinwanichakij2016conformity} analysed data from \cite{Lawrence2007Ukidss,McCracken2012UltraVISTA} and found small scale conformity at high redshifts.

In contrast, the amplitude and the origin of large scale galactic conformity signal has
been a matter of controversy in recent years. \cite{kauffmann2015physical} found a higher fraction of massive galaxies of $\log_{10}(M_{\star}/{\rm M_{\odot}})>11.3$ around quenched centrals in SDSS. They also discovered that massive galaxies near quenched centrals are more likely to host radio-loud AGNs than massive galaxies near star-forming centrals. Furthermore, they analysed the Illustris hydrodynamical simulation \citep{vogelsberger2014Introducing} and argued that the amplitude of the conformity signal in Illustris is much weaker than the observed amplitude. They suggested that, therefore, gas needs to be pushed out of haloes more efficiently in simulations. In contrast, \cite{bray16conformity} found both small and large-scale conformity in the Illustris simulation, with the amplitude of their large scale conformity being higher than reported by \cite{kauffmann2015physical}. Most of  these differences appear to be caused by  different sample selection criteria in the two studies. This highlights the need for the sample selection criteria in the observations and in the models to be closely matched in order to carry out meaningful comparisons. In another study, \cite{Hearin2015conformity,hearin2016physical} speculated that large scale conformity could be evidence for assembly bias and argued that large-scale tidal fields are the origin of the observed signal.

The relatively high amplitude  of the conformity signal has also been challenged in a few studies. \cite{Sin2017Conformity} argued that the large  measured amplitude of the conformity signal on large scales reported by \cite{kauffmann2013re} may not be robust. Possible interpretational problems that were pointed out included miss-classification of satellites as centrals, the contribution of  centrals near massive systems, and the use of the median sSFR instead of its mean value in the Kauffmann studies \footnote {We note, however, that \cite{kauffmann2015physical} provided results for 10, 25, 75, and 90th percentiles of the sSFR  distribution in addition to the median, both in the SDSS data and in their analysis of the observational samples}.  A similar paper by \cite{Sun2018Conformity} argued that the conformity effect was a simple extension of the relation between star formation in galaxies and their large scale environment. \cite{Tinker2018Conformity} reproduced the analysis from \cite{kauffmann2013re} using SDSS data and  recovered the same signal. However, they also argued that the interpretation of the strength of the conformity signal could be influenced by the mis-classification of satellite galaxies as centrals.

More recent studies have focused on a variety of observational data and models to study galactic conformity. \cite{Berti2017Conformity} used the data from the PRism MUlti-object Survey (PRIMUS, \citealt{Coil2011Primus,Cool2013Primus}) and found the conformity signal both on small ($R<\rm 1\, Mpc$) and on large scales (${\rm 1\, Mpc}<R<{\rm 3\, Mpc}$). They reported their large scale signal to be  weaker than observed in SDSS. They also speculated that large-scale tidal fields and assembly bias could be the causes of the signal.

\cite{Calderon2018Conformity} reported strong evidence of large scale galactic conformity out  to $\rm \sim 4\, Mpc$ from the halo centre in SDSS data using marked correlation functions. To further address the origin of large scale conformity, \cite{Kauffmann2018MgIIabsorbers} explored the \cite{Zhu2013MetalAbsorption} sample of \mgii quasar absorption line systems in combination with SDSS galaxy data. They found a higher number of \mgii absorbers out to 10 Mpc near red-low-mass galaxies than near blue-low-mass galaxies. Moreover, out to 5 Mpc of the centres of low-mass galaxies, they reported strong sensitivity of the equivalent width of the distribution of \mgii absorbers to the presence of nearby radio-loud AGNs and argued that this constitutes evidence that AGN were affecting the gas properties around galaxies well outside their own dark matter halos.

\cite{Lacerna2018conformity} analysed a few semi-analytical models, including the \cite{Guo2013Galaxy} version of \textsc{L-Galaxies} and the \cite{Gonzalez-Perez2014How,Lagos2014Galform} versions of \textsc{Galform} to explore the conformity signal. They implemented  several different isolation criteria to identify central galaxies. In each isolation criteria, they chose a "constant" radial distance for all central galaxies, within which no galaxy of comparable mass or more massive should exist.  They reported that the large scale conformity signal is present to some level in these models. However, the signal's amplitude strongly depends on the chosen isolation criteria. 
In another recent work, \cite{Lacerna2021Conformity} analysed the \textsc{SAG} semi-analytical model \citep{cora2018semi} and the IllustrisTNG simulation \citep[][]{Nelson2019public} and argued that central galaxies near massive haloes contribute substantially to the conformity signal, because they find a higher fraction of quenched central galaxies in the vicinity of massive systems.

In this work, we study the strength and the origin of the large scale galactic conformity signal. To do so, we make use of  SDSS and DESI observations in addition to several galaxy formation models, including \textsc{L-Galaxies}, IllustrisTNG, and EAGLE. We define our sample of central galaxies carefully, by devising the \textsc{CenSat} algorithm, to 
optimize our classification of galaxies into satellites and centrals.
In order to consistently compare the models and observations, we make mock galaxy catalogues and apply the same analysis
methods to the mock catalogues and  to the observations. 
Our goal is  to take advantage of the different implementation of physical processes 
in the different models to uncover the origin of the observed conformity signal on large scales.

This paper is structured as follows: In section \ref{sec: Methodology} we explain our definition of the galactic conformity signal and describe the galaxy formation models, observations, and mock galaxy catalogues that are used in this study. In addition, we introduce \textsc{CenSat}, a new algorithm to classify  galaxies as centrals or satellites within their groups, both in simulations and in observations in a uniform way. In section \ref{sec: Results}, we investigate the strength and origin of the galactic conformity signal from both simulations and observations. Finally, we conclude and summarise our results in section \ref{sec: summary}.

%%%%%%%%%%%%%%%%%%%% METHODS %%%%%%%%%%%%%%%%%%

\section{Methodology}
\label{sec: Methodology}
\subsection{Galactic conformity signal}
\label{subsec: methods_conformity}
We define the galactic conformity signal as the difference between the value of a given quantity (e.g., galaxy quenched fraction) at a halocentric distance ($R$) evaluated for star-forming and quenched central (primary) galaxies. We use the phrase "primary galaxy" (or "main central galaxy") whenever we want to distinguish between a central galaxy at $R=0$ (the primary galaxy) and other galaxies (central or satellite) in its vicinity at $R>0$, which we call "secondary" galaxies. In this paper, we investigate a few variations of the conformity signal, which are explained in this subsection.
\subsubsection{Conformity in the quenched fraction of galaxies}
\label{subsubsec: methods_conformity_f_q}
Dividing primary galaxies into two categories based on their specific star formation rate, we define the conformity signal as
\begin{multline}
    \label{eq: conformity_signal_f_q}
    \Delta f_{\rm q}(r) = f_{\rm q}(r){\rm [\text{quenched primary}]}\\
    - f_{\rm q}(r){\rm [\text{star forming primary}]},
\end{multline}
where $\Delta f_{\rm q}(r)$ is the conformity signal, $f_{\rm q}(r){\rm [\text{quenched primary}]}$ is the quenched fraction of galaxies in the vicinity of \textit{quenched primaries} and $f_{\rm q}(r){\rm [\text{star-forming primary}]}$ is the quenched fraction of galaxies in the vicinity of \textit{star-forming primaries}.

In this work, we apply a sSFR cut to identify quenched galaxies. At $z=0$, galaxies with $\log_{10} (\rm sSFR/yr^{-1})<-11$ are considered as quenched. Similarly, at $z=1$ and $z=2$, galaxies with $\log_{10} (\rm sSFR/yr^{-1})<-10$ and $\log_{10} (\rm sSFR/yr^{-1})<-9.8$ are considered quenched, respectively. These values are motivated from \cite{henriques17}, and already used by \cite{Ayromlou2020Comparing,Ayromlou2021Galaxy} for \textsc{L-Galaxies}, TNG, and SDSS. We note that changing the sSFR cuts by 0.3 dex does not influence our results and conclusions significantly.

\subsubsection{Conformity in the stellar mass of galaxies}
\label{subsubsec: methods_conformity_mass}
To explore whether there is a difference between the stellar masses of galaxies in the vicinity of quenched and star forming centrals, we define the stellar mass conformity signal as
\begin{multline}
    \label{eq: conformity_signal_mass}
   \Delta \log_{10}(M_{\rm \star}(r)) = \log_{10}(M_{{\rm \star}, \rm median}(r){\rm [\text{quenched primary}]})\\
   - \log_{10}(M_{{\rm \star}, \rm median}(r){\rm [\text{SF primary}])},
\end{multline}
where $\log_{10}(M_{{\rm \star}, \rm median}(r){\rm [\text{quenched primary}]})$ and $\log_{10}(M_{{\rm \star}, \rm median}(r){\rm [\text{SF primary}])}$ are the median values of the galaxy stellar mass in the vicinity of quenched and star-forming central galaxies, respectively.

\subsubsection{Conformity in the specific star formation rate of galaxies}
\label{subsubsec: methods_conformity_ssfr}
Here, at $z=0$, we divide primary galaxies into three categories: a) highly star forming: ${\rm \log (sSFR_{cen}/yr^{-1})\geq-10}$, b) intermediate star forming: ${\rm  -11\leq\log(sSFR_{cen}/yr^{-1})<-10}$, and c) quenched: ${\rm \log (sSFR_{cen}/yr^{-1})<-11}$. We then capture the conformity signal by taking galaxies in the vicinity of highly star forming and quenched primaries.

\begin{multline}
    \label{eq: conformity_signal_ssfr}
    \Delta {\rm log_{10}(sSFR(r))} = \log({\rm sSFR}_{\rm mean}(r)[{\rm \text{highly SF primary}}])\\
    -  \log({\rm sSFR}_{\rm mean}(r)[{\rm \text{quenched primary} }]).
\end{multline}

We note that there are several caveats and uncertainties in deriving the star formation rates of observed galaxies (e.g., for extremely red galaxies) as well as resolution limits in extracting the star formation rates of simulated galaxies (e.g. zero values for star formation rates in hydrodynamical simulations). Due to the resolution limits in hydrodynamical simulations, the median value of star formation rate and sSFR could be zero in some cases. To avoid this issue, here we use mean sSFR instead of median sSFR. Overall, rather than splitting galaxies into several sSFR bins, it is more robust to simply categorise them into "star forming" and "quenched" as  done in section \ref{sec: conformity}. For completeness, nevertheless, we will show some results based on dividing the samples by their sSFR values in section \ref{subsubsec: conformity_obs_ssfr} as well.

In this paper, wherever we say "the conformity signal" or "the signal", we refer to the conformity signal in the quenched fraction of galaxies. The other definitions described above will be called more specifically by their names, i.e. conformity in sSFR and conformity in the stellar masses of galaxies.

\subsection{Galaxy formation models and relevant definitions}
\label{subsec: simulations_and_model}

\subsubsection{\textsc{L-Galaxies} semi-analytical model of galaxy formation}
\label{subsubsec: L-Galaxies}
The Munich semi-analytical model of galaxy formation and evolution, \textsc{L-Galaxies}, implements a set of
recipes on dark matter-only halo merger trees to describe the evolution of gas and stars in dark matter
halos and to make predictions about the properties of galaxies \citep{kauffmann1993formation,kauffmann1999clustering,springel2001populating}.
Current versions of  \textsc{L-Galaxies}  run on top of the dark matter halo merger trees extracted from the Millennium and Millennium-II simulations \citep{springel2005simulations,croton2006many,de2006formation,bertone2007recycling,boylan2009resolving,guo2011dwarf,henriques2015galaxy}. Employing the method introduced by \cite{angulo2010one,angulo2015cosmological}, both simulations are mapped to the $\rm \Lambda CDM$ Planck cosmology \citep{planck2015_xiii} from their original sets of cosmological parameters.

The recent versions of \textsc{L-Galaxies} split baryonic matter associated to each subhalo into seven main components: hot gas, cold gas (divided into HI and $\rm H_2$), stars (divided into stellar disc, bulge stars, and halo stars), supermassive black holes, and ejected material. As soon as a subhalo is formed, \textsc{L-Galaxies} tracks a variety of processes affecting the gas in the subhalo. 
These include the infall of diffuse hot gas into the subhalo as it grows, cooling of the gas, star formation, stellar and black hole feedback, mergers, environmental effects, and several other processes (see the supplementary material of LGal-A21 for a full description).

In this work, we use two recent versions of \textsc{L-Galaxies} \citep{henriques2020galaxies,Ayromlou2021Galaxy}  \footnote{\href{https://lgalaxiespublicrelease.github.io}{https://lgalaxiespublicrelease.github.io}}, to investigate galactic conformity. We 
analyse the models run on the Millennium simulation. The two model versions have a very similar implementations of most physical processes,
with the important exception of their treatment of environmental processes, which is described in section \ref{subsubsec: methods_relevant_processes}. In both the models, we take galaxies with $\log_{10}(M_{\star}/{\rm M_{\odot}})>9.5$, where $M_{\star} = M_{\rm \star,disc} + M_{\rm \star,bulge}$. The stellar mass interval is chosen to represent galaxies that are resolved properly both in the simulations and in the real data. The main parameters of the \textsc{L-Galaxies} outputs are given in Table \ref{tab: data_stat}. The side-length of these two simulations is $\sim 714 \rm \, Mpc$.
\begin{table*}
	\centering
	\caption{The main parameters of the observations and simulations used in this paper. The stellar mass interval is chosen to represent galaxies that are resolved properly both in the simulations and in the real data.}
	\label{tab: data_stat}
	\begin{tabular}{|*{6}{c|}}
		\hline \hline
		\textbf{Observation/Model} & \textbf{Stellar mass field in catalogue} & \textbf{Stellar mass interval} & \textbf{(s)SFR field in catalogue} & \textbf{Redshift} & \textbf{${\rm N_{gal}}$} \\
		\hline \hline
		SDSS & Median stellar mass & $\log_{10}(M_{\star}/{\rm M_{\odot}})>9.5$ & Median sSFR & $0\leq z_{\rm spec} \leq 0.04$ & 29553 \\
		\hline
		DESI & MASS_BEST & $\log_{10}(M_{\star}/{\rm M_{\odot}})>9.5$ & SSFR_BEST & $0\leq z_{\rm photo} \leq 0.1$ & 974128 \\ 
		\hline \hline
		 & & $\log_{10}(M_{\star}/{\rm M_{\odot}})>9.5$ & & $z=0$ & 2638377 \\
		LGal - A21 & StellarMass & $\log_{10}(M_{\star}/{\rm M_{\odot}})>9.5$ & StarFormationRate & $z=1$ & 2056649 \\
		 & & $\log_{10}(M_{\star}/{\rm M_{\odot}})>9.5$ & & $z=2$ & 1191608 \\
		\hline
		 & & $\log_{10}(M_{\star}/{\rm M_{\odot}})>9.5$ & & $z=0$ & 3168793 \\
		LGal - H20 & StellarMass & $\log_{10}(M_{\star}/{\rm M_{\odot}})>9.5$ & Sfr & $z=1$ & 2226392 \\
		 & & $\log_{10}(M_{\star}/{\rm M_{\odot}})>9.5$ & & $z=2$ & 1128076 \\
		\hline
		 & & $\log_{10}(M_{\star}/{\rm M_{\odot}})>9.5$ & & $z=0$ & 175572 \\
		TNG 300 & SubhaloMassType [star particles] & $\log_{10}(M_{\star}/{\rm M_{\odot}})>9.5$ & SubhaloSFR & $z=1$ & 152089 \\
		 & & $\log_{10}(M_{\star}/{\rm M_{\odot}})>9.5$ & & $z=2$ & 95803 \\
		\hline
		 & & $\log_{10}(M_{\star}/{\rm M_{\odot}})>9.5$ & & $z=0$ & 7335 \\
		EAGLE & MassType [star particles] & $\log_{10}(M_{\star}/{\rm M_{\odot}})>9.5$ & StarFormationRate & $z=1$ & 4625 \\
		 & & $\log_{10}(M_{\star}/{\rm M_{\odot}})>9.5$ & & $z=2$ & 3585 \\
		\hline
	\end{tabular}
\end{table*}
\subsubsection{IllustrisTNG Simulation}
\label{subsubsec: TNG}
The IllustrisTNG simulations \citep[TNG  hereafter;][]{nelson18a,springel2018first,pillepich2018First,marinacci2018first,naiman2018first}\footnote{\href{https://www.tng-project.org/}{https://www.tng-project.org/}} are the next generation of the Illustris simulation \citep[][]{vogelsberger2014Introducing} with improved implementation of physical processes. Employing the \textsc{AREPO} code \citep[][]{springel2010pur}, TNG provides solutions to the equations of gravity and magnetohydrodynamics \citep[][]{pakmor2011magnetohydrodynamics,pakmor2013simulations} and implements a set of physical processes to model galaxy formation and evolution on cosmological scales. These physical processes include gas cooling, star formation, the evolution of stars, supernova feedback \citep{pillepich2018Simulating}, and supermassive black hole relevant processes such as seeding, evolution, and AGN feedback \citep[][]{weinberger17}.

To date, the model has been performed on three different cubic boxes with side lengths of $\sim$ 50 Mpc (TNG50, \citealt[][]{nelson2019First,pillepich19}), 100 Mpc (TNG100), and 300 Mpc (TNG300). The resolution of the simulation decreases with the box size. The smallest box (TNG50) provides the highest resolution while the largest box (TNG300) provides the best statistics. In this work, we use the publicly available TNG300 simulation \citep{Nelson2019public} with a dark matter resolution of $m_{\rm DM} = 5.9\times 10^{7}\rm M_{\odot}$ and average gas cell mass of $m_{\rm gas} \simeq 1.1\times 10^{7}\rm M_{\odot}$.

The model parameters of the TNG simulations are calibrated at the TNG100 resolution employing several observations (e.g. the stellar mass function and the stellar-to-halo mass ratio at z = 0). The derived model parameters are kept unchanged for the other runs, including TNG300 and TNG50. Consequently, the TNG model has complex numerical convergence behaviour, which must be evaluated for each galaxy property at different resolution levels \citep[see][]{pillepich2018Simulating}. We address the conformity signal convergence between TNG300 (used here) and the calibrated TNG100 model in Appendix \ref{app: TNG_consistency}.

TNG follows a $\Lambda \rm CDM$ cosmology with parameters taken from \cite{planck2015_xiii}.  
Here we take galaxies with $\log_{10}(M_{\star}/{\rm M_{\odot}})>9.5$, where $M_{\star}$ is the total stellar mass of each galaxy. The star formation rate of each galaxy is also taken as the total star formation rate of the galaxy. The main parameters of the TNG outputs are given in Table \ref{tab: data_stat}.
\subsubsection{EAGLE Simulation}
\label{subsubsec: EAGLE}
The EAGLE (Evolution and Assembly of GaLaxies and their Environments) simulations \citep{schaye2015eagle,Crain2015TheEagle}\footnote{\href{http://icc.dur.ac.uk/Eagle/}{http://icc.dur.ac.uk/Eagle/}} are a set of hydrodynamical simulations that employ a modified version of the \textsc{GADGET-3} smoothed particle hydrodynamics
(SPH) code \citep{Springel2005Gadget} to solve the coupled equations of gravity and hydrodynamics. 
Similar to TNG, EAGLE produces galaxies in cosmological scale boxes by implementing recipes
for astrophysical processes including gas cooling, star formation, stellar feedback, formation and evolution of supermassive black holes and AGN feedback. 

In this work, we use a version of EAGLE with a cubic box volume of $(100\rm \, Mpc)^3$, dark matter particle mass of $m_{\rm DM} = 9.7\times 10^6 {\rm M_{\odot}}$, and initial gas particle mass of $m_{\rm gas} = 1.81 \times 10^6 {\rm M_{\odot}}$. The EAGLE simulation follows a $\Lambda \rm CDM$ cosmology with parameters taken from \cite{Planck2014}. For our analysis in this work, we take galaxies with $\log_{10}(M_{\star}/{\rm M_{\odot}})>9.5$, where $M_{\star}$ is the total stellar mass of each galaxy. The star formation rate of each galaxy is also taken as the total star formation rate of the galaxy. The parameters of the EAGLE outputs are given in Table \ref{tab: data_stat}.

\subsubsection{Identifying haloes and galaxies with FOF and SUBFIND}
\label{subsubsec: FOF_SUBFIND}
In all galaxy formation models and simulations used in this paper, the Friends of Friends (FOF) algorithm \citep[][]{Davis1985TheEvolution} is applied to identify groups of particles, which are called "FOF haloes" (or simply haloes) in 3D space. The gravitationally bound substructures of a FOF halo are called subhaloes and are detected using the \textsc{SUBFIND} algorithm \citep[][]{springel2001populating} in 3D space, enforcing a resolution condition of $\geq 20$ particles per subhalo. For each FOF halo, there can be only one central subhalo, which is usually the FOF halo's most massive substructure. The rest of the subhaloes of the FOF halo are classified as satellite subhaloes.

FOF haloes typically do not have a well-defined shape. Nevertheless, it is common to consider a halo boundary, $R_{200}$, the radius within which the density of the halo is 200 times the critical density of the Universe. The mass within $R_{200}$ is called $M_{200}$, and the circular velocity that is computed based on these values is $V_{200}$\footnote{$R_{200}$, $M_{200}$, and $V_{200}$ are often taken as the halo virial radius, mass, and velocity in the literature, though they are not exactly the same thing.}. These values are computed directly from the particle data of each simulation, and we use them in the initial part of our analysis where we do not perform any comparison with observations (section \ref{subsec: conformity_sims}). When comparing with observations, we compute these values based on the galaxy stellar mass, both in the models and in the observations, as explained in section \ref{subsubsec: methods_halo_mass_rad_vel}. Finally, we note that satellite galaxies of a FOF halo can reside both within and beyond $R_{200}$.

\subsubsection{Physical processes most relevant to this work}
\label{subsubsec: methods_relevant_processes}
The quenching of galaxies happens through a combination of several physical processes. In all models described above, supernova and black hole feedback make key contributions to controlling star formation in low-mass and massive galaxies, respectively.

More importantly for this study, in addition to intrinsic physical processes, galaxies are also subject to environmental effects such as tidal and ram-pressure stripping, which can influence their star formation dramatically. These processes happen naturally in hydrodynamical simulations such as TNG and EAGLE, although their efficiency may depend on the resolution of the simulation and hydrodynamics scheme. Environmental processes have been shown to influence galaxies out to several Megaparsecs from the centres of massive haloes in hydrodynamical simulations (e.g. see \citealt{Ayromlou2020Comparing}). In semi-analytic models such as \textsc{L-Galaxies}, the baryonic environmental effects need to be modelled explicitly. The main difference between the two versions of \textsc{L-Galaxies} used in this work, is their treatment of environmental processes, in particular, the ram-pressure stripping of hot gas in subhalos.

The most recent version of \textsc{L-Galaxies} \citep[][LGal-A21 hereafter]{Ayromlou2021Galaxy} uses the local background environment (LBE) measurements introduced by \cite{ayromlou2019new} to implement ram-pressure stripping, locally and uniformly, for all galaxies in the simulation. These LBE measurements are made by directly employing the particle data of the underlying dark matter-only simulation on which \textsc{L-Galaxies} is run. The measurements include the local density of the environment of each galaxy as  well as the velocity of the galaxy relative to its local environment. In contrast, the \cite{henriques2020galaxies} version of the model (LGal-H20 hereafter) resolves ram-pressure stripping only for satellites within $R_{200}$ of massive clusters ($M_{200}>5\times10^{14}{\rm M_\odot}$). This
model assumes an average intracluster medium (ICM) isothermal density profile ($\rho_{\rm ICM}\propto r^{-2}$) and takes the velocity of the galaxy relative to its environment to be the galaxy's host halo virial velocity, $V_{200}$. Both LGal-A21 and LGal-H20 include tidal stripping of the hot gas for satellite galaxies, with LGal-H20 limiting tidal stripping to satellites within the halo virial radius while LGal-A21 extends 
tidal stripping to all satellites of FOF haloes, including satellites at radii beyond $R_{200}$. Neither of the two models  include stripping of the cold gas in the interstellar medium (ISM).

LGal-A21 predicts that galaxies lose hot gas due to ram-pressure stripping not only within the virial radii of the massive haloes, but also in the outskirts of haloes, when moving through the warm–hot intergalactic medium (WHIM). This results in a higher fraction of quenched galaxies around intermediate-mass and massive haloes
that extends out to several Megaparsecs from the halo centre. These trends are in better agreement with observations than LGal-H20, which fails to reproduce the observed trends at large halocentric radii (see \citealt[][]{Ayromlou2021Galaxy}).

\subsection{Observational data}
\label{subsec: obs_data}
\subsubsection{SDSS galaxy catalogue with spectroscopic redshifts}
\label{subsubsec: methods_SDSS}
We use the publicly available MPA-JHU galaxy catalogues that were released as a part of data release 7 (DR7) of the Sloan Digital Sky Survey (SDSS, \citealt{Abazajian2009SDSS})\footnote{\href{https://www.mpa-garching.mpg.de/SDSS/DR7/}{https://www.mpa-garching.mpg.de/SDSS/DR7/}}. The stellar masses and star formation rates were estimated employing the methodologies described in \citet{Kauffmann2003Stellar} \citep[see also][]{Brinchmann2004Physical, Salim2007UV}. Briefly, the stellar masses were derived using model fits to the spectrophotometry of galaxies, adopting the \citet{bruzal03} stellar population models and the \citet{kroupa01} initial mass function (IMF). The star formation rates were measured with scaling relations connecting star formation rates with the emission line properties of galaxies. The star formation rates were also corrected for aperture size of the SDSS fibres \citep[see][for more details]{Brinchmann2004Physical}. 

In this paper, we take galaxies with $\log_{10}(M_{\star}/{\rm M_{\odot}})>9.5$ within the spectroscopic redshift interval $0 < z \leq 0.04$. The statistics of the galaxy sample are noted in Table \ref{tab: data_stat}.

\subsubsection{DESI photo$-z$ galaxy catalogue}
\label{subsubsec: methods_DESI}
We also use the photo$-z$ galaxies detected in the legacy imaging survey \citep{dey19} of the Dark Energy Spectroscopic Instrument (DESI). Recently, \cite{zou19} constructed an extensive photo$-z$ galaxy catalogue\footnote{\href{http://cdsarc.u-strasbg.fr/viz-bin/cat/J/ApJS/242/8}{http://cdsarc.u-strasbg.fr/viz-bin/cat/J/ApJS/242/8}} based on DR7 of the legacy imaging survey, which contains $\sim 3\times 10^8$ galaxies with $0<z_{\rm photo}<1$. The photometric redshifts are estimated based on a linear regression method described in \cite{beck16}. The physical properties of galaxies such as stellar masses and star formation rates are measured with the Le Phare code \citep{ilbert09} which adopts the \cite{bruzal03} stellar population models and the \cite{chabrier03} IMF. The completeness of the catalogue is $\gtrsim 90\%$ for galaxies with an $r-$ band magnitude of $r<23$ \citep{zou19}, which is also the magnitude cut for galaxies in our sample.

In this paper, we take a sample of galaxies with $\log_{10}(M_{\star}/{\rm M_{\odot}})\geq9.5$ within the photometric redshift interval $0\leq \,z_{\rm photo}\leq\,0.1$. We compile the sample statistics in Table \ref{tab: data_stat}.

\subsubsection{The applications of SDSS and DESI galaxy catalogues}
The SDSS galaxy data were those initially used to introduce the concept of galactic conformity on small \citep{weinmann2006properties} and large scales \citep{kauffmann2013re}. The precise spectroscopic redshifts of SDSS galaxies enable us to make robust measurements of the conformity signal at $z\sim 0$. On the other hand, DESI provides us with much better statistics. Therefore, in addition to investigating the conformity signal in SDSS, we carry out the first analysis of galactic conformity with the DESI data and compare it to results for mock galaxy catalogues. Nevertheless, we note that the downside of our analysis using the DESI data is the comparatively large uncertainty in the photometric redshifts of galaxies ($\sigma_{\Delta z} \sim 0.017$ \footnote{Defined as the standard deviation of $\Delta z_{\rm norm}=\frac{z_{\rm phot}-z_{\rm spec}}{1+z_{\rm spec}}$.}, see \citealt{zou19}). Due to these uncertainties, we have to project DESI galaxies in a thicker slice of redshift/line-of-sight velocity compared to SDSS. Although this influences the amplitude of the signal in DESI compared to SDSS, it should not affect our scientific conclusions.

\subsection{Estimating the halo mass, radius, and velocity}
\label{subsubsec: methods_halo_mass_rad_vel}
The properties of haloes, such as mass and radius, cannot be observed directly and must be estimated from observables such as the galaxy stellar mass.
Following \cite{Ayromlou2021Galaxy}, we estimate the halo $M_{200}$, $R_{200}$, and $V_{200}$ using LGal-A21 using the central galaxy stellar to halo mass relation, by fitting to simulation results using a power law relation of the form:
\begin{equation}
\label{eq: Mstellar_to_M200_conversion}
\log_{10} (M_{200}/{\rm M_{\odot}}) = \alpha_1  \log_{10} (M_{\star}/{\rm M_{\odot}}) + \beta_1.
\end{equation}
Here, $\alpha_1$ and $\beta_1$ are free parameters that are calibrated against the simulation results: $\alpha_1 = 1.65$, $\beta_1 = -5.16$ for $\log_{10}(M_{\star}/{\rm M_{\odot}})\geq10.5$, and $\alpha_1 = 0.80$, $\beta_1 = 3.70$ for $\log_{10}(M_{\star}/{\rm M_{\odot}})<10.5$.
With the above estimated $M_{200}$, we calculate $R_{200}$ and $V_{200}$:
\begin{equation}
    \label{eq: R200_M200_conversion}
    R_{200} = \left( \frac{3M_{200}}{4\pi \times 200\rho_{\rm crit}} \right)^{1/3},
\end{equation}
\begin{equation}
    \label{eq: V200}
    V_{\rm 200,halo} = \sqrt{GM_{\rm 200,halo}/R_{\rm 200,halo}}.
\end{equation}
Note that to estimate the properties of a halo, the stellar mass of its central galaxy is the only quantity that we use. In the simulations, halo mass scatters around the prediction of this power law relation with an {\it rms} of 0.3 dex. For all our comparisons with the SDSS and DESI observations in this paper, we estimate the halo properties of both simulations and observations from the stellar mass of the central galaxy using equations \ref{eq: Mstellar_to_M200_conversion} to \ref{eq: V200}.

%%%%%%%%%%%%%%%%%%%% METHODS: CenSat %%%%%%%%%%%%%%%%%%
\subsection{\textsc{CenSat}: An algorithm to identify central and satellite galaxies/subhaloes}
\label{subsec: censat}
\begin{figure}
    \centering
    \includegraphics[width=1\columnwidth]{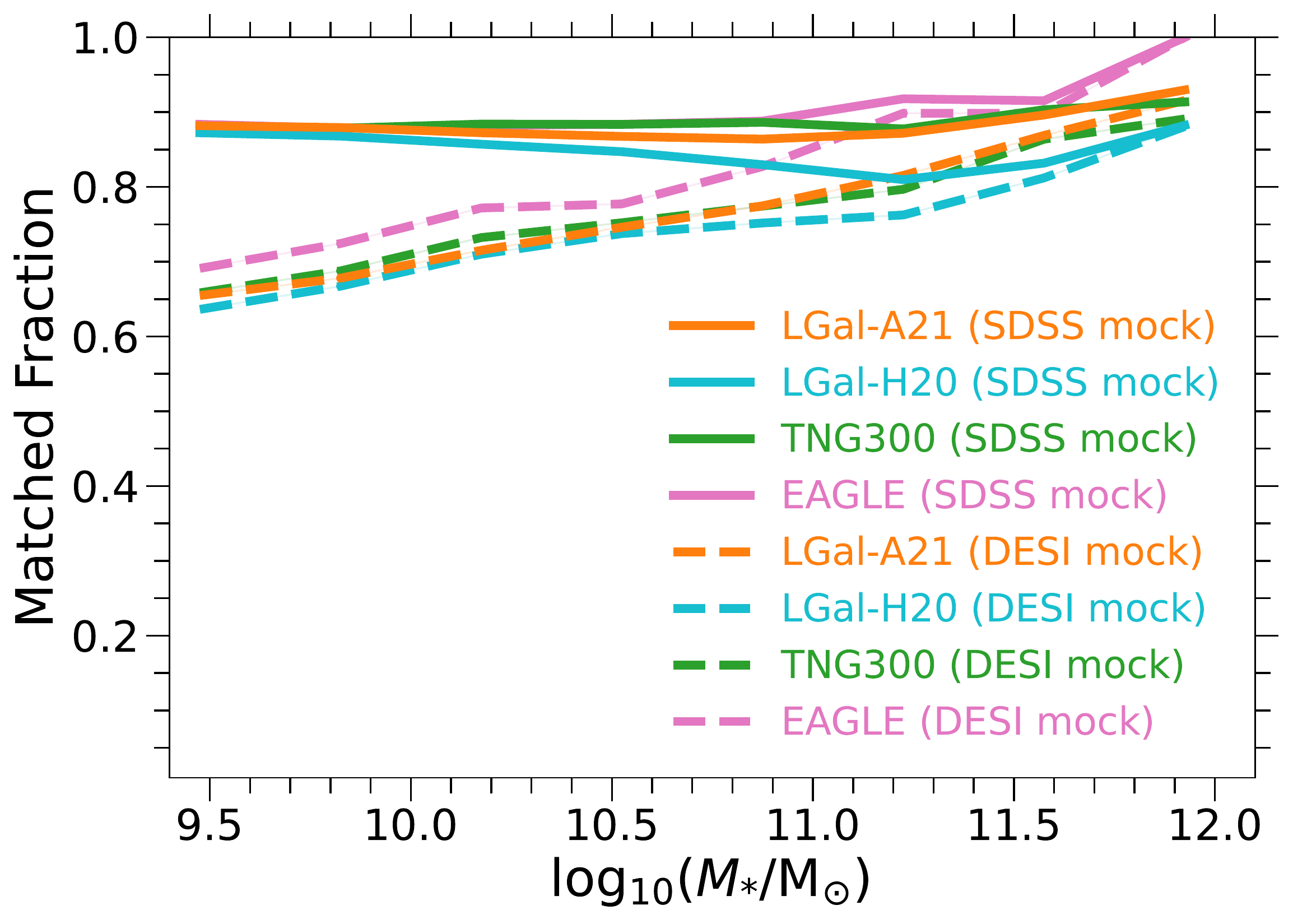}
    \caption{The fraction of  galaxies where the status (central or satellite) is the same
    for \textsc{CenSat} (performed in 2D projected space) and for \textsc{Subfind} (performed in 3D space).
    Results are shown for different galaxy formation models and simulations. The solid lines correspond to the \textsc{CenSat}'s calibration using observations with spectroscopic redshifts (SDSS in this work), for which  $\alpha_{\rm R} = 1.5$ and $\alpha_{\rm V} = 2$ (see section \ref{subsubsec: methods_CenSat_SDSS}). The matched fraction is nearly $90\%$ for this adjustment. The dashed lines show the second adjustment for observations with photometric redshift (DESI in this work, section \ref{subsubsec: methods_CenSat_DESI}).}
\label{Fig: Matched_Fraction2}
\end{figure}

We devise a novel algorithm, \textsc{CenSat} (\textsc{Cen}tral and \textsc{Sat}ellite identifier), to classify galaxies as either central or satellite systems within their groups. The classifier operates in two spatial and one velocity/redshift dimensions. \textsc{CenSat} is based on an \textit{adaptive isolation criterion} and can be applied both to simulations and to observations uniformly. In the following, we explain how the algorithm works in six steps:
\begin{enumerate}
    \item Sort the galaxies in the sample based on their stellar masses in descending order.
    \item Assign the "status" of all galaxies to "unknown".
    \item Assign the "status" of the most massive galaxy in the sample to "Central".
    \item Find all galaxies with $0<r_{\rm proj}\leq R_{\rm c}$ and $|\Delta v_{\rm LOS}|\,\leq V_{c}$ and assign their "status" to "Satellite".
    \item Exclude all galaxies that are "Central" or "Satellite" from the sample, i.e. only keep galaxies with "status" = "unknown".
    \item If there is any galaxy with "status" = "unknown" go to step (iii). Else, the operation is over and every galaxy in the sample is labelled as either "Central" or "Satellite".
\end{enumerate}

Here, for each galaxy in the vicinity of a central galaxy, $r_{\rm proj}$ is its projected distance to the central galaxy and $|\Delta v_{\rm LOS}|$ is its line of sight velocity relative to the line of sight velocity of the central galaxy. Additionally, $R_{\rm c}$ and $V_{\rm c}$ are critical values that should be interpreted as the halo boundary and line of sight projection depth, respectively. Later in this subsection, we will introduce two different choices for $R_{\rm c}$ and $V_{\rm c}$, which correspond to observations with spectroscopic (in this work SDSS) and photometric (in this work DESI) redshifts and their respective mock galaxy catalogues (see section \ref{subsubsec: methods_CenSat_SDSS} and \ref{subsubsec: methods_CenSat_DESI}).

The \textsc{CenSat} algorithm has several advantages. First of all, no galaxy remains unclassified as either "Central" or  "Satellite". Moreover, it is applicable both to simulations and to observations. Most importantly, \textsc{CenSat}  is calibrated to match the output of \textsc{Subfind} (see sections \ref{subsubsec: methods_CenSat_SDSS} and \ref{subsubsec: methods_CenSat_DESI}). This is particularly important because the conformity signal strongly depends on the star formation rates of central galaxies and misclassification of satellites as centrals would result in misleading values for the conformity signal. Finally, by
definition, more massive galaxies cannot be satellites of less massive galaxies. In other words, a central detected by \textsc{CenSat} is always more massive than its satellites (see the first two steps of the algorithm).

\subsubsection{Running \textsc{CenSat} on SDSS data and its mock catalogues}
\label{subsubsec: methods_CenSat_SDSS}
For data such as SDSS, with precise spectroscopic redshifts, and its respective mock galaxy catalogues, we take $R_{\rm c}$ and $V_{\rm c}$ as a function of the properties of the central galaxy's halo:
\begin{equation}
\begin{aligned}
R_{\rm c} = \alpha_{\rm R}\, R_{\rm 200,cen},\\
V_{\rm c} = \alpha_{\rm V}\, V_{\rm 200,cen},
\end{aligned}
\end{equation}
where $R_{\rm 200,cen}$ and $V_{\rm 200,cen}$ are derived directly from the galaxy stellar mass as described in Section \ref{subsubsec: methods_halo_mass_rad_vel}. Note that the isolation criterion  used here is adaptive and depends only on the galaxy stellar mass. In order to calibrate \textsc{CenSat}, we vary $\alpha_{\rm R}$ and $\alpha_{\rm V}$, which are two adjustable parameters. For LGal-A21, we compare the status of galaxies as estimated in redshift space by \textsc{CenSat} against the status of the same galaxies as identified by \textsc{SUBFIND}. If a galaxy has the same status in the two cases, we call it a match. We define the best values of $\alpha_{\rm R}$ and $\alpha_{\rm V}$ as those which output the highest matched fraction of galaxies with $\log_{10}(M_{\star}/{\rm M_{\odot}})>9.5$ between \textsc{CenSat} and \textsc{Subfind}. Locating the highest matched fraction is done via a grid search, which results in $\alpha_{\rm R} = 1.5$ and $\alpha_{\rm V} = 2$.

With these parameters, we apply \textsc{CenSat} both to the simulations and to the SDSS observations employed in this study. Fig. \ref{Fig: Matched_Fraction2} shows the fraction of galaxies where the status is successfully matched. The solid lines correspond to the matched fraction of SDSS mock catalogues (see section \ref{subsec: methods_mock_catalogues}). A "match" happens only when the \textsc{CenSat} status of a galaxy is the same as that of the original halo finder algorithm, \textsc{Subfind}. For most stellar mass ranges, the matched fraction is nearly $90\%$ in all the simulations. The remaining $10\%$ difference is mainly caused by projection effects.

\subsubsection{Running CenSat on DESI data and its mock catalogues}
\label{subsubsec: methods_CenSat_DESI}
Although DESI provides us with promising statistics, it does not include spectroscopic redshifts for most galaxies, and only photometric redshifts are reported, with typically large uncertainties (see section \ref{subsubsec: methods_DESI}). To maximise the functionality of \textsc{CenSat} on such data, we take a constant $V_{\rm c} = 6000\, {\rm km\,s^{-1}}$ ($\Delta z \sim 0.02$)  for all central galaxies, regardless of their stellar mass. This ensures that our projection depth is comparable to the typical photometric redshift error of the sample. For $R_{\rm c}$, we simply take the adaptive value that we have calibrated before (see section \ref{subsubsec: methods_CenSat_SDSS}). The matched fraction for this adjustment of \textsc{CenSat} is shown in Fig. \ref{Fig: Matched_Fraction2}, where the dashed lines illustrate the matched fraction of the galaxies from DESI mock catalogues (see section \ref{subsec: methods_mock_catalogues}). In this adjustment, the agreement with \textsc{Subfind} is weaker than the previous adjustment (solid lines in Fig. \ref{Fig: Matched_Fraction2}), although it is still acceptable. This discrepancy is inevitable and is mainly caused by a deeper projection as well as high uncertainties in DESI's photometric redshifts.

We note that small changes to $R_{\rm c}$ and/or $V_{\rm c}$ do not change our results significantly. Nevertheless, applying a strict cut on redshift separation (smaller than the typical photo$-z$ error), would lead to mis-identification of central galaxies. Moreover, a very deep projection (e.g. projecting the whole sample) not only results in a lower matched fraction, but also makes the conformity signal almost completely vanish. That is because too many foreground/background galaxies will be contributing to the estimate of the quenched fraction of galaxies in the vicinity of each central.

\subsection{Mock catalogues}
\label{subsec: methods_mock_catalogues}
In order to make a fair comparison between models and observations, we create mock catalogues from the galaxy formation simulations used in this work. We undertake the following steps to generate the mock catalogues:
\begin{enumerate}
    \item Transform the positions and peculiar velocities of the simulated galaxies into redshift/velocity space to mimic observations. 
    \item Add an error equal to the observation error to the redshifts/velocities to reproduce the uncertainties in the observed redshifts of galaxies.
    \item Find $M_{200}$, $R_{200}$, and $V_{200}$ for each galaxy based on its stellar mass, as described in section \ref{subsubsec: methods_halo_mass_rad_vel}.
    \item Classify galaxies into central/satellite using \textsc{CenSat} with  model parameters calibrated for the particular observations the models will be compared with (see section \ref{subsubsec: methods_CenSat_SDSS} for SDSS and section \ref{subsubsec: methods_CenSat_DESI} for DESI).
    \item For each central galaxy, find its neighbours out to 10 Mpc from the halo centre.
\end{enumerate}

We note that the error mentioned in step "ii" is considerable for our DESI mock catalogues due to the uncertainties in DESI's photometric redshifts, but is set to zero for SDSS mocks due to negligible error of SDSS spectroscopic redshifts. To add the error consistently, we first bin the observed galaxies based on their stellar masses and take the mean and standard deviation of the redshift errors associated with galaxies in each stellar mass bin. A random Gaussian error with the same mean and standard deviation as the observed galaxies is then added to the simulated galaxies in each stellar mass bin. The generated mock catalogues can now be consistently compared with the SDSS and DESI observations.

The remaining caveats, which concern our comparison with DESI observations, are listed below.
\begin{itemize}
    \item The photometric redshifts of DESI galaxies results in considerable uncertainties in their line of sight velocities as well as both their distances to us and their projected distances to each other. Although by adding a random error (step two of making mock catalogues) we mimic a redshift and line of sight velocity uncertainty similar to DESI in the simulations, we do not attempt to change the projected distances between galaxies based on these uncertainties.
    \item The projection depth we employ for DESI and its mock catalogues, $V_{\rm c} = 6000\, {\rm km\,s^{-1}}$ ($\Delta z \sim 0.02$), corresponds to $\sim 90 \,\rm Mpc$ at $z\sim0$. Adding a random error to the redshifts in the mock catalogues (step two of making mock catalogues) could also add a similar value, which sums up to on average $\sim 200 \, \rm Mpc$ spatial projection depth. Such a projection is consistent with the large box size of LGal-A21 and LGal-H20 ($\sim \rm 714 \, Mpc$). On the other hand, it is comparable to TNG300's box size ($\sim 300 \, \rm Mpc$), although its considerable scatter, coming from DESI's redshift uncertainties, could exceed the box size. Note that the projection depth significantly exceeds EAGLE's box size ($\sim 100 \, \rm Mpc$). This may influence the DESI mocks of TNG and EAGLE.
\end{itemize}

%%%%%%%%%%%%%%%%%%%% RESULTS %%%%%%%%%%%%%%%%%%

\section{Results: Galactic Conformity}
\label{sec: conformity}
In this section we analyse the conformity signal both in the simulations and in the observations. Initially, we focus on simulation outputs. We then investigate the conformity signal in the SDSS and DESI observations, comparing it with mock galaxy catalogues. We also study the influence of backsplash and fly-by primary galaxies, as well as central and satellite secondary galaxies, on the signal.
\label{sec: Results}
\subsection{Conformity in simulations}
\label{subsec: conformity_sims}
\begin{figure*}
    \centering
    \includegraphics[width=0.33\textwidth]{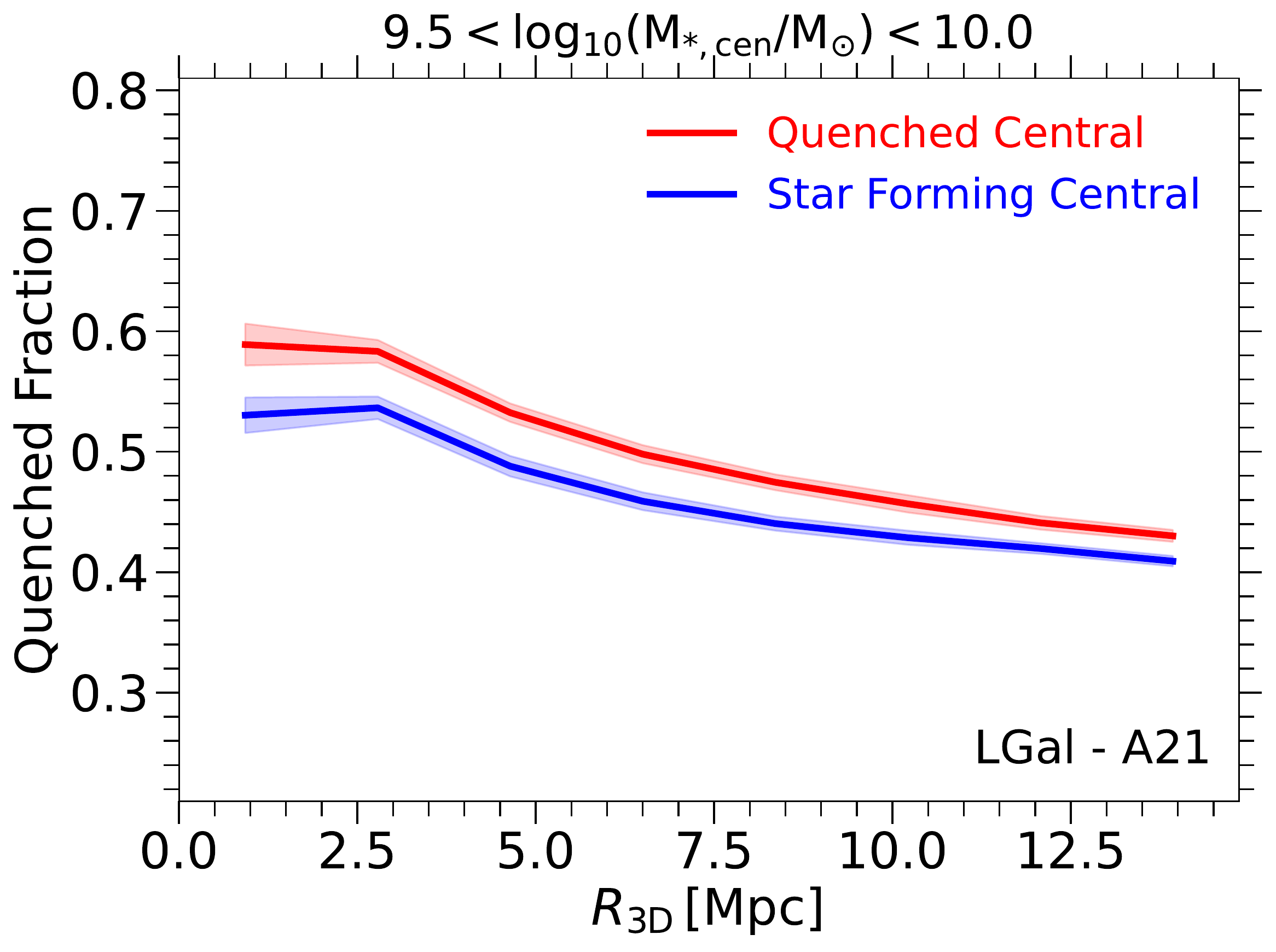}
    \includegraphics[width=0.33\textwidth]{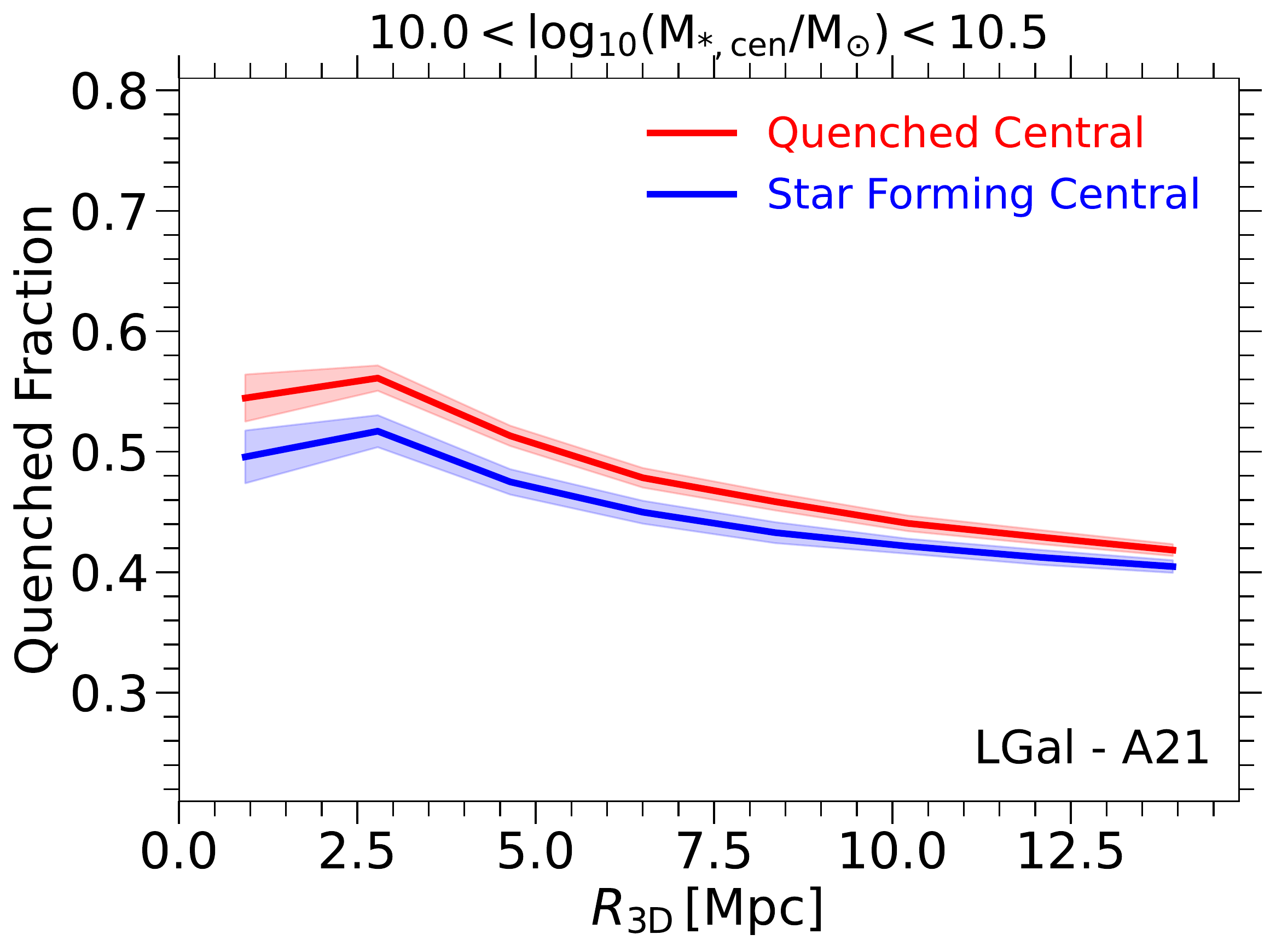}
    \includegraphics[width=0.33\textwidth]{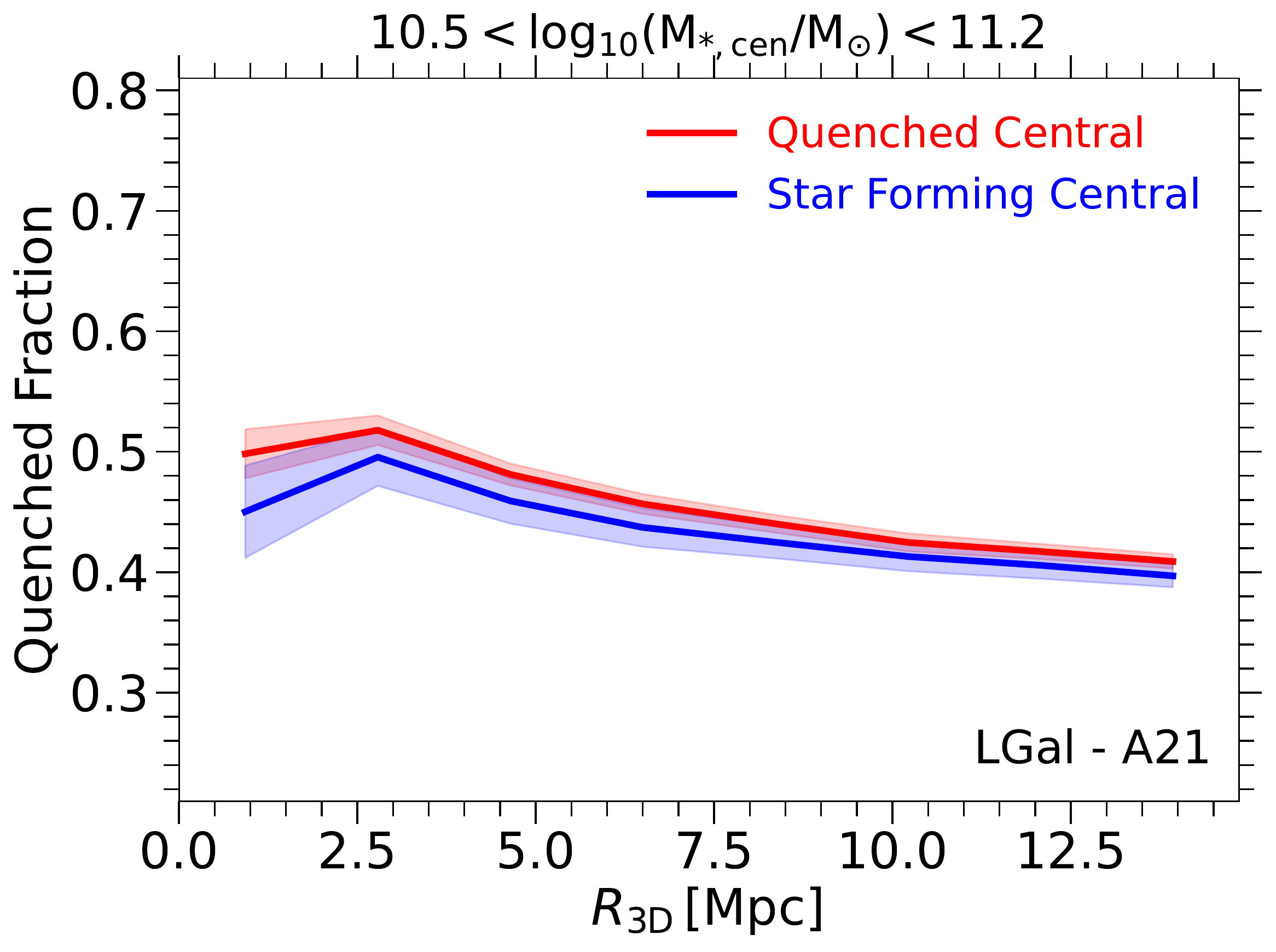}
    \includegraphics[width=0.33\textwidth]{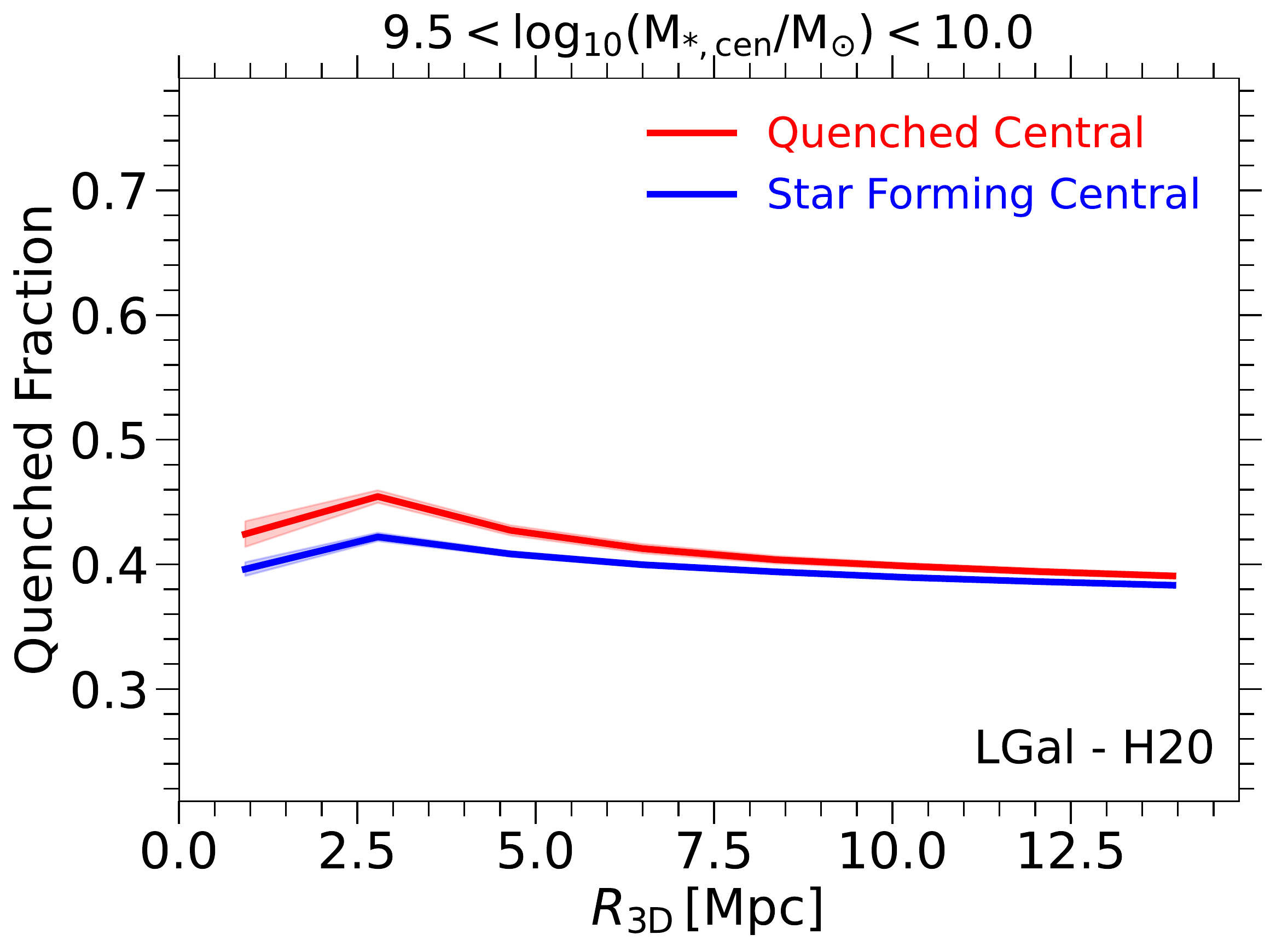}
    \includegraphics[width=0.33\textwidth]{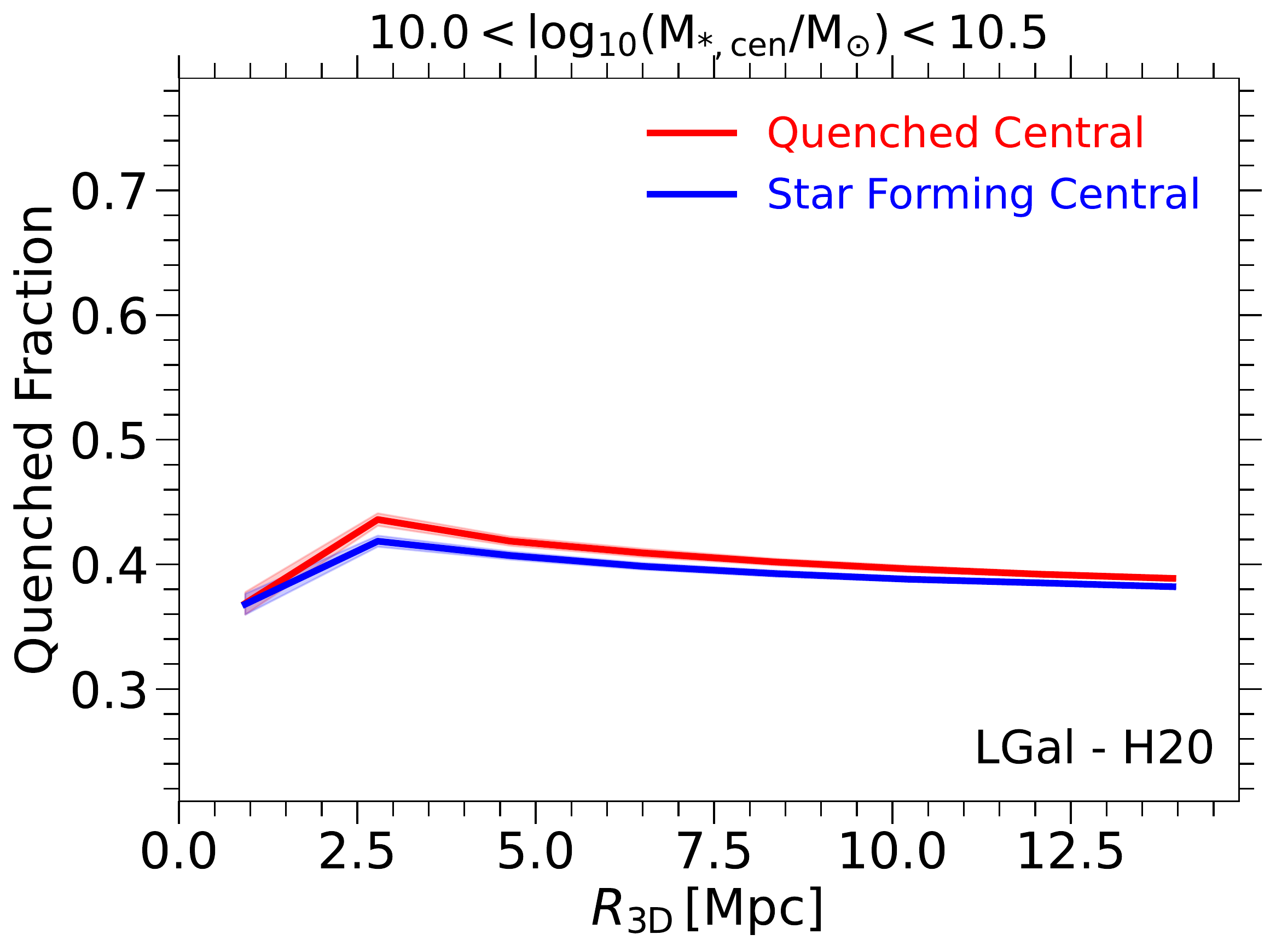}
    \includegraphics[width=0.33\textwidth]{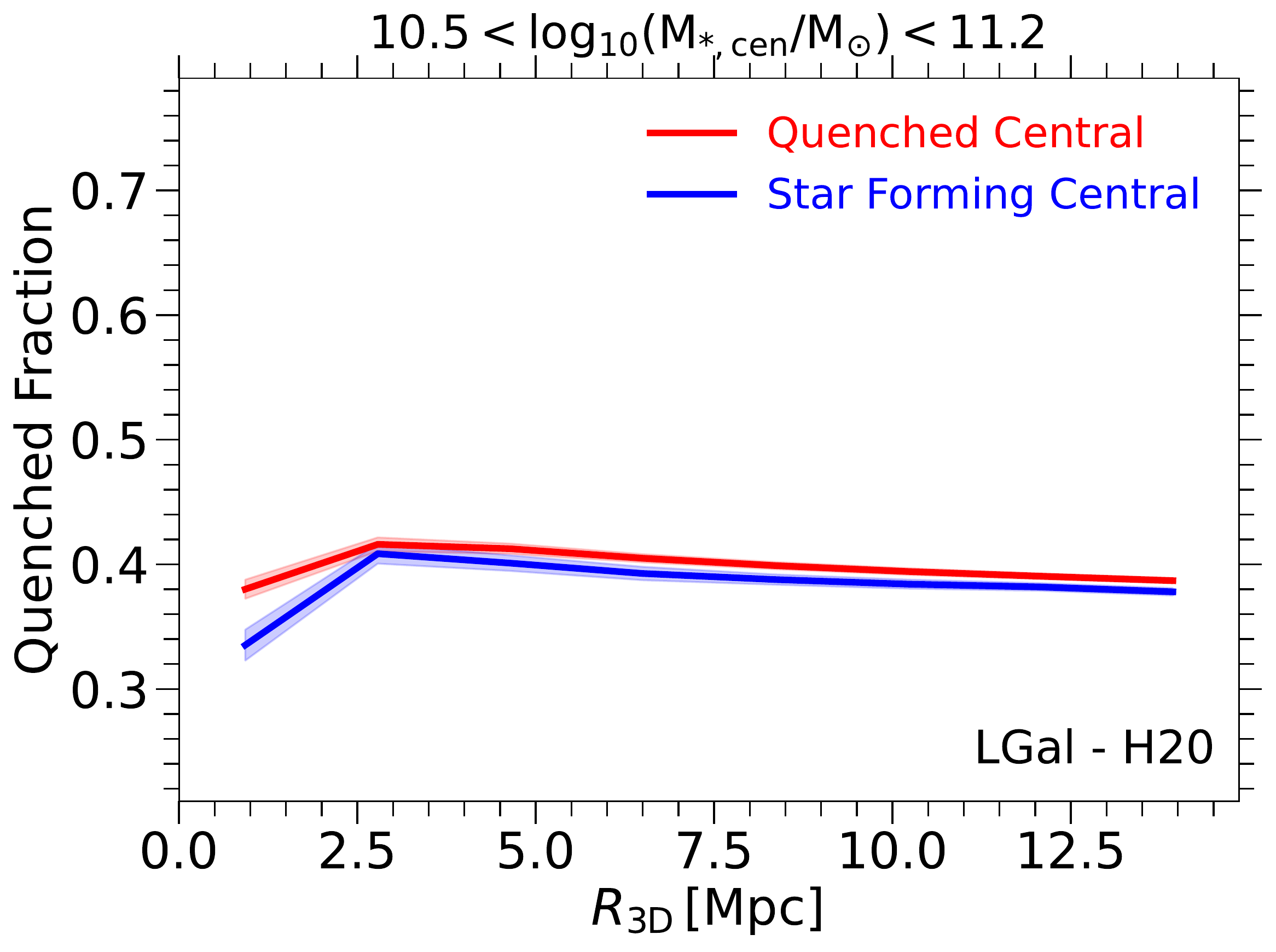}
    \includegraphics[width=0.33\textwidth]{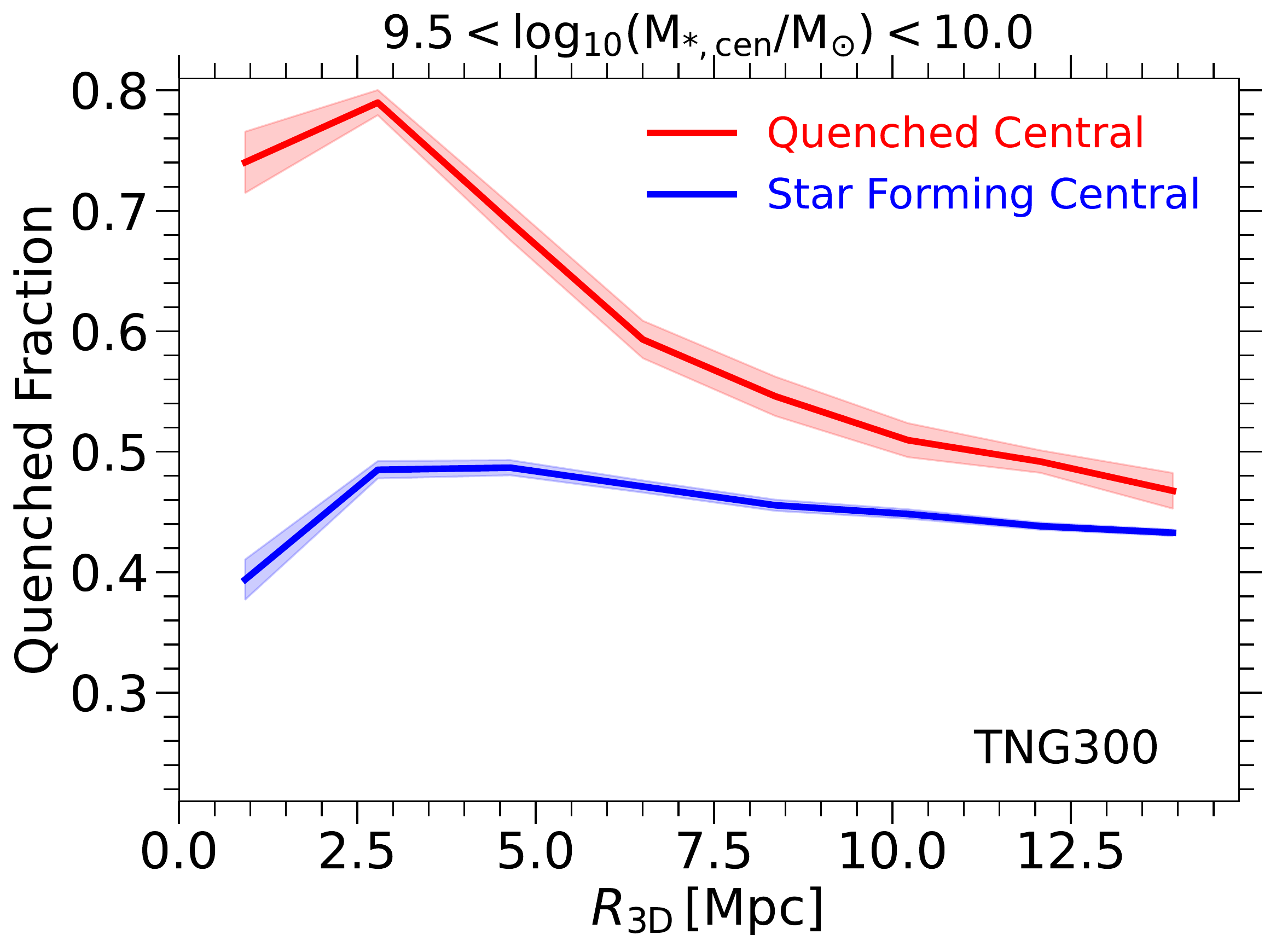}
    \includegraphics[width=0.33\textwidth]{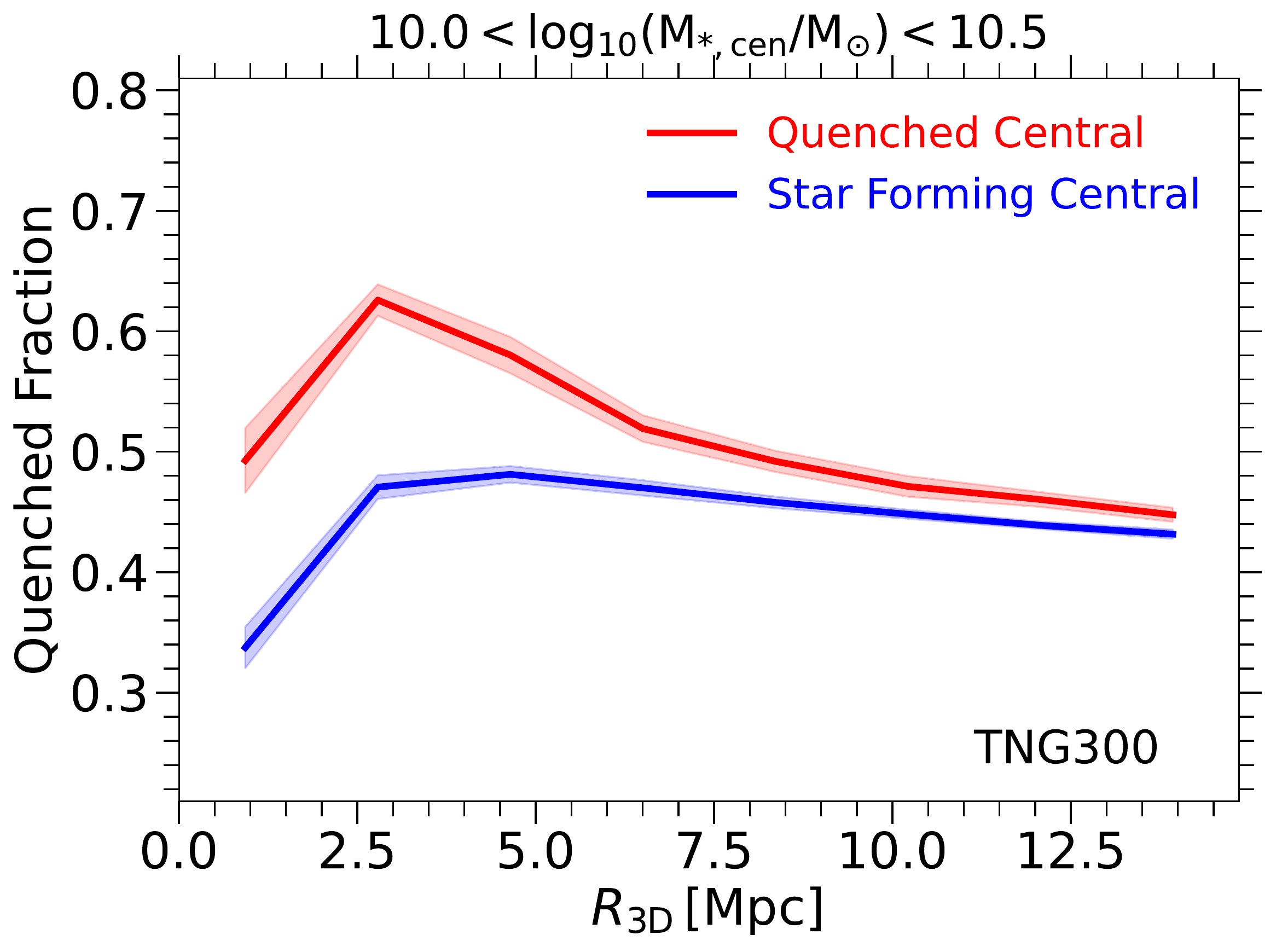}
    \includegraphics[width=0.33\textwidth]{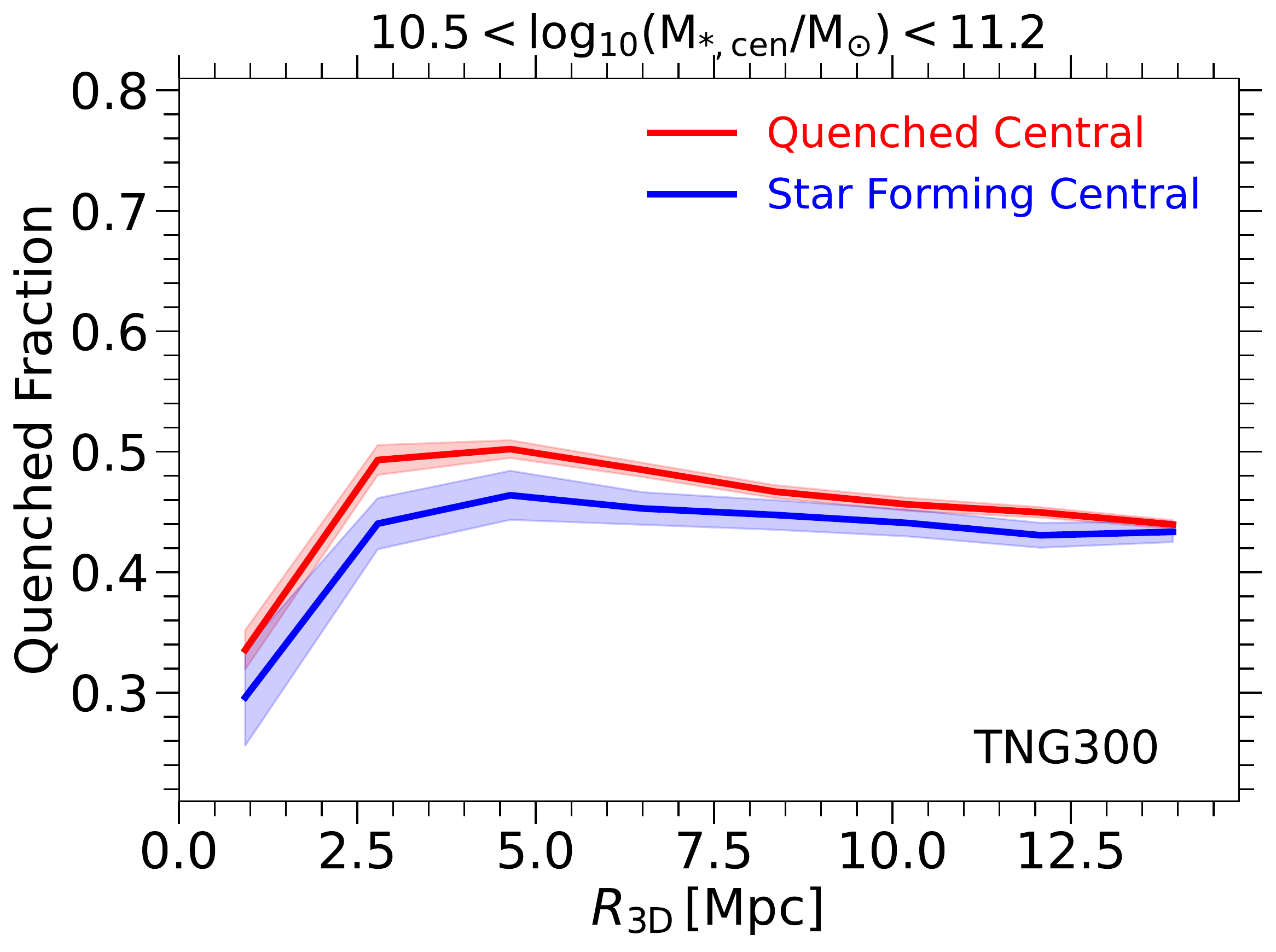}
    \includegraphics[width=0.33\textwidth]{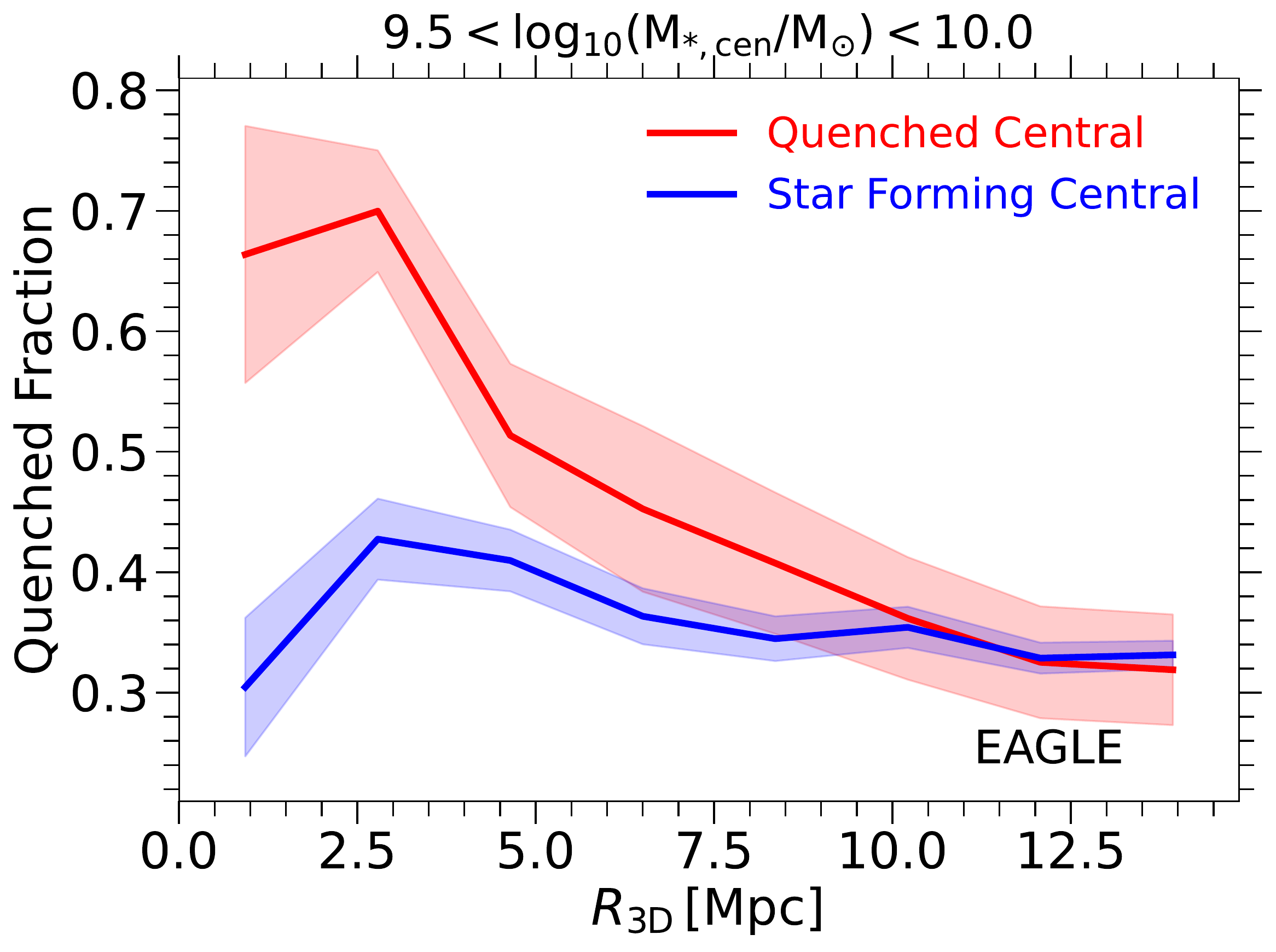}
    \includegraphics[width=0.33\textwidth]{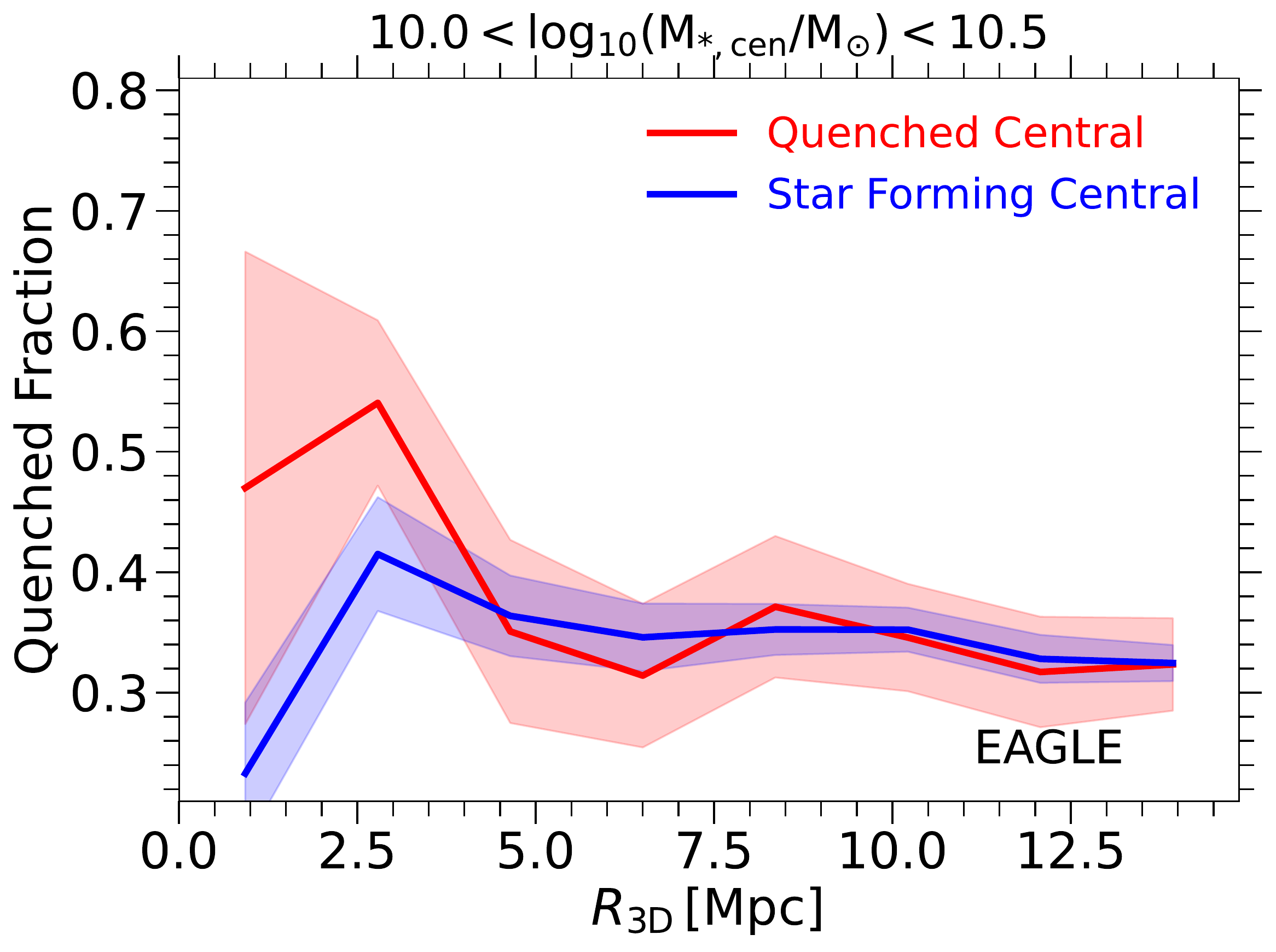}
    \includegraphics[width=0.33\textwidth]{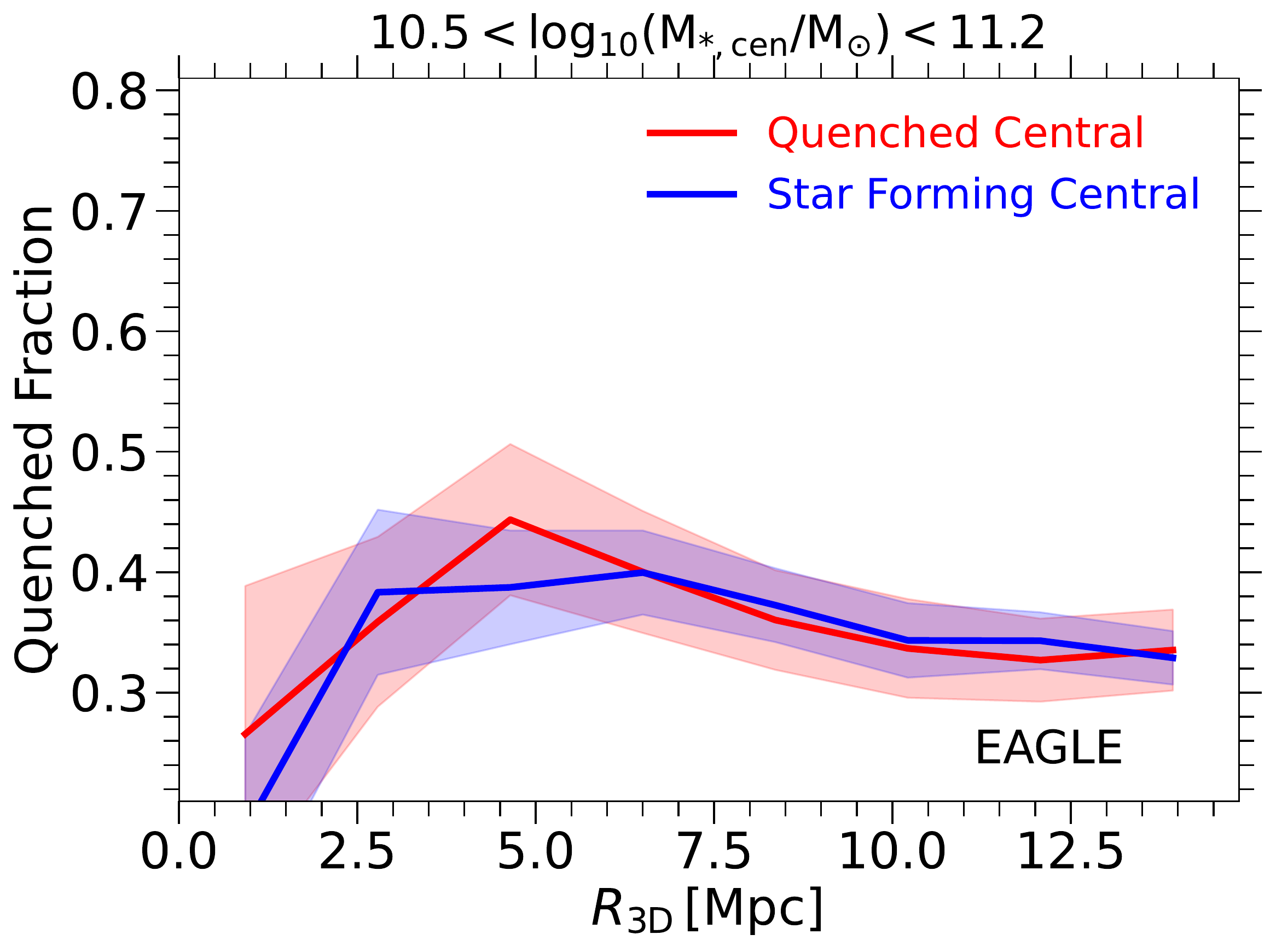}
    \caption{The fraction of quenched galaxies as a function of 3D halocentric distance in the LGal-A21 (top row), LGal-H20 (second row), TNG (third row), and EAGLE (bottom row) models at $z=0$. Red and blue lines correspond to galaxies in the vicinity of quenched and star-forming primary galaxies, respectively. The shaded regions show the uncertainty of the results, derived using the bootstrap method. Different columns show the primary galaxies' stellar masses. The galactic conformity signal (the difference between the red and blue lines) is seen in the LGal-A21, TNG, and EAGLE models out to several Megaparsecs, but is absent in LGal-H20.}
\label{Fig: Conformity_quenching_Mpc_sim}
\end{figure*}
In this subsection, we take the direct output of simulations to investigate galactic conformity. 
Rather than running \textsc{CenSat}, we adopt the satellite/central 
classifications as given by the simulation's halo finder algorithm. This is done to
investigate the strength of the signal without dilution  by projection effects and mis-classification of satellites as centrals (and vice versa).

\begin{figure*}
    \centering
    % {\LARGE Redshift = 0}\par\medskip
    \includegraphics[width=0.33\textwidth]{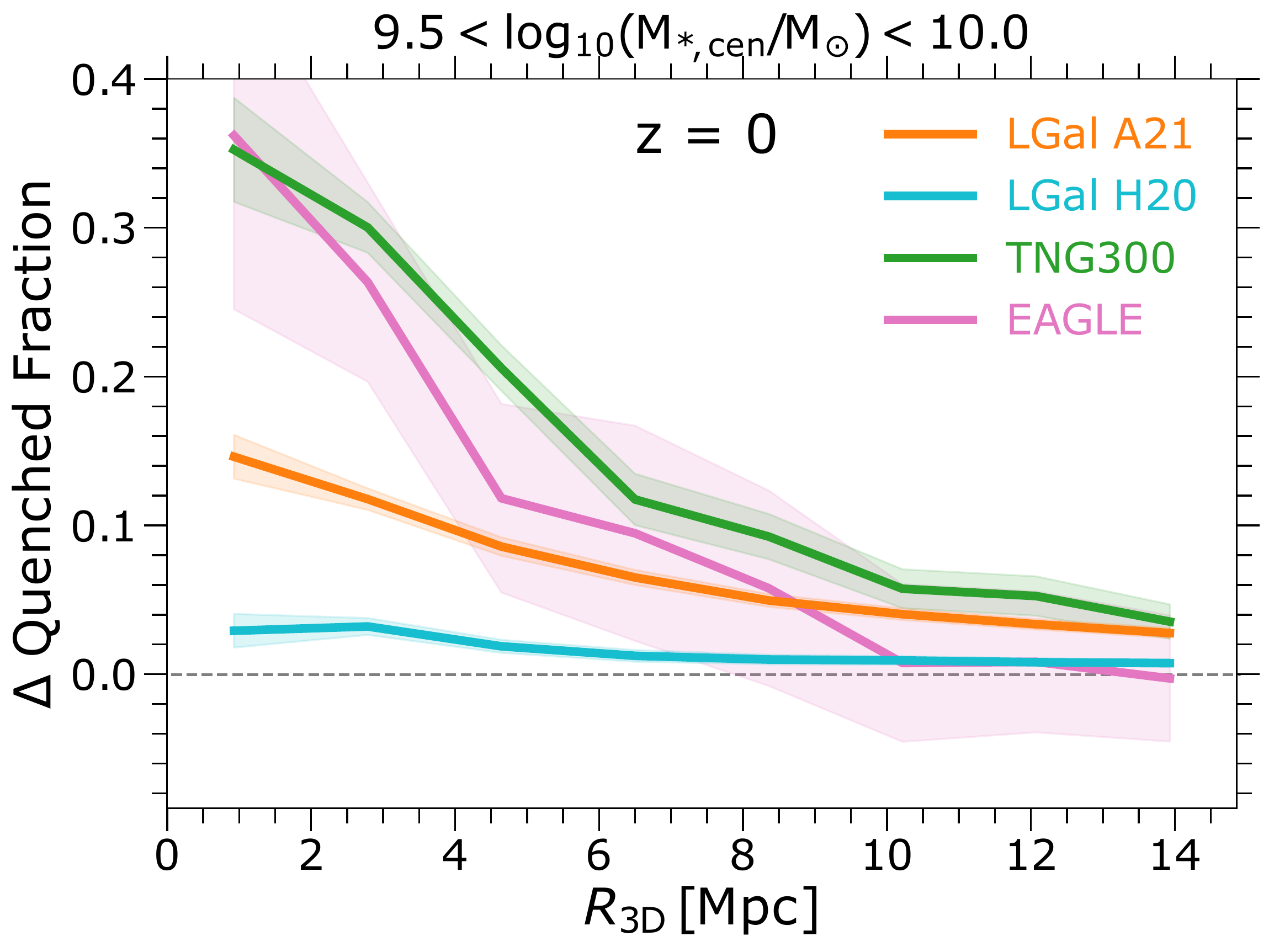}
    \includegraphics[width=0.33\textwidth]{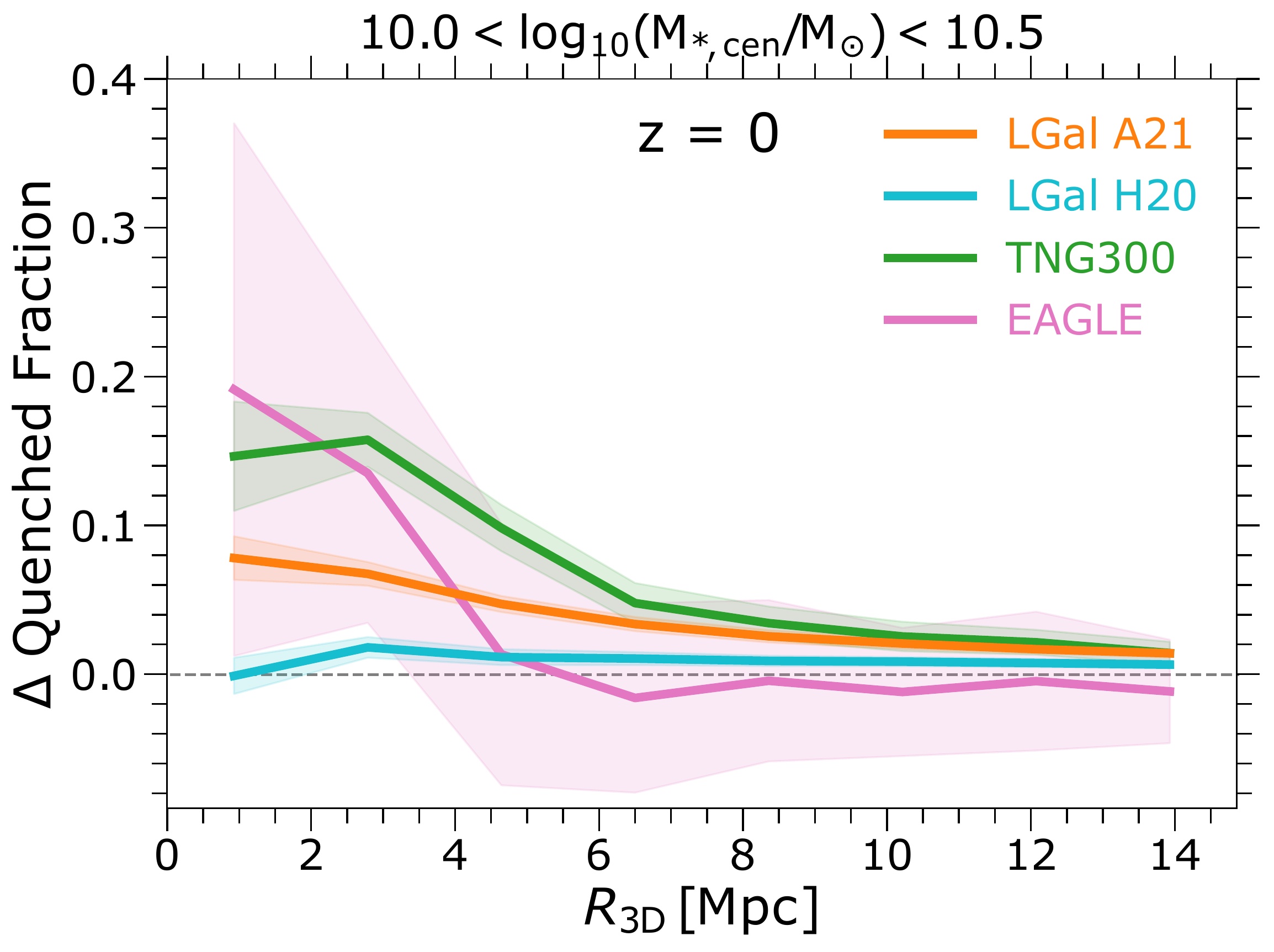}
    \includegraphics[width=0.33\textwidth]{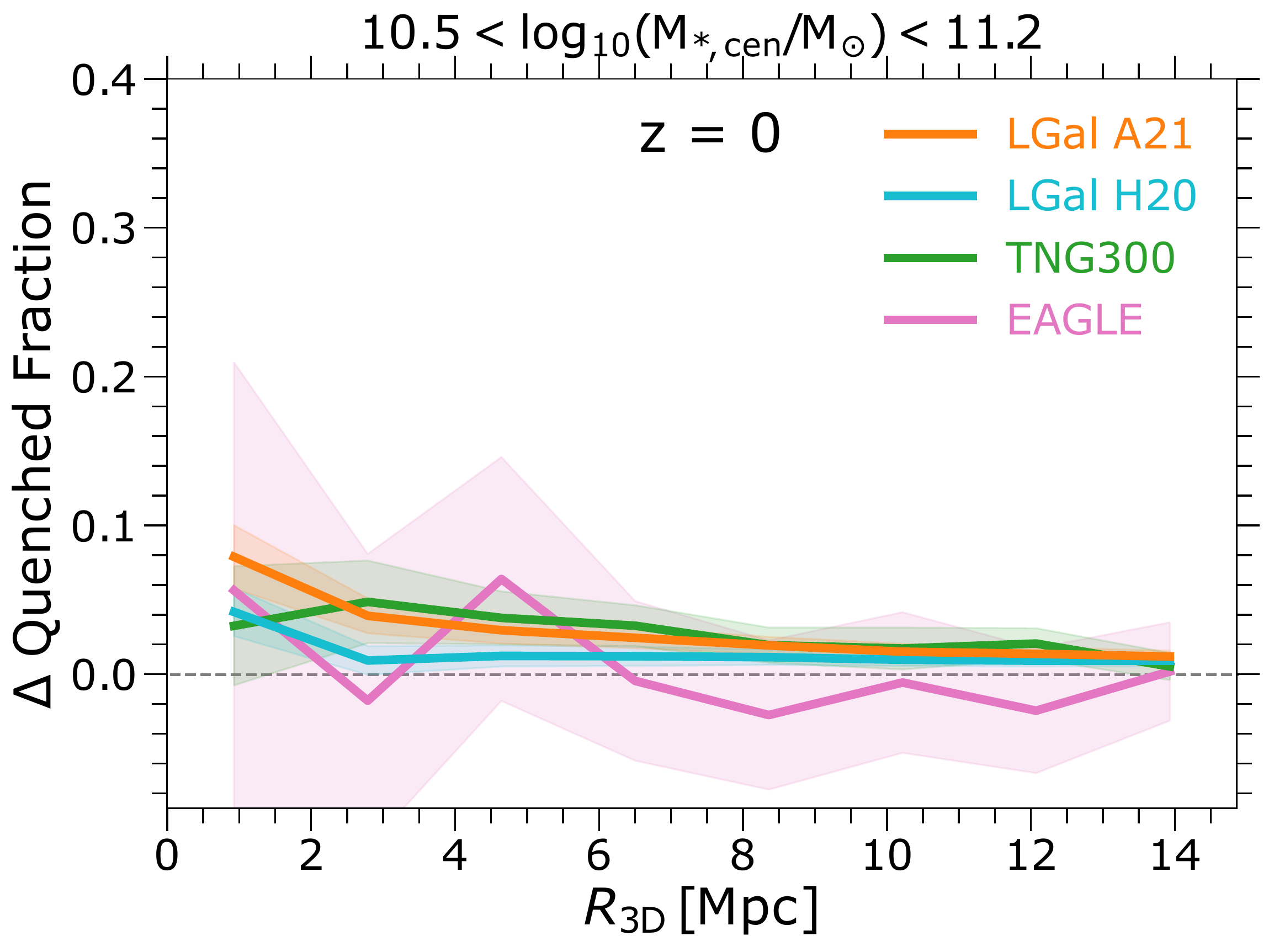}
    
    % {\LARGE Redshift = 1}\par\medskip
    \includegraphics[width=0.33\textwidth]{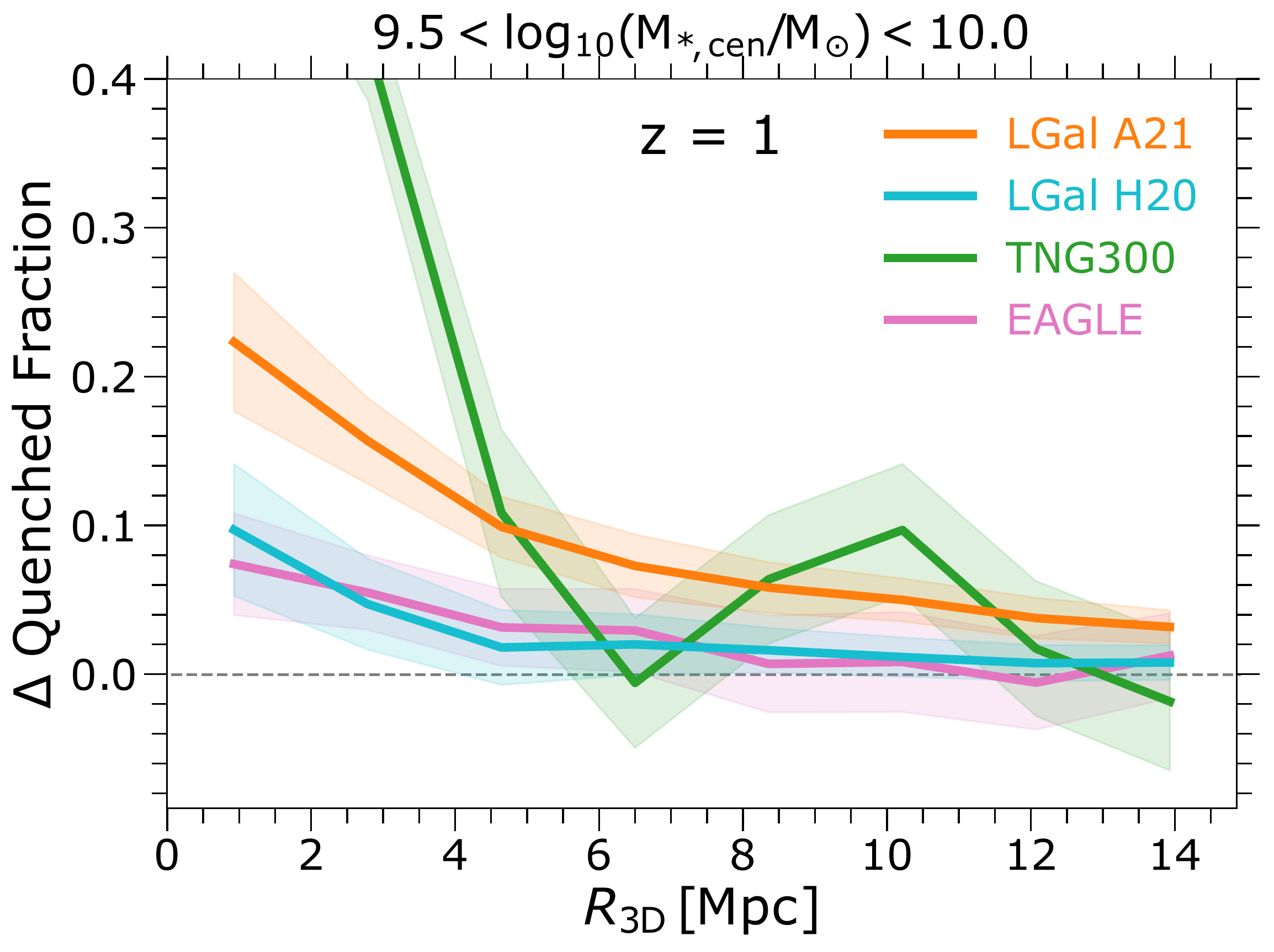}
    \includegraphics[width=0.33\textwidth]{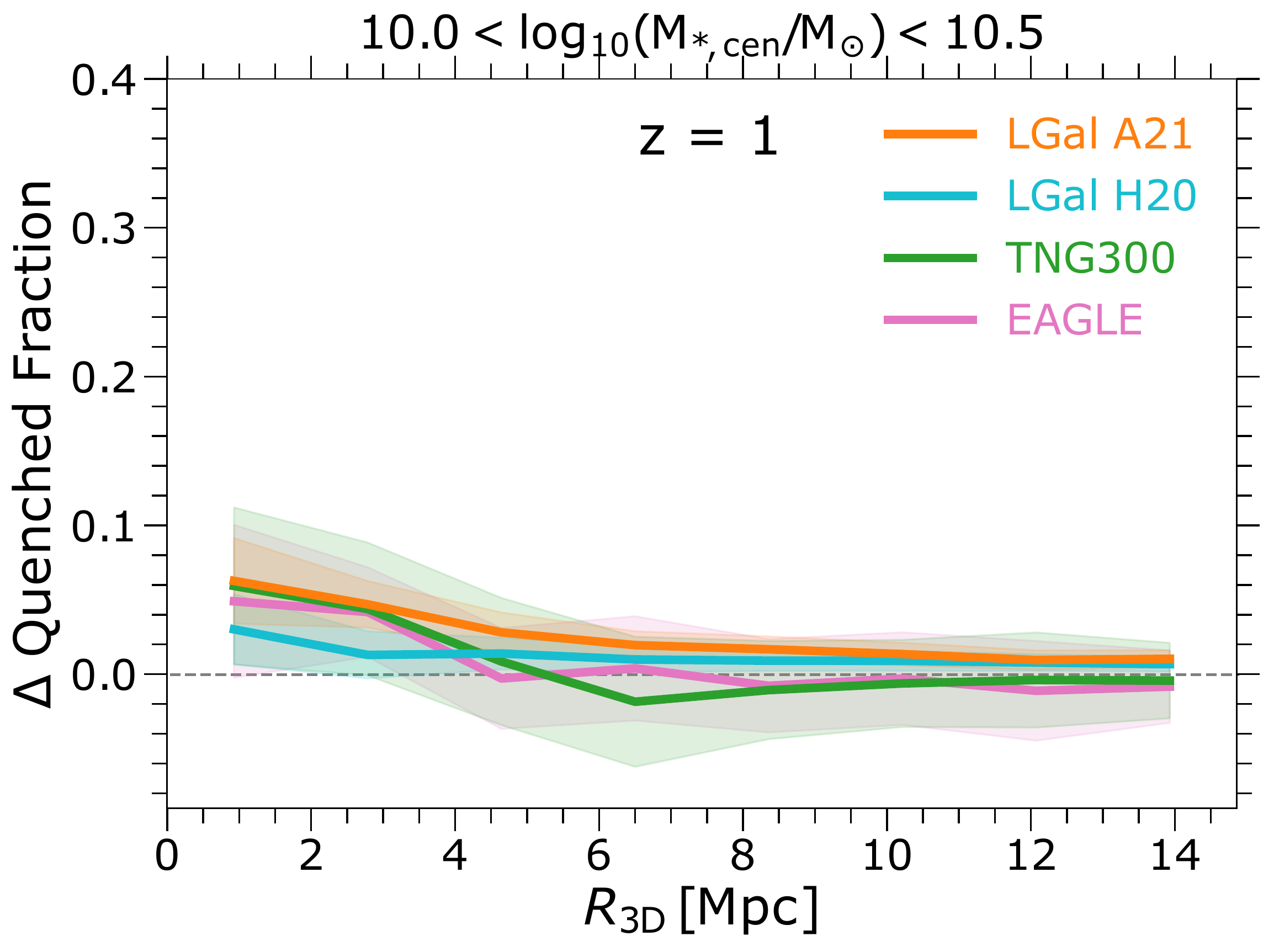}
    \includegraphics[width=0.33\textwidth]{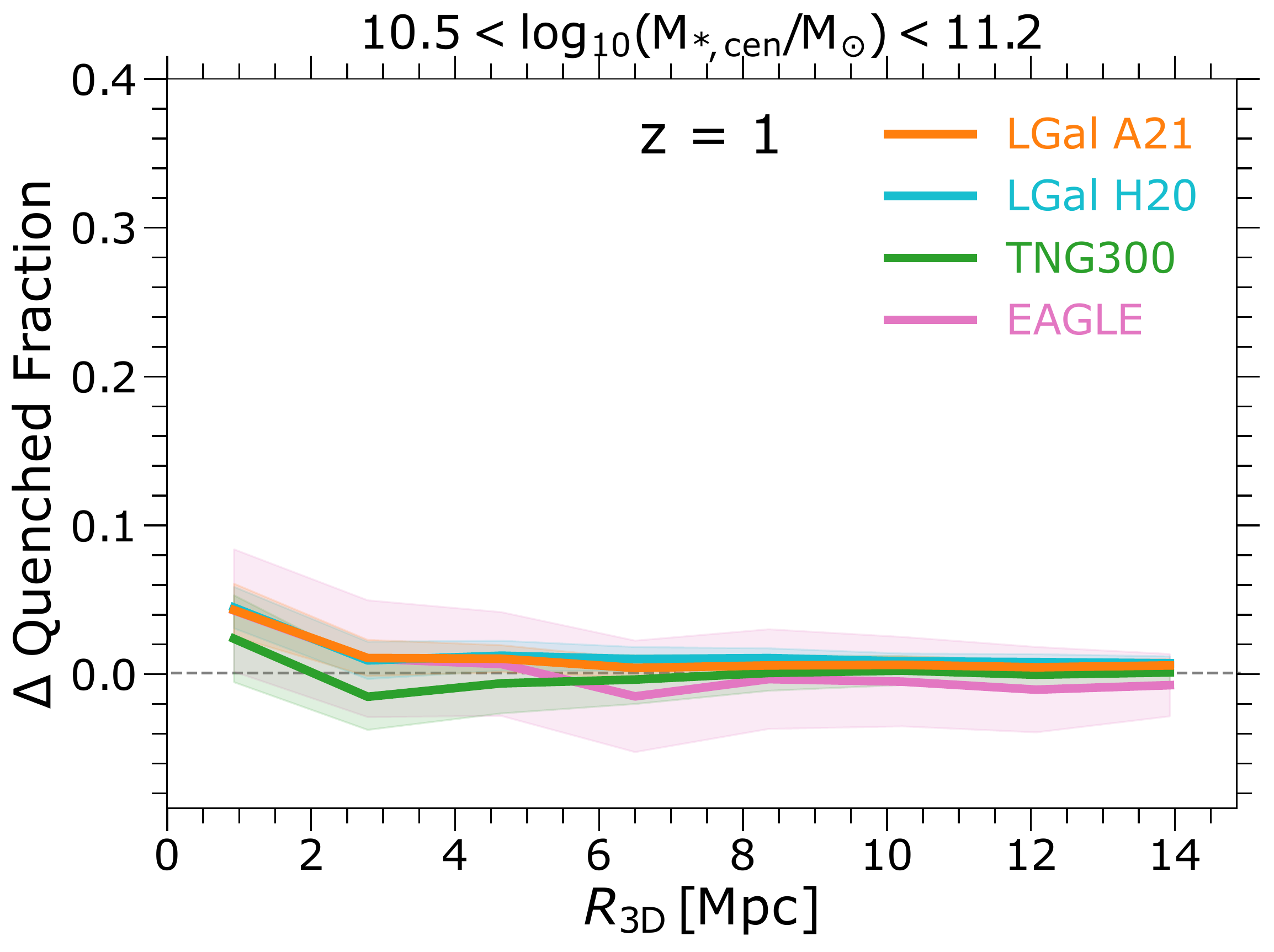}

    % {\LARGE Redshift = 2}\par\medskip
    \includegraphics[width=0.33\textwidth]{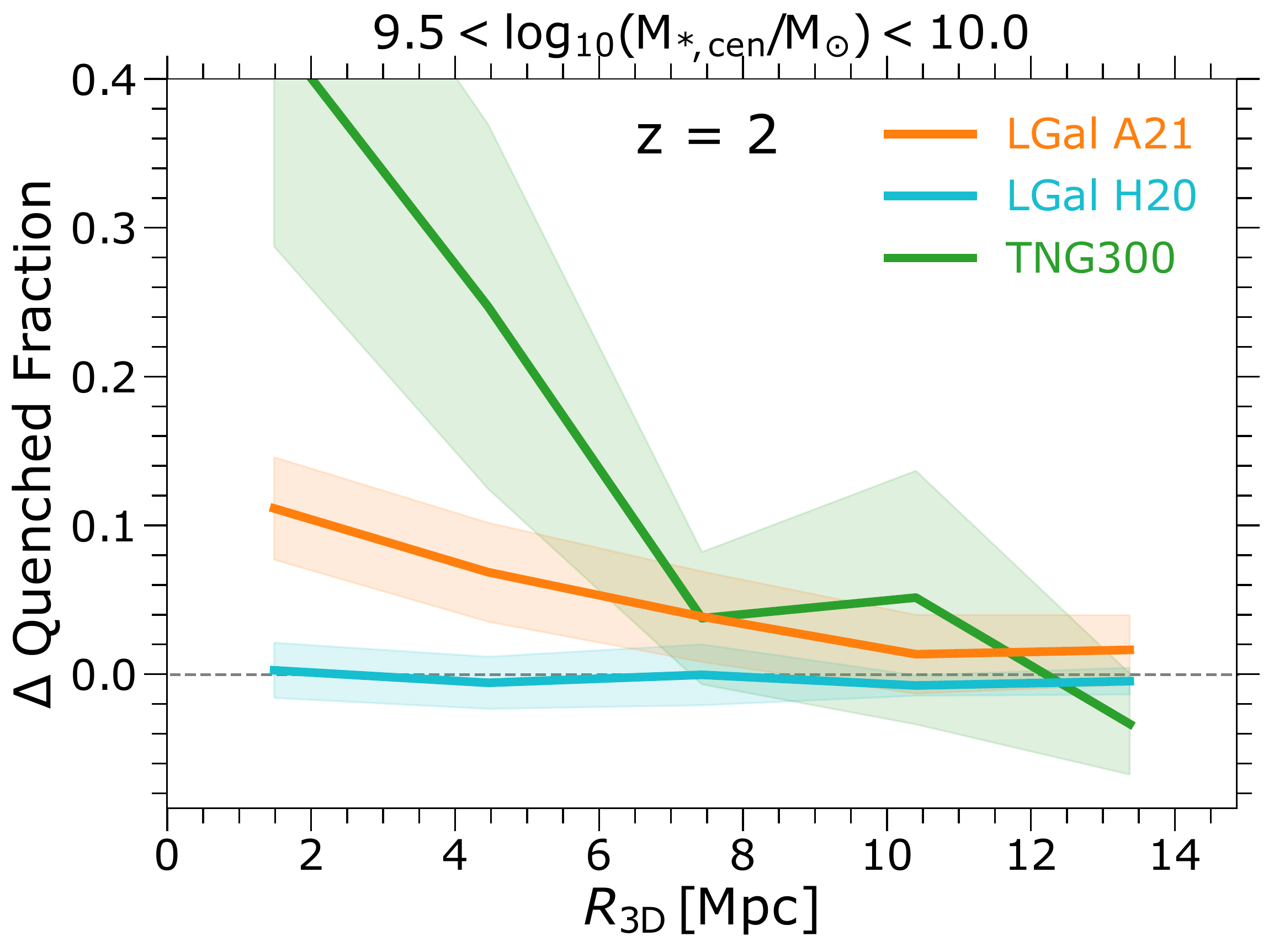}
    \includegraphics[width=0.33\textwidth]{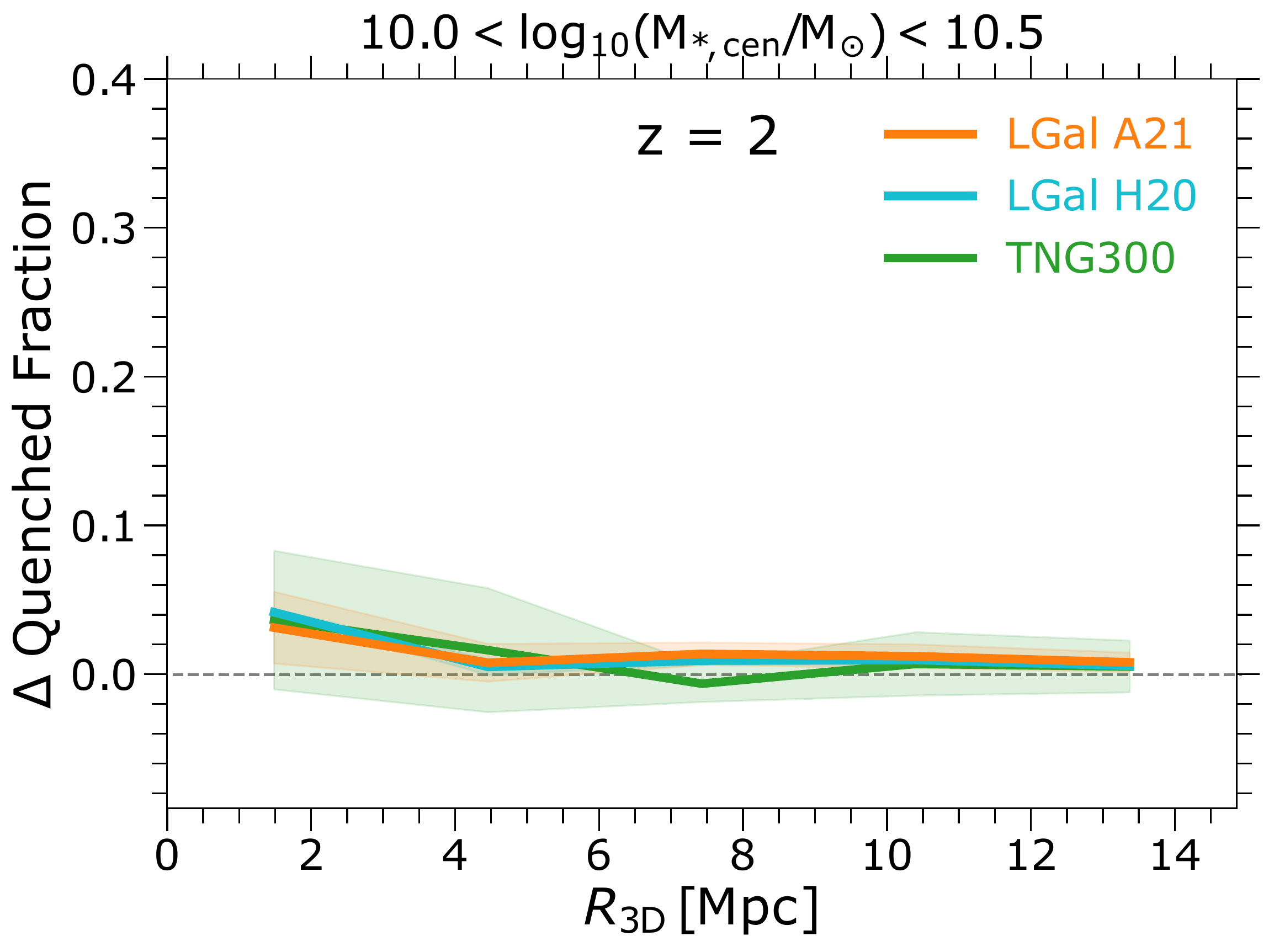}
    \includegraphics[width=0.33\textwidth]{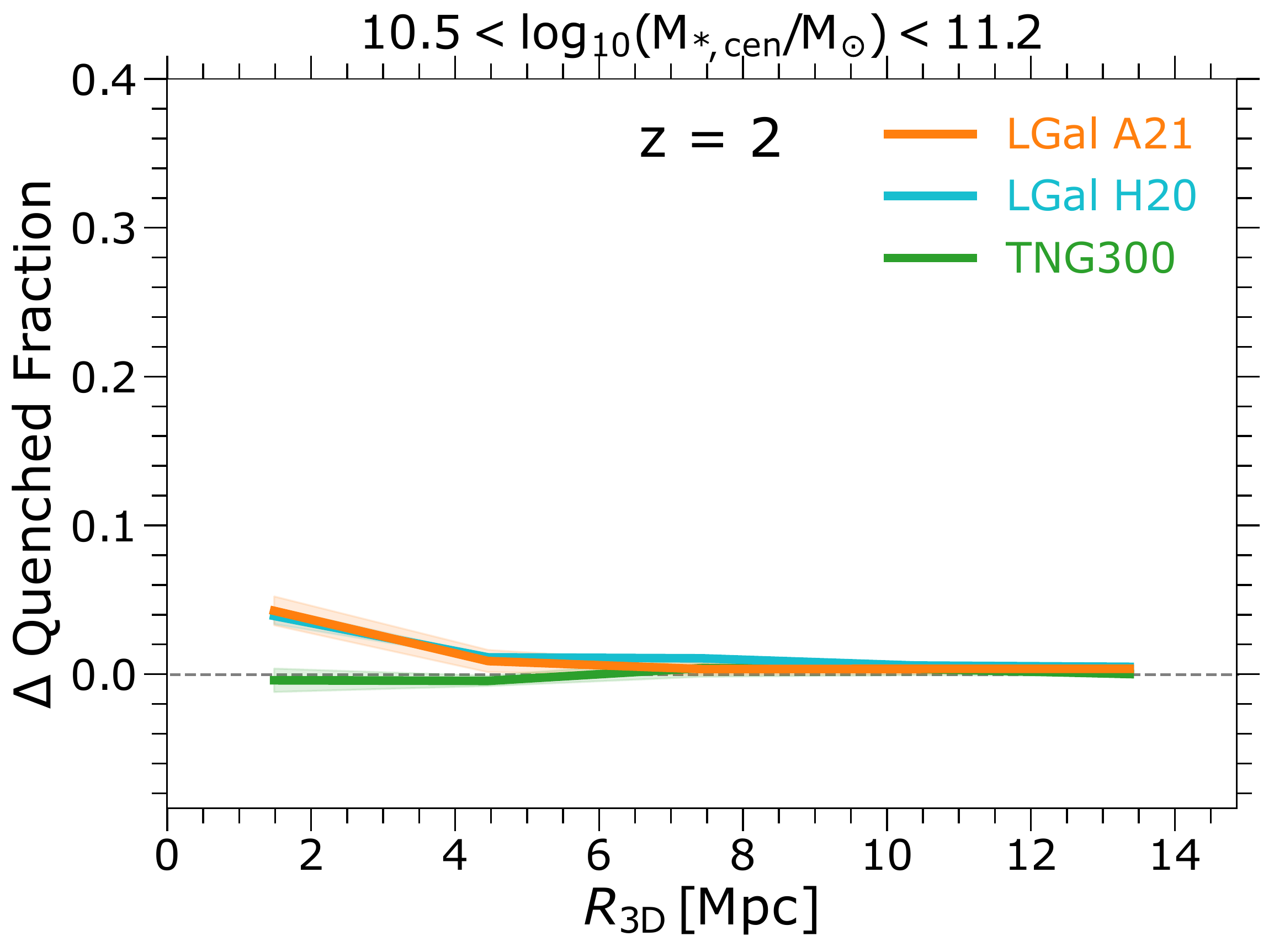}

    \caption{The conformity signal versus comoving 3D halocentric distance. The signal is defined as the difference between the quenched fraction of galaxies in the vicinity of quenched primaries and those in the vicinity of star forming primaries. The shaded regions show the uncertainty of the results, derived using the bootstrap method. The figure is divided into three columns based on the primary galaxies' stellar masses. The results are presented at three different redshifts: $z=0$ (top row), $z=1$ (middle row), and $z=2$ (bottom row). Due to the low number of galaxies in EAGLE at $z=2$, we do not show the signal from EAGLE in the bottom panel.}
\label{Fig: Conformity_signal_quenching_models_redshift}
\end{figure*}

Fig. \ref{Fig: Conformity_quenching_Mpc_sim} shows the fraction of quenched galaxies in the vicinity of quenched (red lines) and star-forming (blue lines) primary galaxies in LGal-A21 (top tow), LGal-H20 (second row), TNG300 (third row), and EAGLE (bottom row) at $z=0$. The results are split into three columns based on the stellar masses of the primary galaxies. Looking at the difference between the red and blue lines, the galactic conformity signal is strongly present out to several Megaparsecs, in all the models, with the exception of LGal-H20, where the signal is significantly weaker. This is more clearly shown in Fig. \ref{Fig: Conformity_signal_quenching_models_redshift}, where we illustrate the conformity signal as defined in Eq. \ref{eq: conformity_signal_f_q}, which is the difference between the red and blue lines from Fig. \ref{Fig: Conformity_quenching_Mpc_sim}. For brevity, in the rest of this paper we only show the conformity signal rather than the actual quenched fractions in the vicinity of quenched and star-forming centrals.

\begin{figure*}
    \centering
    \includegraphics[width=0.33\textwidth]{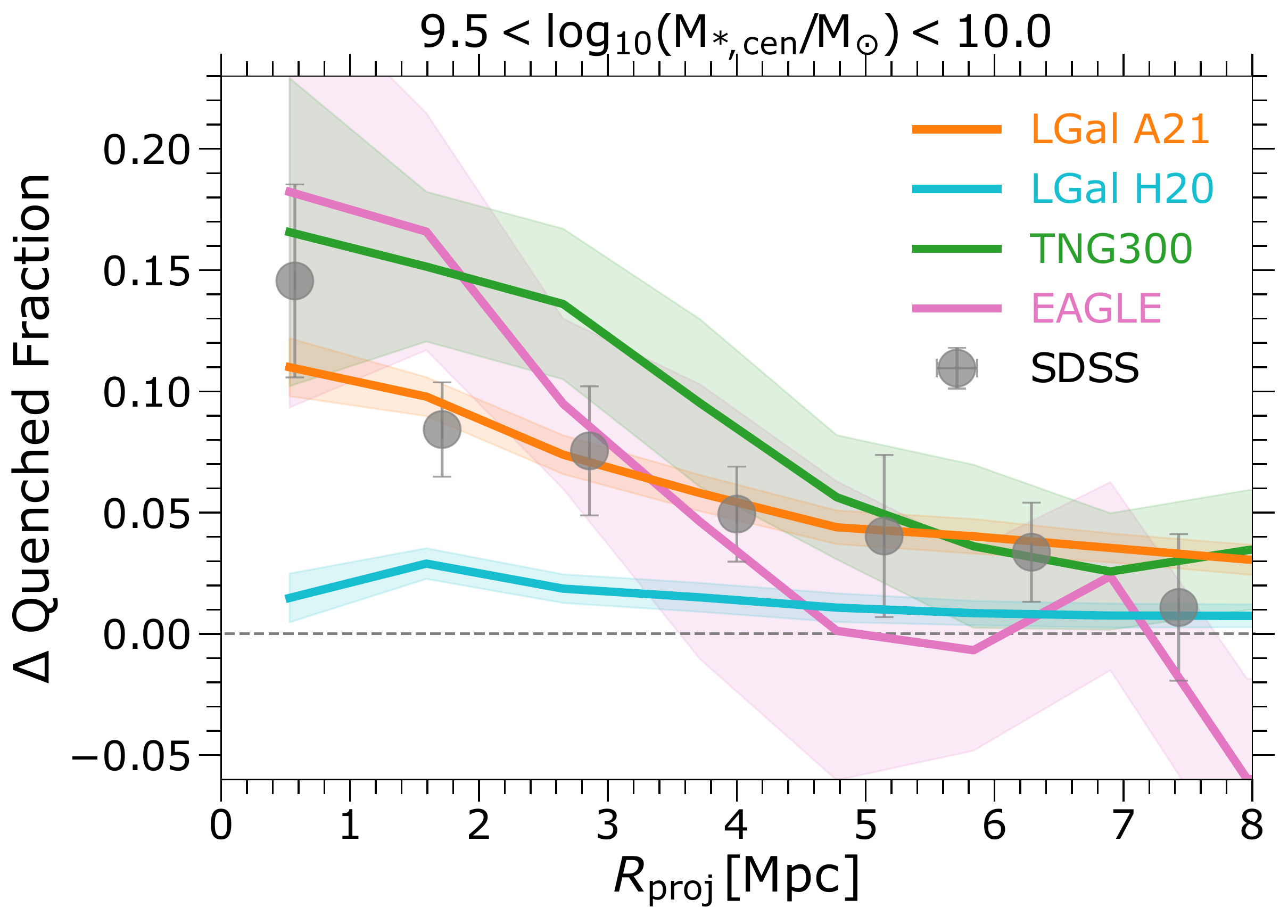}
    \includegraphics[width=0.33\textwidth]{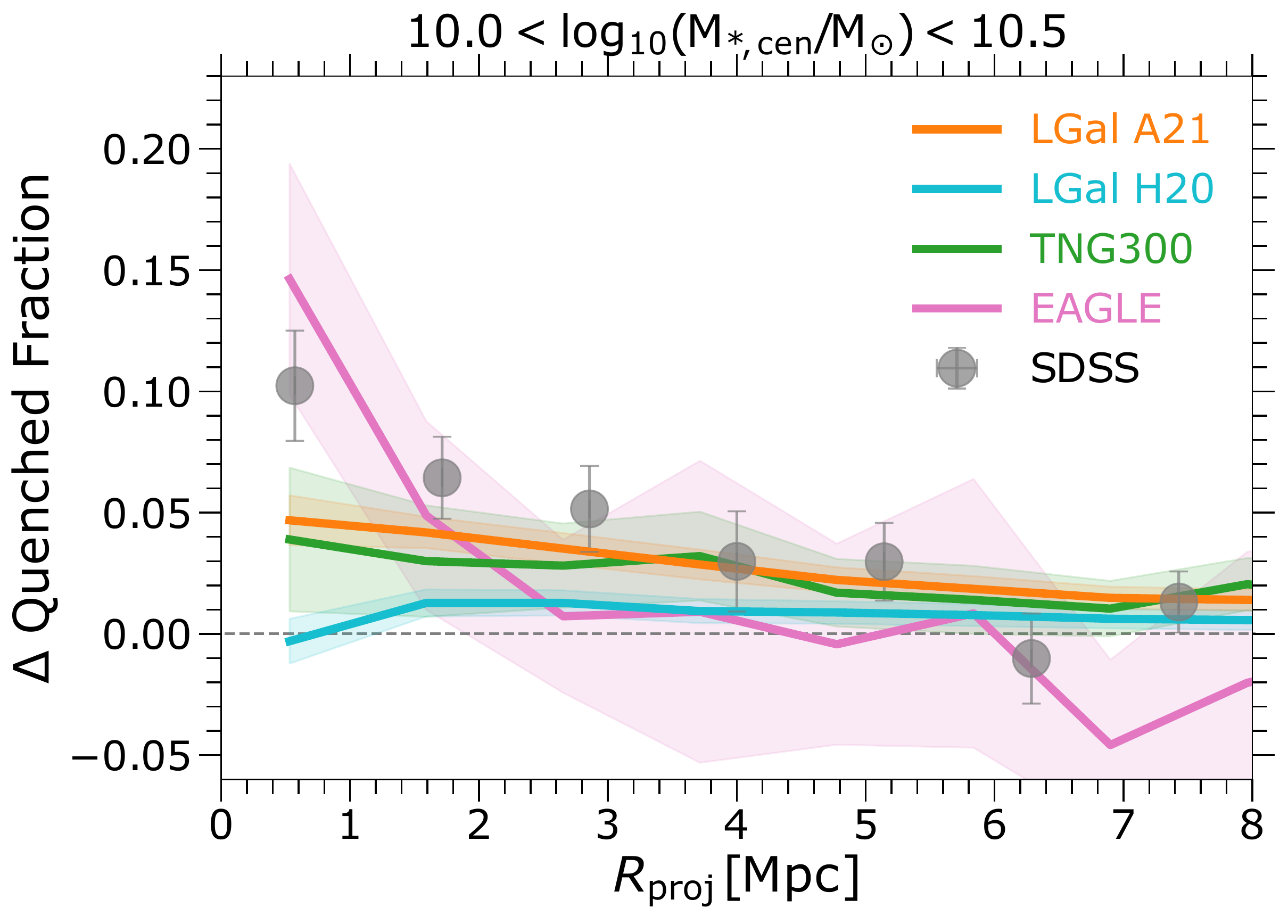}
    \includegraphics[width=0.33\textwidth]{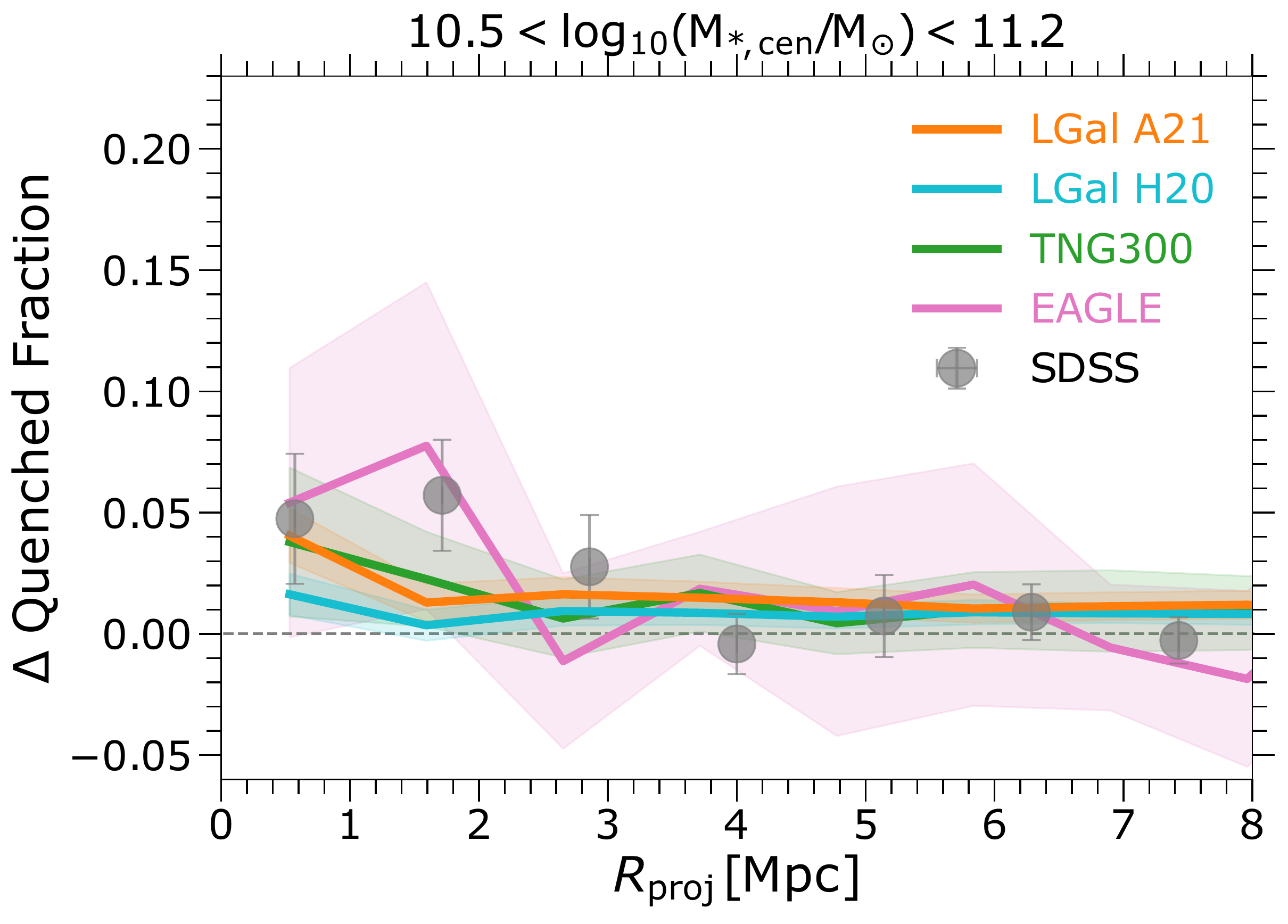}

    \includegraphics[width=0.33\textwidth]{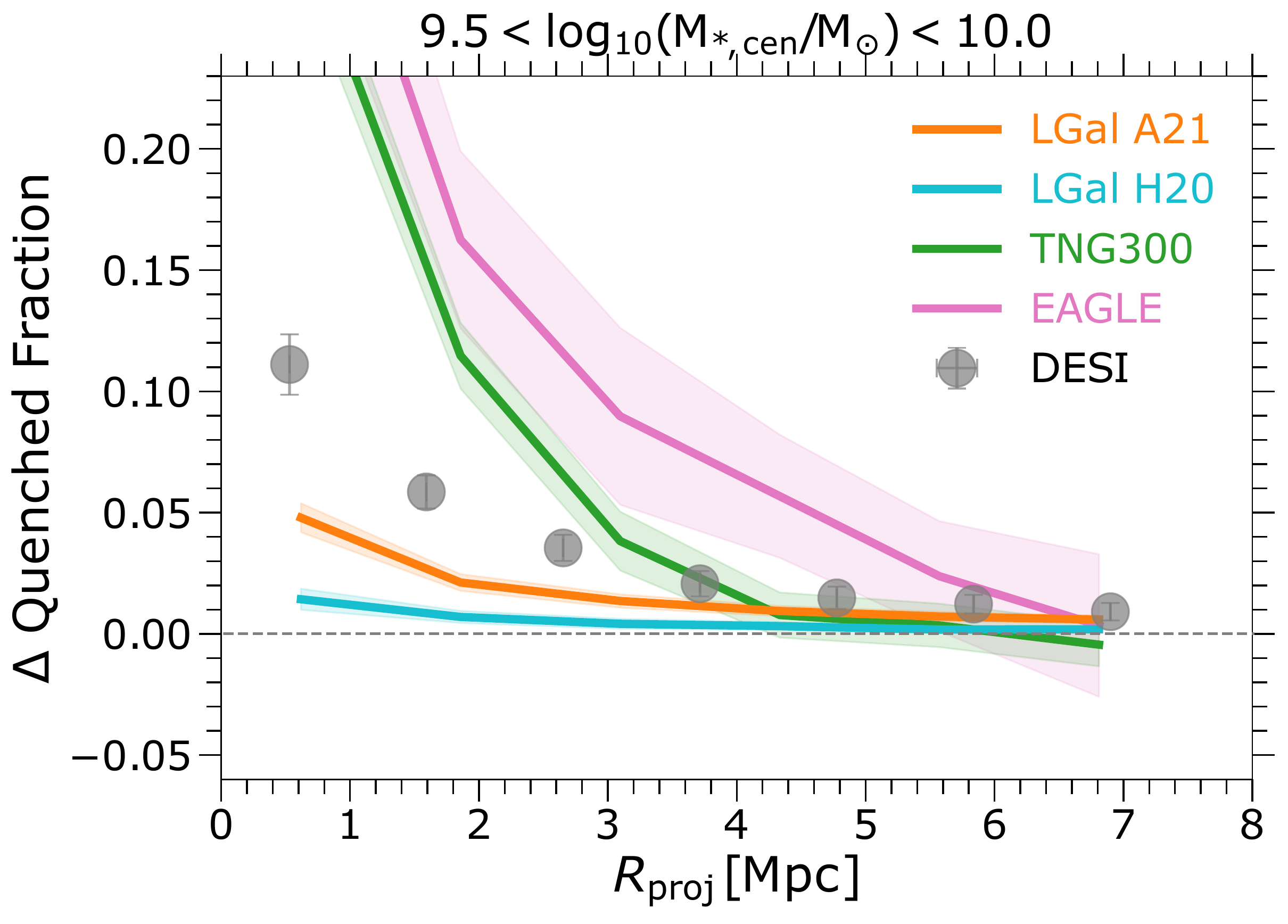}
    \includegraphics[width=0.33\textwidth]{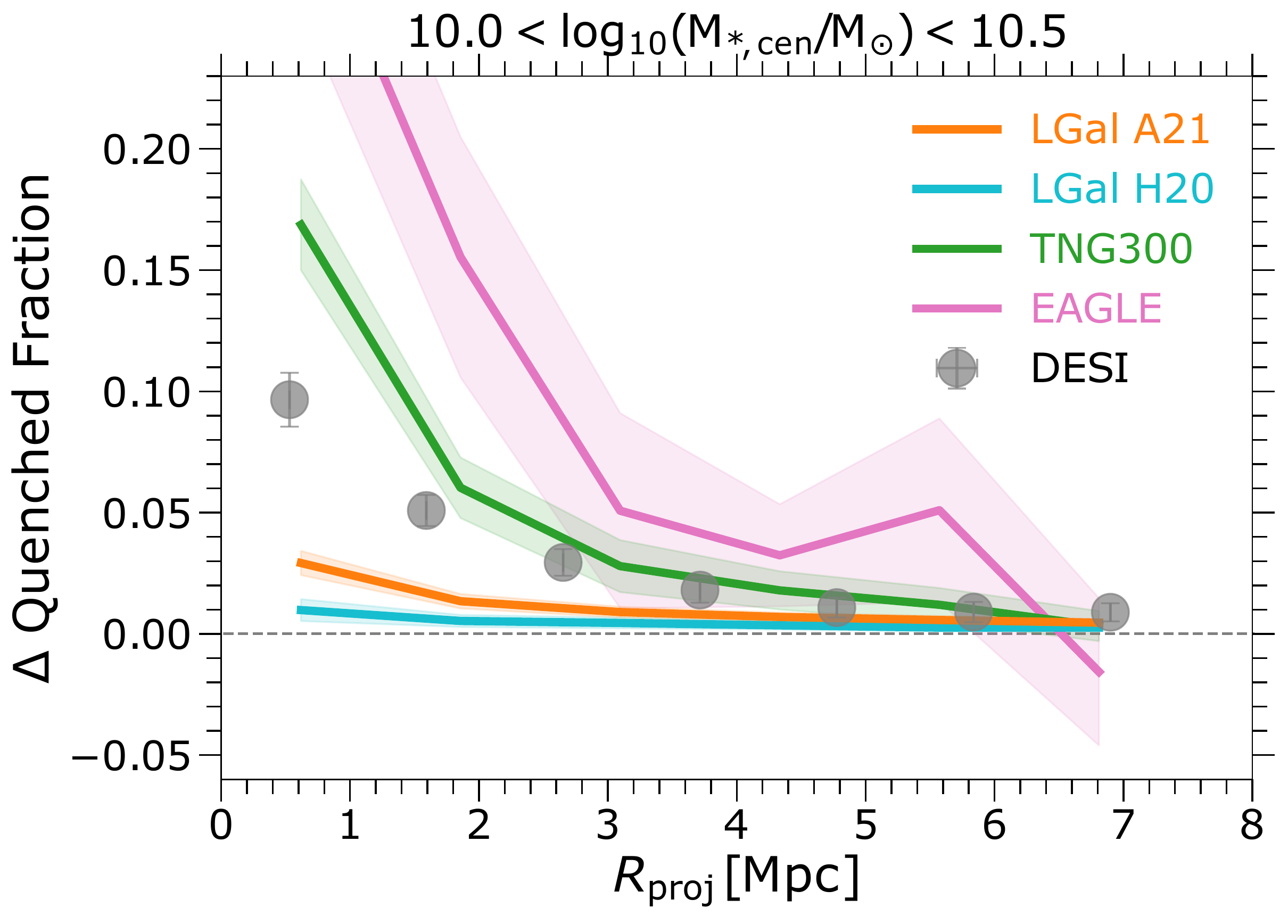}
    \includegraphics[width=0.33\textwidth]{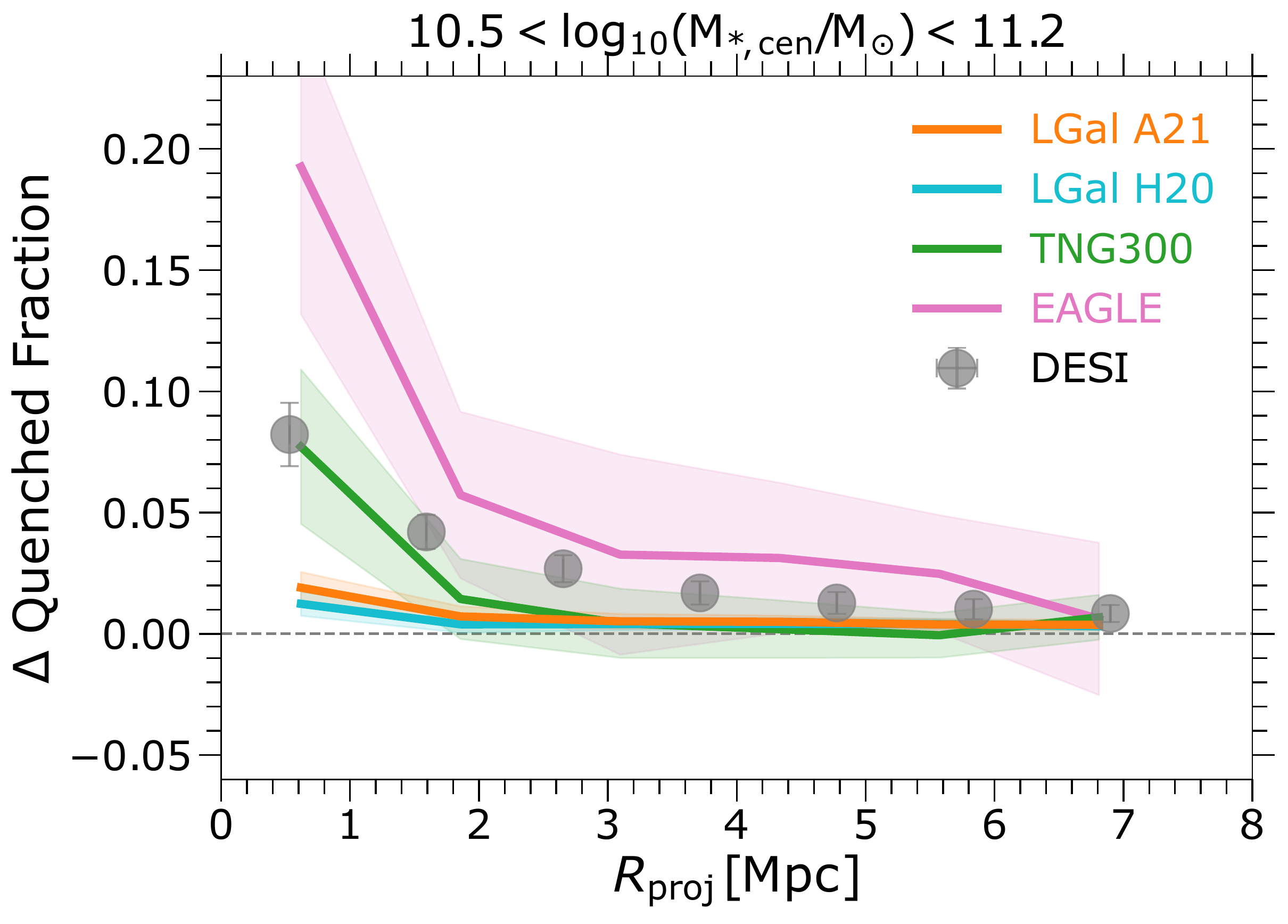}
    
    \caption{The conformity signal (Eq. \ref{eq: conformity_signal_f_q}) as a function of projected halocentric distance at $z\sim 0$. The top (bottom) panel shows the result from SDSS (DESI) and mock galaxy catalogues that are made specifically to compare with SDSS (DESI). For SDSS, we take galaxies with spectroscopic redshifts in the range $0<z<0.04$, while for DESI we take galaxies with photometric redshifts in the range $0<z<0.1$.
    The observational error bars and the shaded regions around the simulation results show the uncertainty of the results, derived using the bootstrap method. Galaxy catalogues are made using the outputs of the simulations at $z=0$. The figure is divided into three columns based on the primary galaxies' stellar masses. The galactic conformity signal is seen in the LGal-A21, TNG, and EAGLE models and both the SDSS and DESI observations up to several Megaparsecs, while it is absent in LGal-H20. The differences between the strength of the signal in the top and bottom panels mainly come from projection effects as well as the redshift uncertainties in DESI galaxies (see sections \ref{subsec: obs_data} and \ref{subsec: methods_mock_catalogues})}
\label{Fig: Conformity_signal_quenching_SDSS}
\end{figure*}

Each row in Fig. \ref{Fig: Conformity_signal_quenching_models_redshift} corresponds to a different redshift, from $z=0$ (top panel) out to $z=2$ (bottom panel). In LGal-A21 (orange lines), TNG (green lines), and EAGLE (magenta lines), the conformity signal is significant both on small and on large scales. The signal decreases with the halocentric radius on large scales ($R\gtrsim 2\rm Mpc$) at all redshifts from $z=0$ up to $z=2$. In contrast to the other models, LGal-H20 (blue lines) shows almost no signal on large scales at any  redshift. For reference, at $z=0$ and at $R_{\rm 3D} \sim 3\, \rm Mpc$ from the halo centre, the signal in the vicinity of low-mass centrals in different models is: $\sim 0.28$ in TNG, $\sim 0.25$ in EAGLE, $\sim 0.12$ in LGal-A21, and $\sim 0.03$ in LGal-H20.

The signal also depends on the stellar masses of the primary galaxies. Comparing different columns of Fig. \ref{Fig: Conformity_signal_quenching_models_redshift}, the signal is strongest near low-mass primary galaxies (the leftmost column, $9.5<\log_{10}(M_{\rm \star,cen}/{\rm M_{\odot}})<10$) and decreases with increasing primary galaxy mass. On large scales ($R \gtrsim 2\rm Mpc$), and for more massive primary galaxies (the right most column, $10.5<\log_{10}(M_{\rm \star,cen}/{\rm M_{\odot}})<11.2$), the signal almost vanishes at all redshifts. We note that the noise in TNG's and EAGLE's signals is due to the relatively small number of galaxies in these simulations (see Table \ref{tab: data_stat}).

\subsection{Large-scale conformity in observations and mock galaxy catalogues}
\label{subsec: conformity_obs}
\subsubsection{Conformity in the quenched fraction of galaxies}
\label{subsubsec: conformity_obs_quenching}
Fig. \ref{Fig: Conformity_signal_quenching_SDSS} shows the large-scale conformity signal, defined by Eq. \ref{eq: conformity_signal_f_q}, in the SDSS (top panel) and DESI (bottom panel) observations and the mock catalogues. We made each mock catalogue specifically for the observation it is compared to (see section \ref{subsec: methods_mock_catalogues}).

The conformity signal is apparent both in SDSS and in DESI and is maximum in the vicinity of low-mass primary galaxies (left panels, $9.5<\log_{10}(M_{\star}/M_{\odot})<10$). In general, the observed signal is slightly weaker in DESI, which is due to its deeper projection (see section \ref{subsec: methods_mock_catalogues}). The signal is also present in LGal-A21, TNG, and EAGLE, while almost absent in LGal-H20. Comparing the models with observations, all models are in relatively good agreement with observations, except LGal-H20, whose signal is much below the observed value. For reference, at $R_{\rm proj} \sim 3\rm \, Mpc$ in SDSS and its mock galaxy catalogues, the signal in the vicinity of low-mass galaxies ($9.5<\log_{10}(M_{\star}/{\rm M_{\odot}})<10$) is: $\sim 0.13$ in TNG, $\sim 0.08$ in EAGLE, $\sim 0.07$ in SDSS, $\sim 0.07$ in LGal-A21, and $\sim 0.01$ in LGal-H20\footnote{In our further analysis, we find that on small scales $(R_{\rm proj}<1\, \rm Mpc)$, the conformity signal increases both in the observations and in the models, although we do not show it here.}.

Comparing SDSS with its mock galaxy catalogues (top panel of Fig. \ref{Fig: Conformity_signal_quenching_SDSS}), in the vicinity of low-mass centrals ($9.5<\log_{10}(M_{\star}/M_{\odot})<10$), LGal-A21 makes better predictions, whereas TNG and EAGLE overestimate the signal at distances smaller than 4 Mpc from the centres of haloes. This overestimation of the signal in TNG and EAGLE is mainly originated from satellite secondary galaxies (see section \ref{subsec: secondary_cens_sats}). On the other hand, EAGLE's predictions are in better agreement with observations in the vicinity of intermediate-mass primaries ($10<\log_{10}(M_{\star}/M_{\odot})<10.5$), where both LGal-A21 and TNG underestimate the signal by a few per cent at $R<4\, \rm Mpc$. In the bottom panel of Fig. \ref{Fig: Conformity_signal_quenching_SDSS}, where we compare DESI with its mock catalogues, LGal-A21 underestimates the signal by a factor of two. In contrast, TNG and EAGLE significantly overestimate the observed signal, which is likely influenced by the limitations in their simulation box sizes (see section \ref{subsec: methods_mock_catalogues}).

Both in the observations (SDSS and DESI) and in the models (LGal-A21, TNG, EAGLE), the signal decreases with halocentric distance. The signal also considerably decreases with increasing the stellar masses of primary galaxies.

One scenario to explain the origin of the signal is that low- and intermediate- mass primary galaxies and their neighbours lose gas due to environmental processes when moving through the warm-hot intergalactic medium and in the outskirts of massive haloes, leading to spatially correlated suppression of star-formation in relatively low-mass galaxies. The absence of signal in LGal-H20, which is the only model without gas stripping implemented for such galaxies (see \ref{subsubsec: methods_relevant_processes}), is strong evidence to support this idea. 

\subsubsection{Conformity in the specific star formation rates of galaxies}
\label{subsubsec: conformity_obs_ssfr}
\begin{figure*}
    \centering
    % \textbf{\LARGE SDSS and its mock catalogues}\par\medskip
    \includegraphics[width=0.33\textwidth]{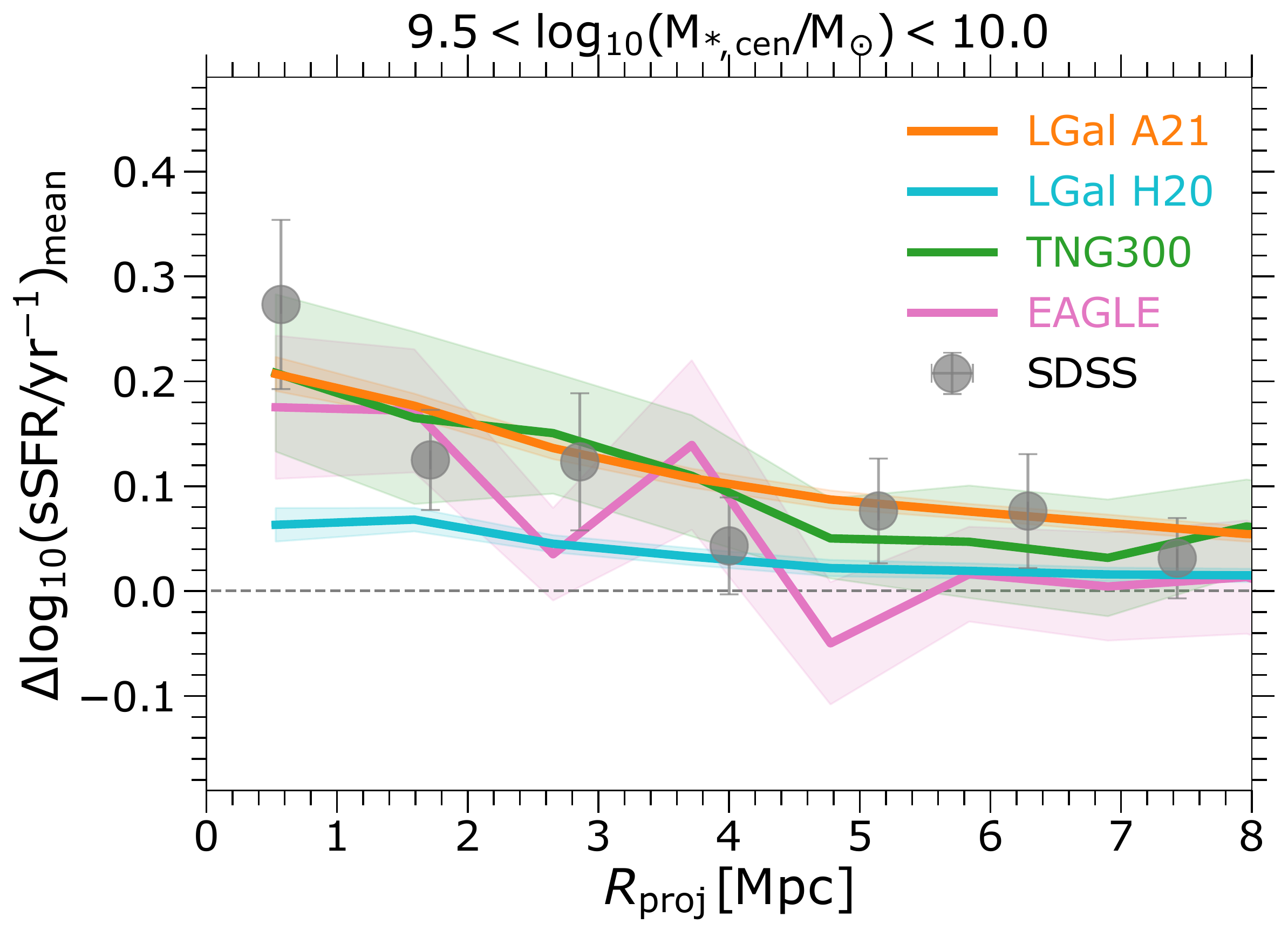}
    \includegraphics[width=0.33\textwidth]{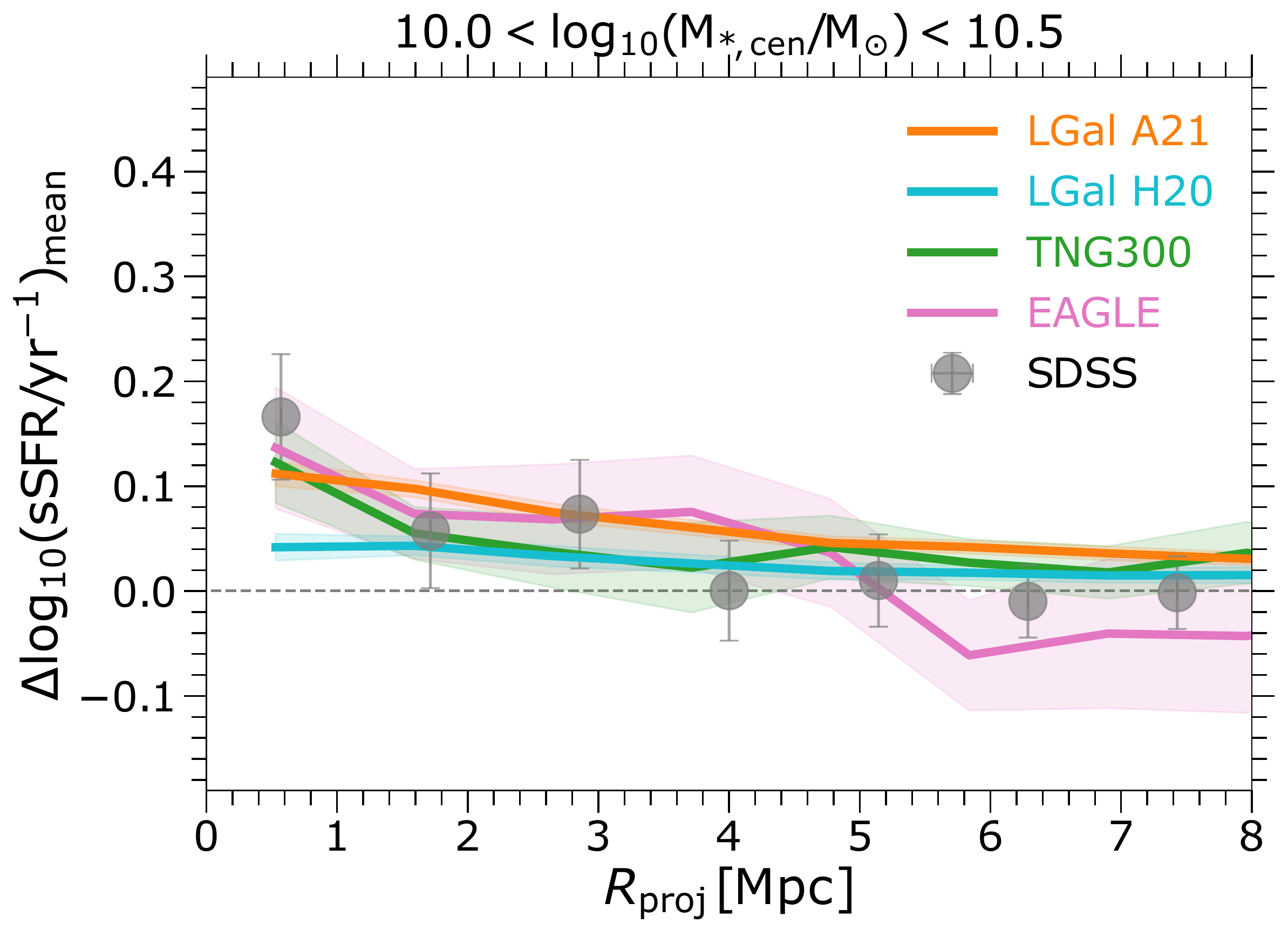}
    \includegraphics[width=0.33\textwidth]{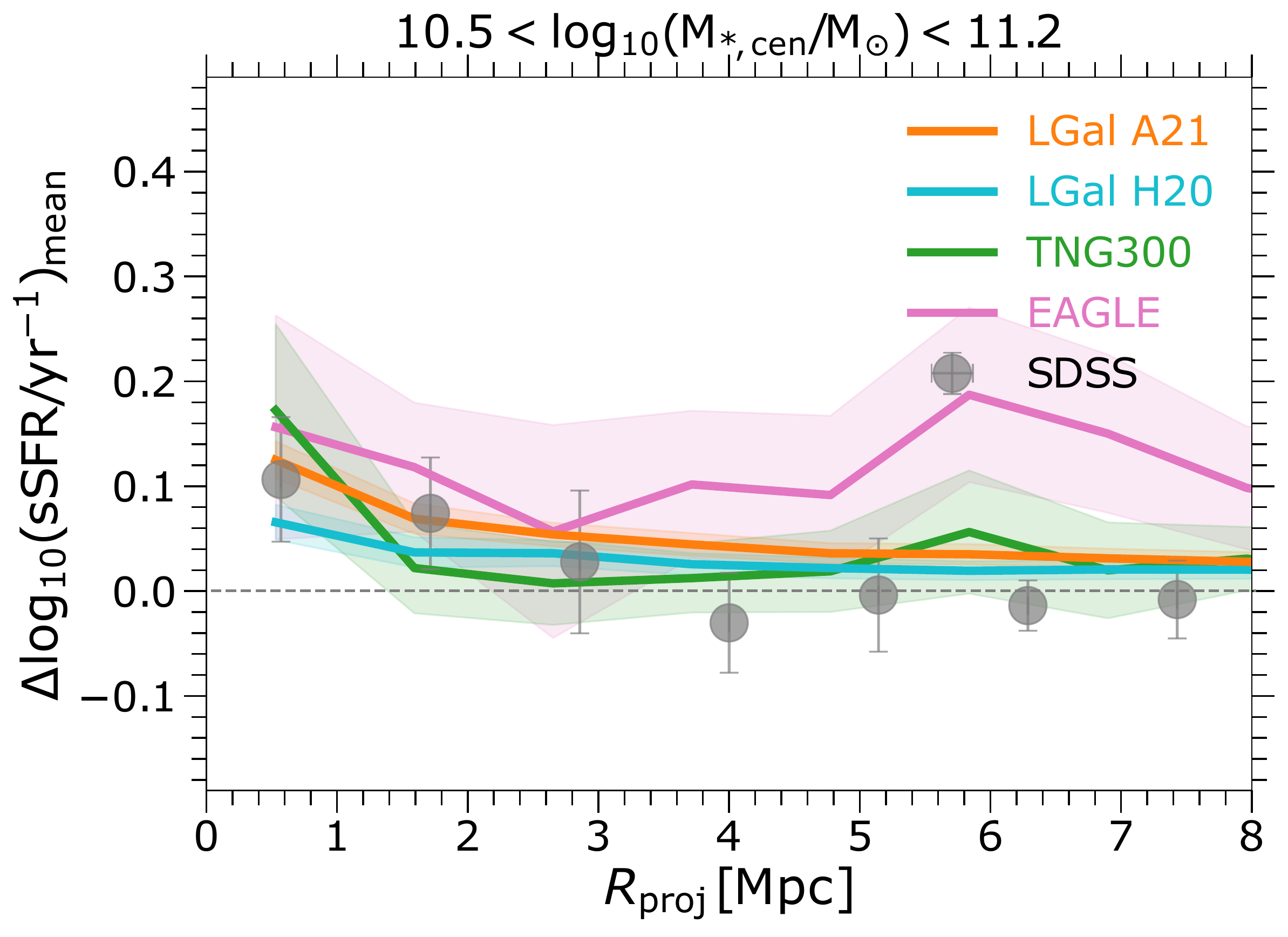}
    
    % \textbf{\LARGE DESI and its mock catalogues}\par\medskip
    \includegraphics[width=0.33\textwidth]{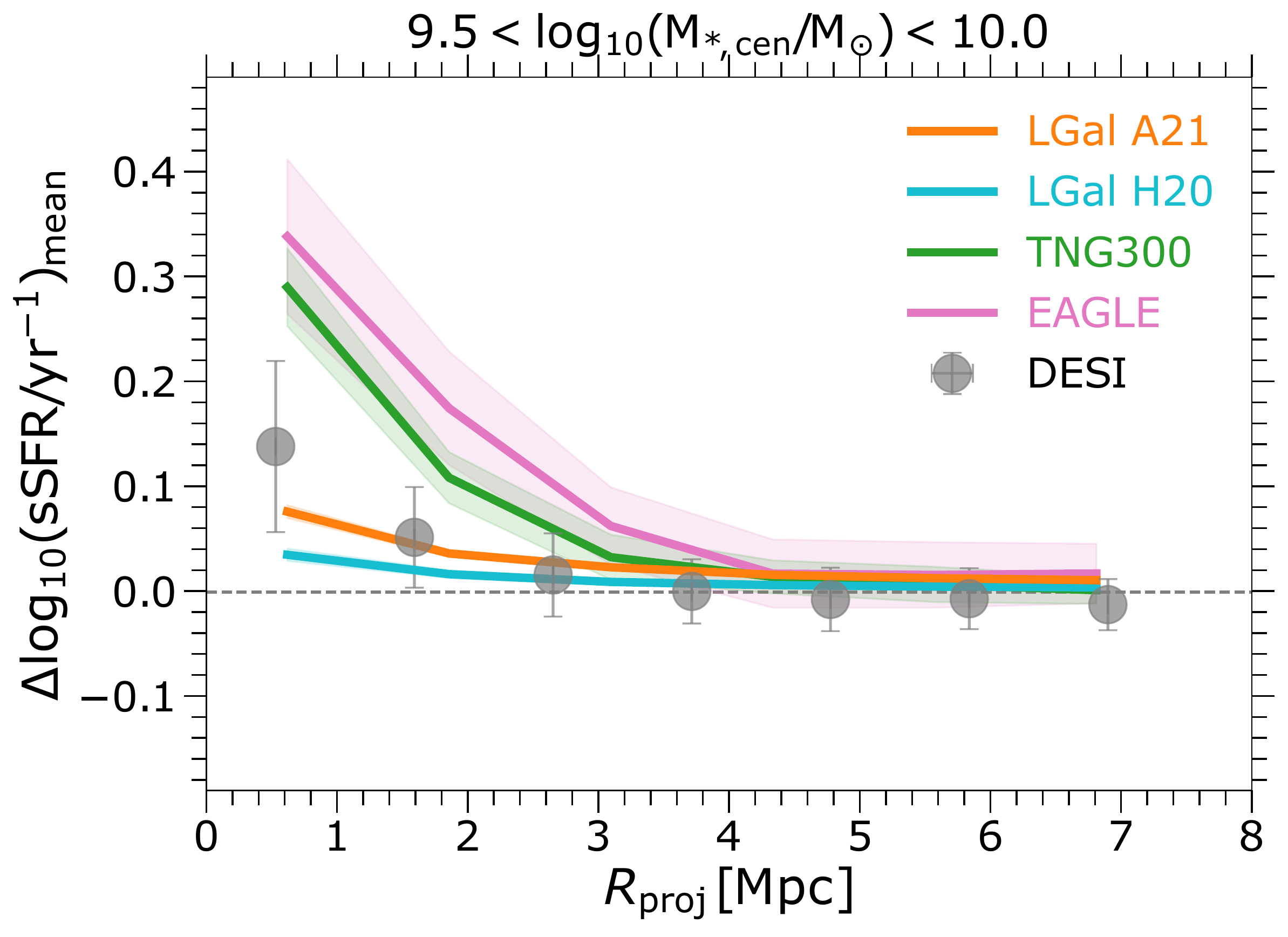}
    \includegraphics[width=0.33\textwidth]{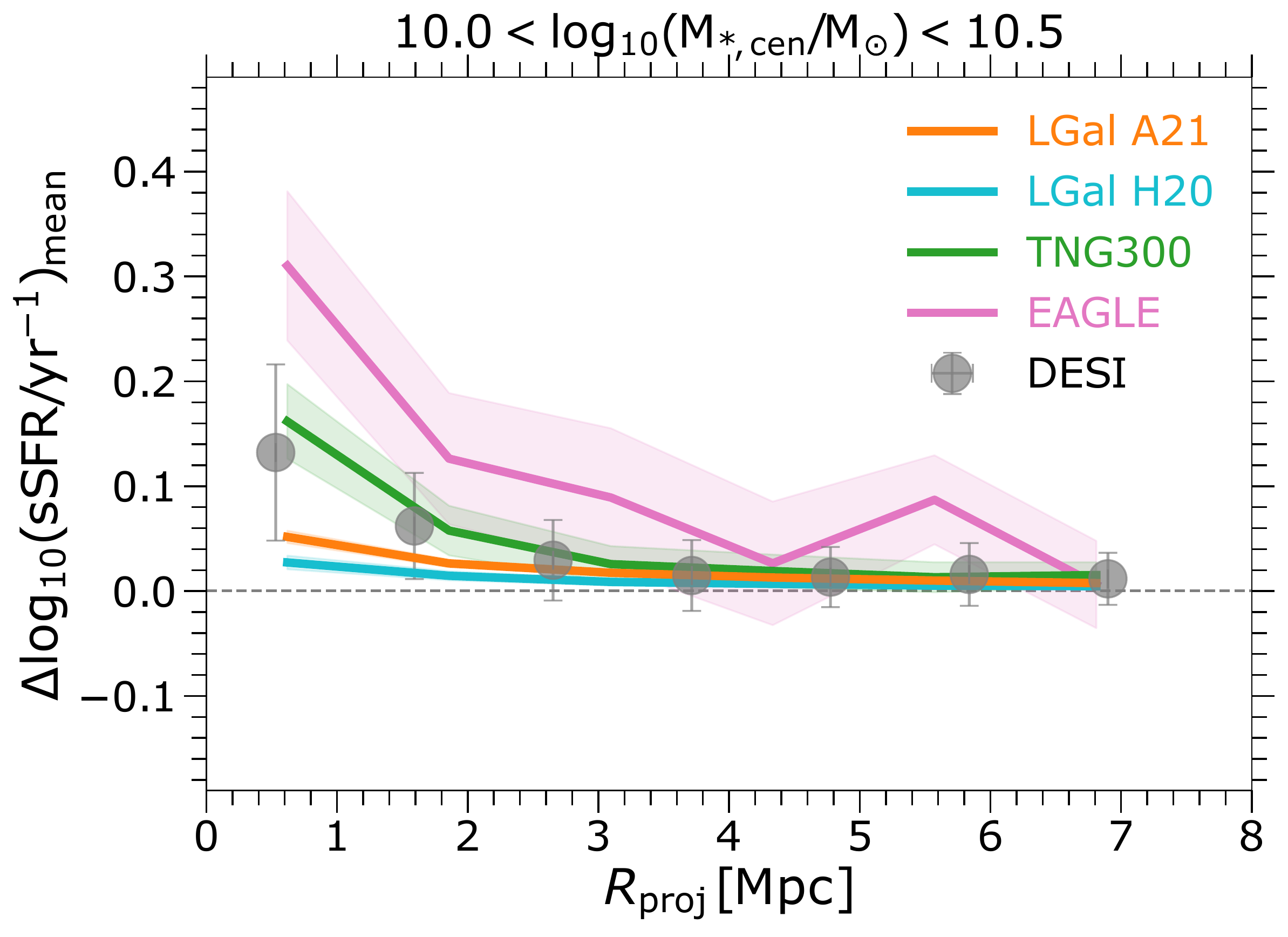}
    \includegraphics[width=0.33\textwidth]{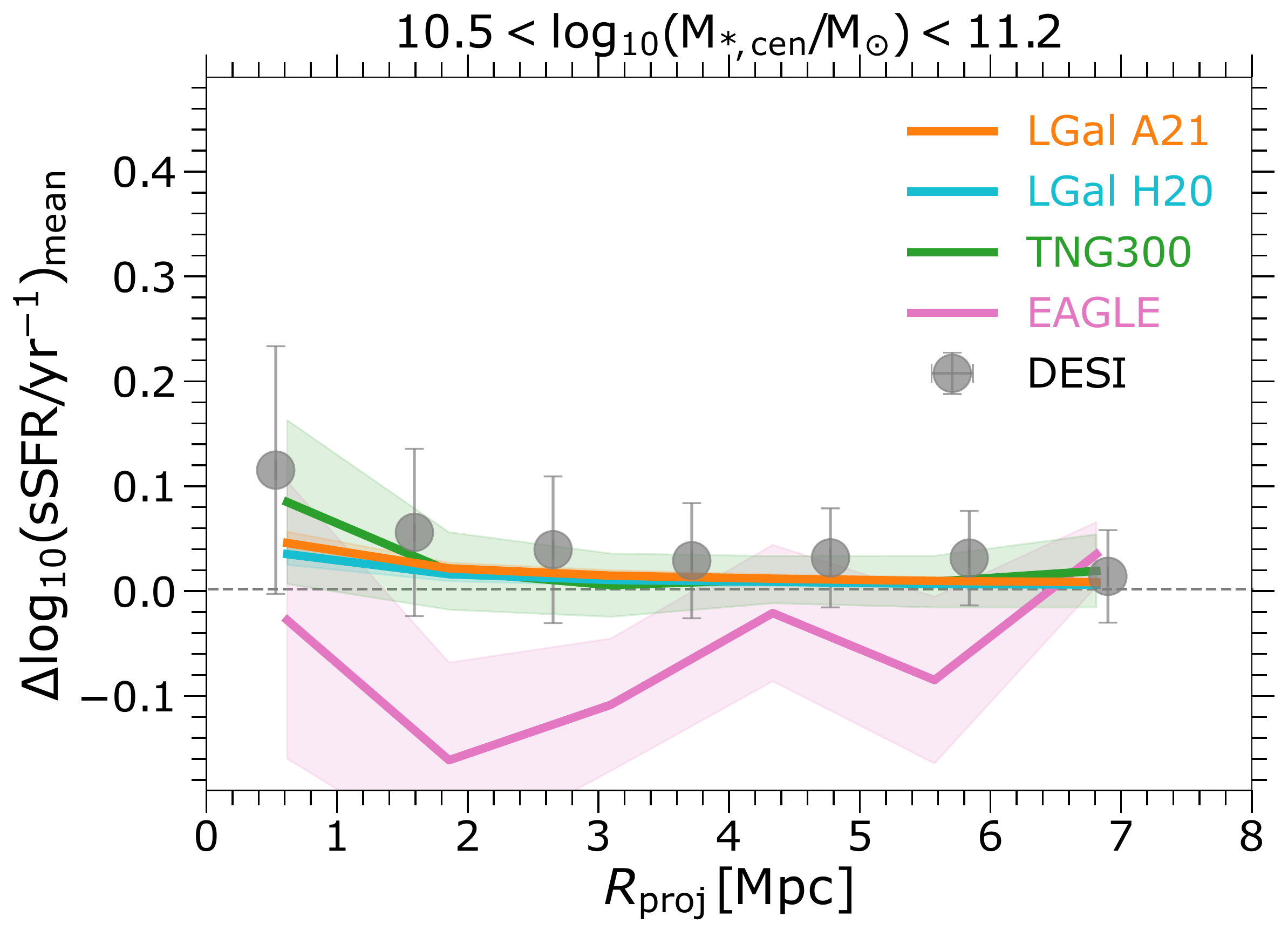}

    \caption{The conformity signal in sSFR: The difference between the mean sSFR of galaxies in the vicinity of star forming primaries and those in the vicinity of quenched primaries as function of projected halocentric distance. The top (bottom) panel shows the results from SDSS (DESI) and its respective mock galaxy catalogues. The observational error bars and the shaded regions around the simulation results show the uncertainty of the results, derived using the bootstrap method. The figure is divided into three columns based on the primary galaxies' stellar mass. All the results are presented at $z=0$.}
\label{Fig: Conformity_signal_ssfr_SDSS}
\end{figure*}

In addition to the conformity signal extracted from the fraction of quenched galaxies, we explore the signal derived from the sSFRs of galaxies. Fig. \ref{Fig: Conformity_signal_ssfr_SDSS} shows the conformity signal based on Eq. \ref{eq: conformity_signal_ssfr}. Overall, the conformity signal in sSFR is present both in observations and simulations except for LGal-H20, and follows similar trends as the signal in the quenched fraction of galaxies (see \ref{subsubsec: conformity_obs_quenching}). The absence of the signal in LGal-H20 confirms our finding in section \ref{subsubsec: conformity_obs_quenching} that the signal is caused by environmental effects like ram-pressure stripping, possibly acting on large-scales (also see section \ref{sec: summary} for a discussion).

Comparing SDSS with its mock galaxy catalogues (top panel), LGal-A21, TNG, and EAGLE are in relatively good agreement with the observations. The bottom panel shows the conformity signal in DESI and its respective mock catalogues. In the vicinity of low-mass primary galaxies (bottom left, $9.5<\log_{10}(M_{\star}/M_{\odot})<10$), LGal-A21 is in a relatively good agreement with DESI, while both TNG and EAGLE overestimate the signal. In the vicinity of intermediate-mass primaries ( $10<\log_{10}(M_{\star}/M_{\odot})<10.5$), TNG's signal agrees relatively well with the observations, whereas LGal-A21 and EAGLE underestimate and overestimate the signal, respectively. Moreover, near more massive primaries (bottom right, $10.5<\log_{10}(M_{\star}/M_{\odot})<11.2$), TNG's signal is close to DESI, whereas LGal-A21 underestimates the signal.

We note that EAGLE's signal becomes very noisy and unreliable in the vicinity of massive primaries (right column, $10.5<\log_{10}(M_{\star}/M_{\odot})<11.2$), due to its small box size (see section \ref{subsec: methods_mock_catalogues}). Therefore, we will not discuss it further here, although we show it for completeness.

\subsubsection{Conformity in the stellar masses of galaxies}
\label{subsubsec: conformity_obs_Mstar}
\begin{figure*}
    \centering
    \includegraphics[width=0.33\textwidth]{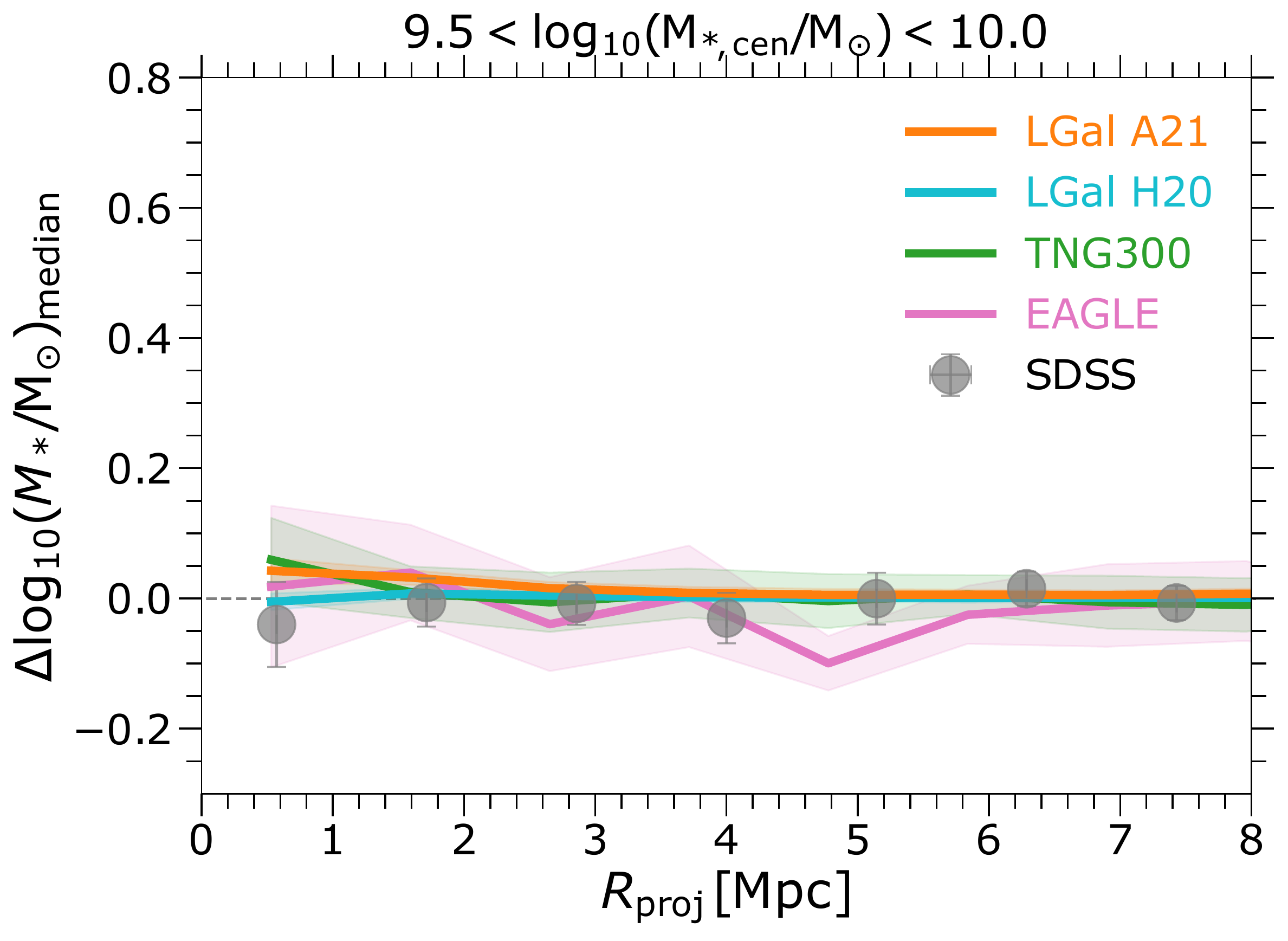}
    \includegraphics[width=0.33\textwidth]{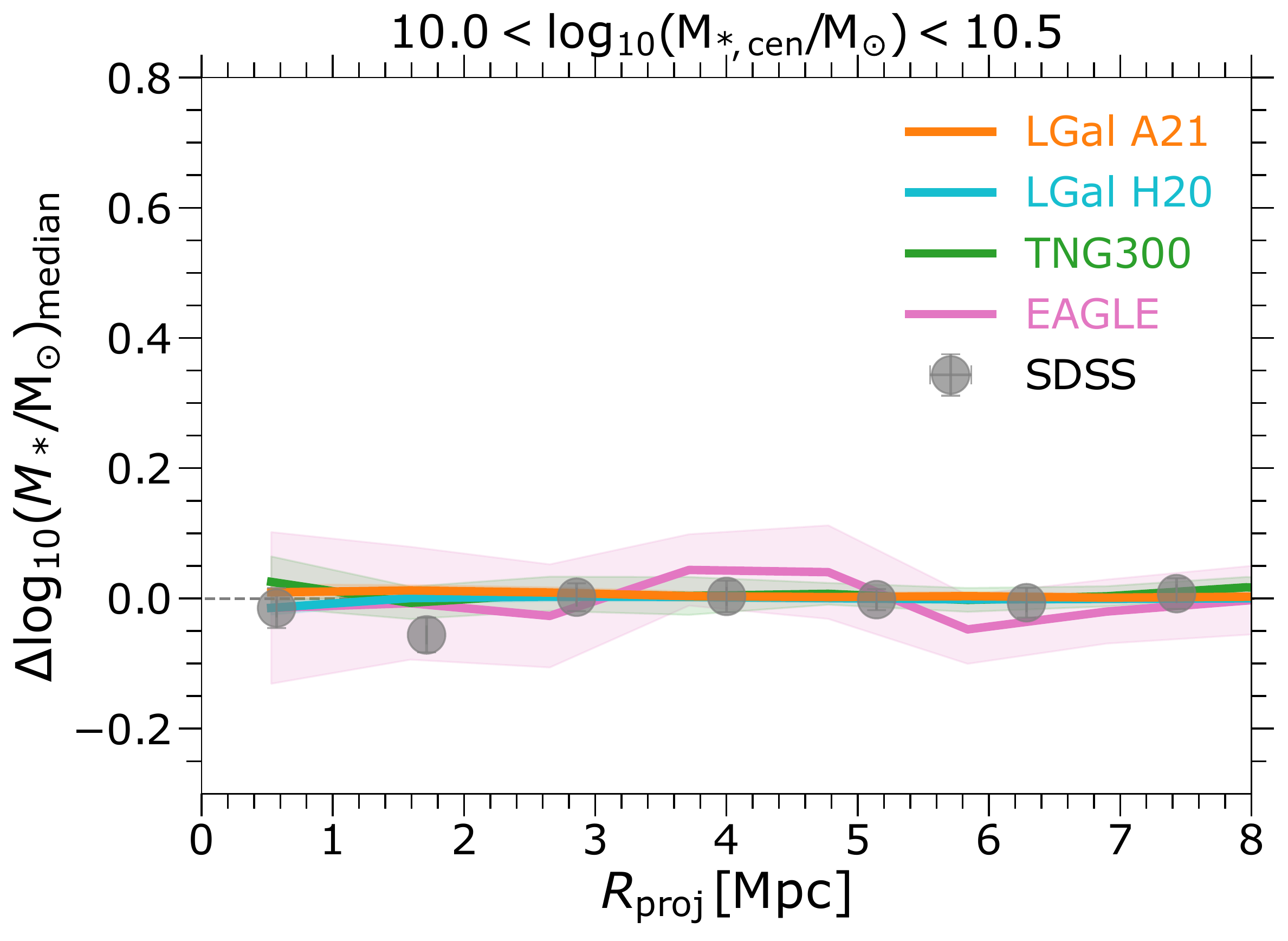}
    \includegraphics[width=0.33\textwidth]{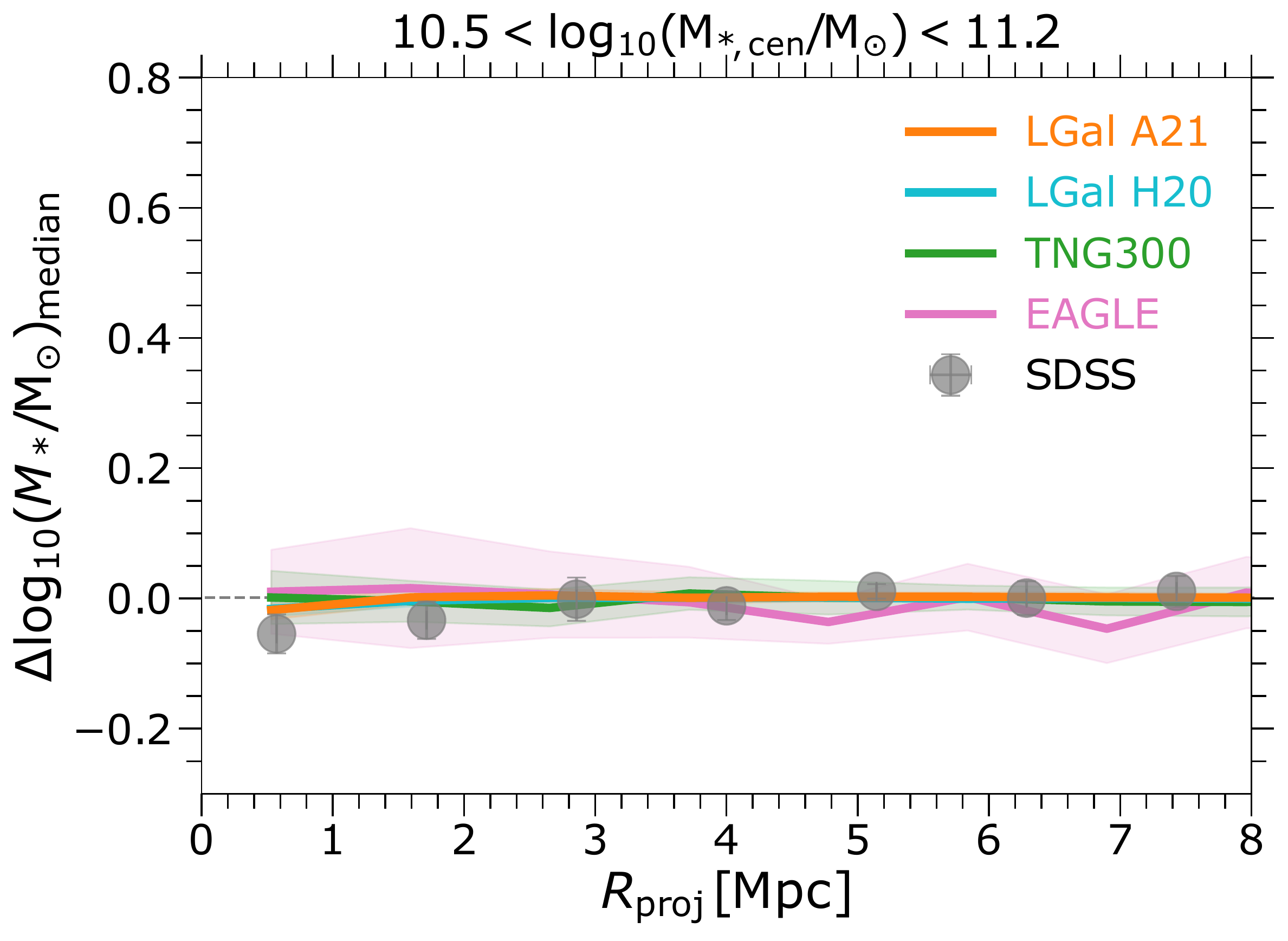}
    
    \includegraphics[width=0.33\textwidth]{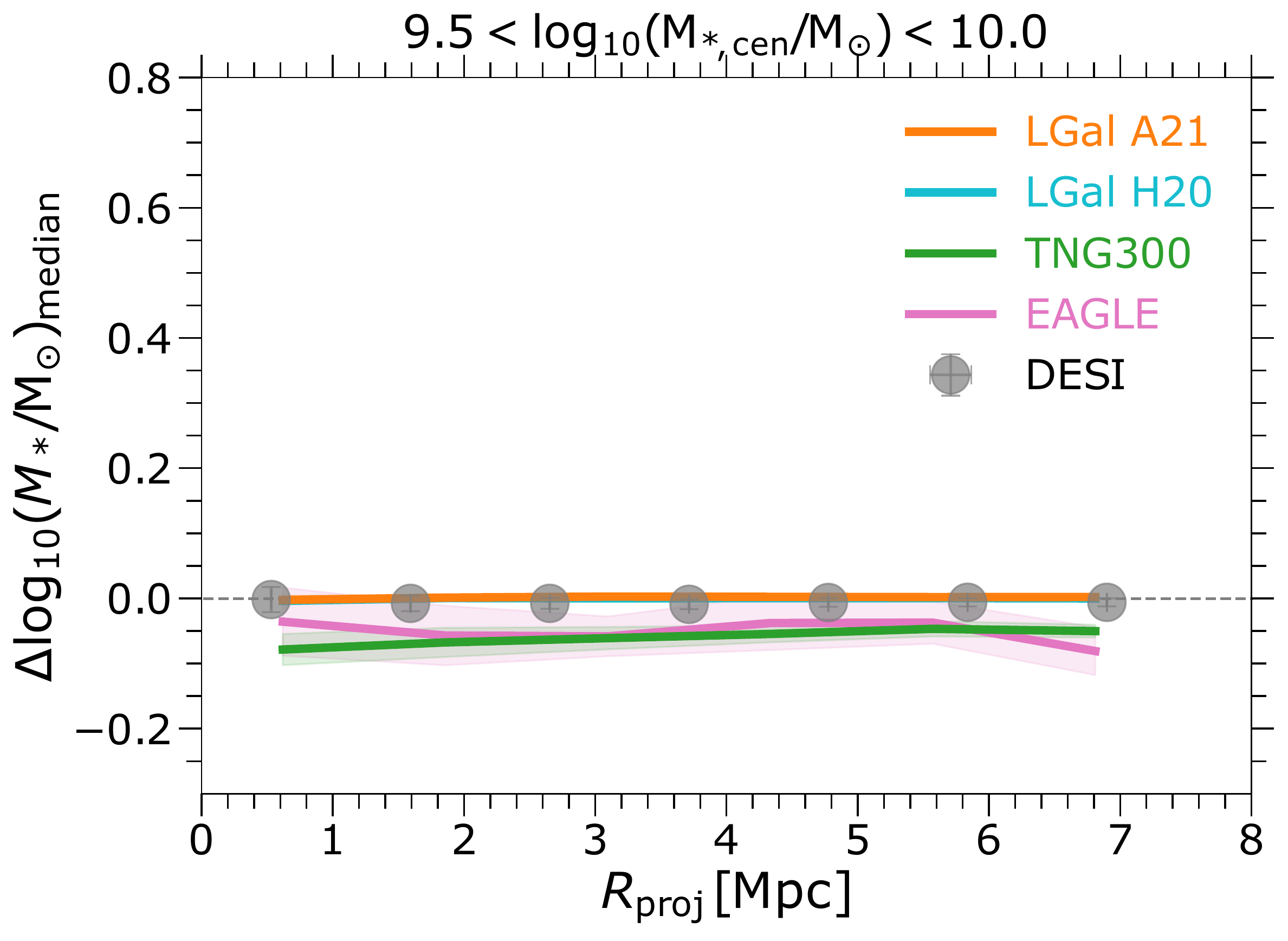}
    \includegraphics[width=0.33\textwidth]{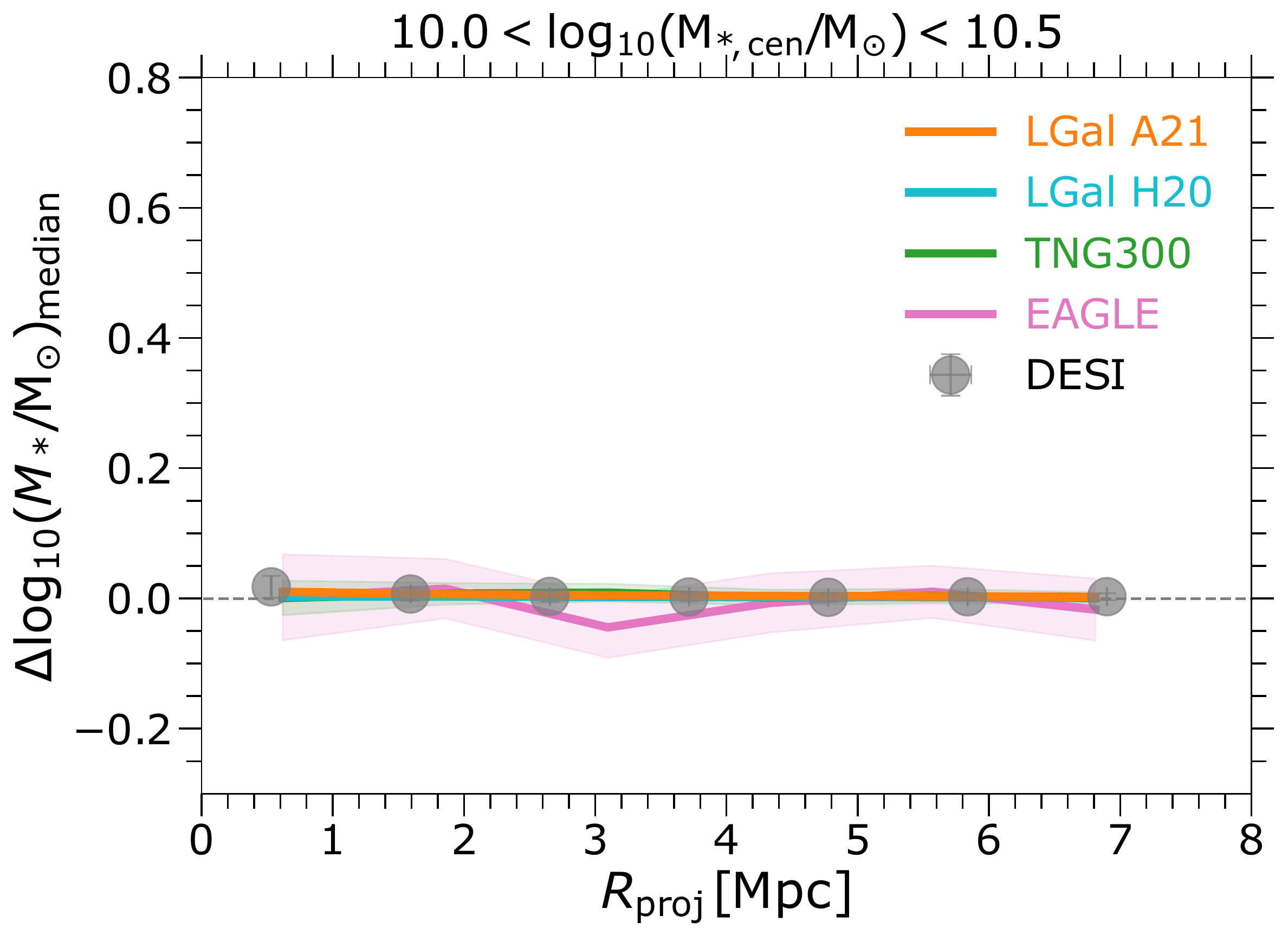}
    \includegraphics[width=0.33\textwidth]{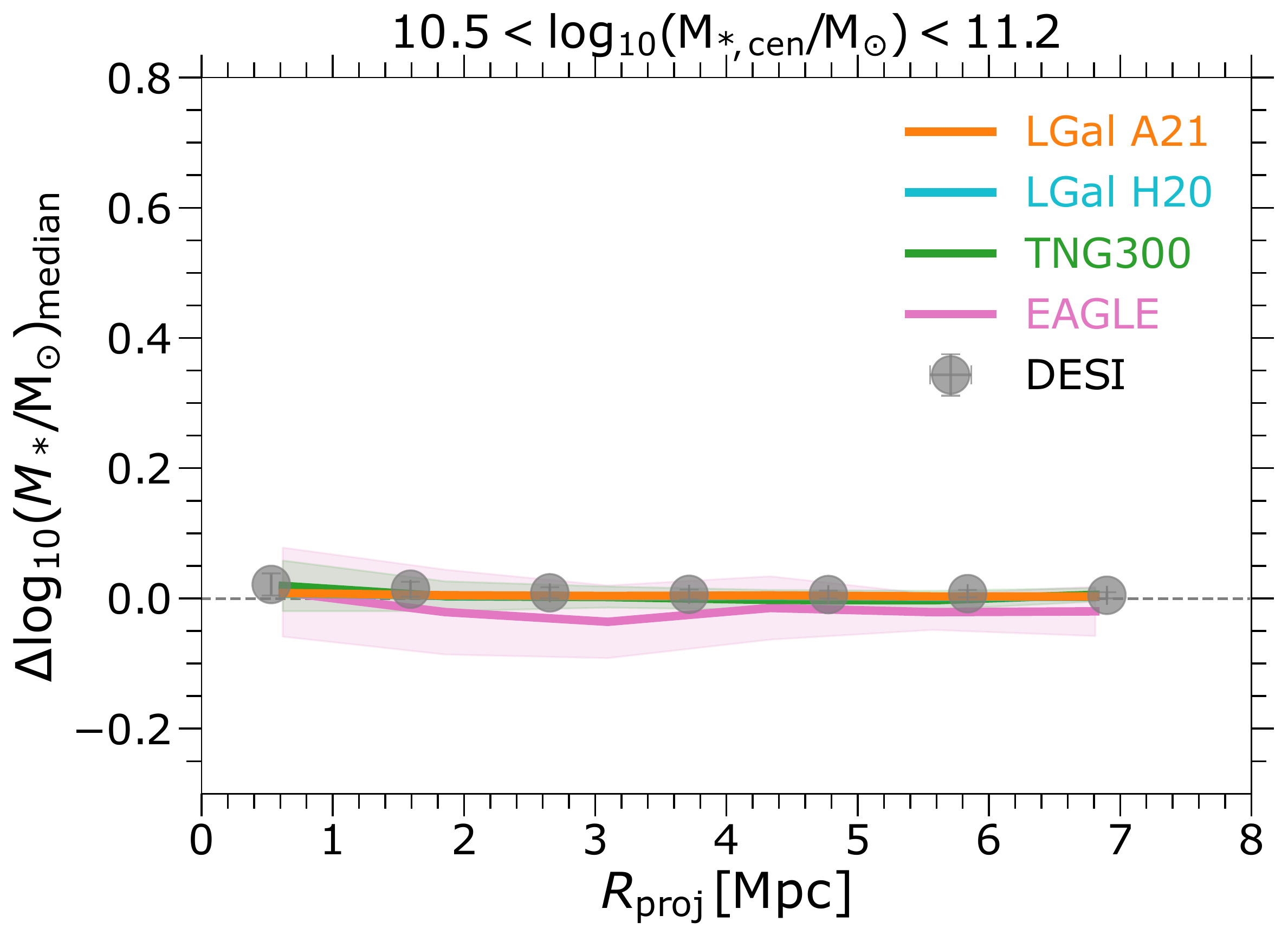}

    \caption{The conformity signal in the galaxy stellar mass: The difference between the median stellar mass of galaxies in the vicinity of star forming centrals and those in the vicinity of quenched centrals as a function of halocentric distance. The top (bottom) panel shows the results from SDSS (DESI) and its respective mock galaxy catalogues. The observational error bars and the shaded regions around the simulation results show the uncertainty of the results, derived using the bootstrap method. The figure is divided into three columns based on the primary galaxies' stellar masses. All the results are presented at $z=0$. The galactic conformity signal in stellar masses is not present at any scale.}
\label{Fig: Conformity_signal_Mstar_SDSS}
\end{figure*}
\begin{figure*}
    \centering
    {\LARGE Only \textbf{central} secondary galaxies included}\par\medskip
    \includegraphics[width=0.33\textwidth]{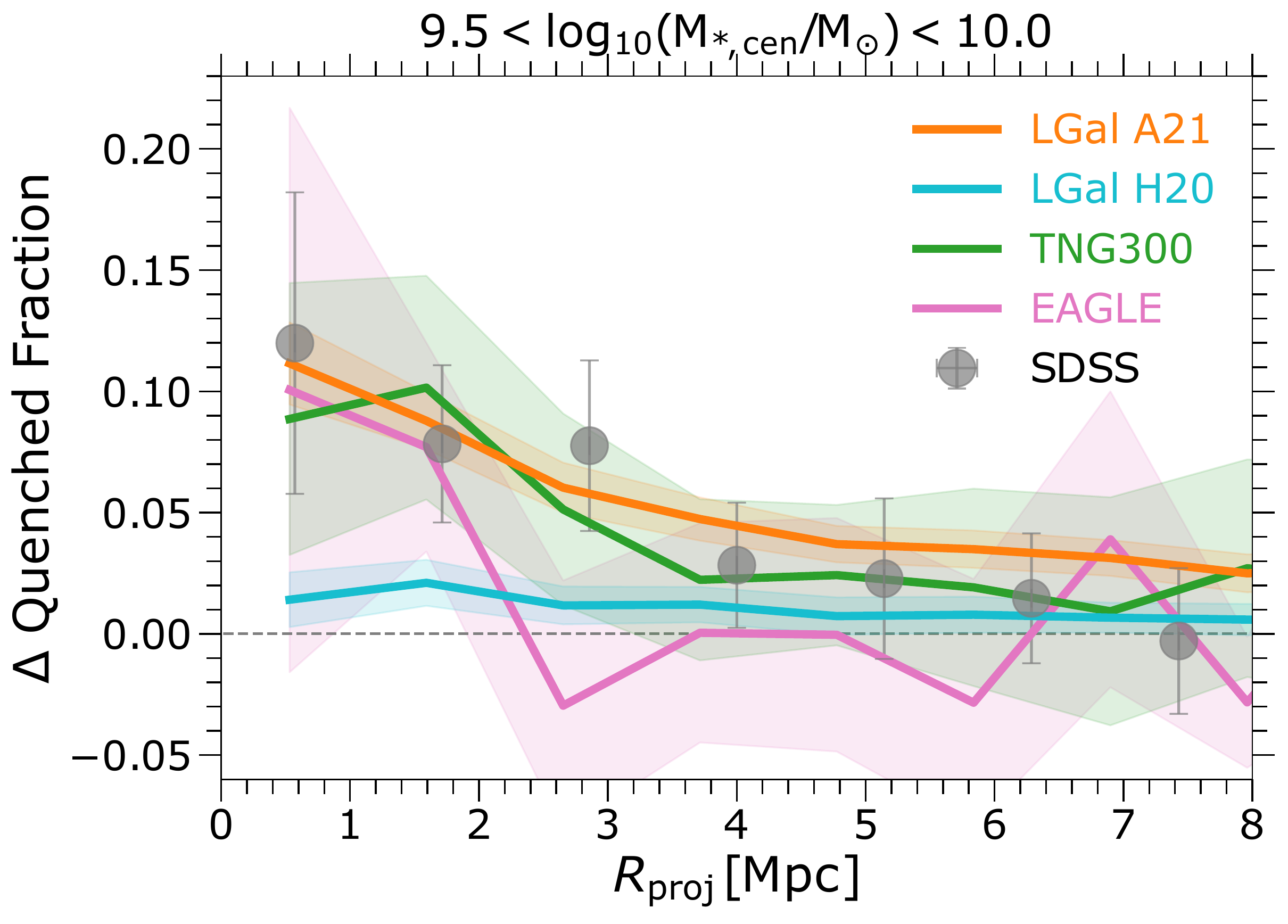}
    \includegraphics[width=0.33\textwidth]{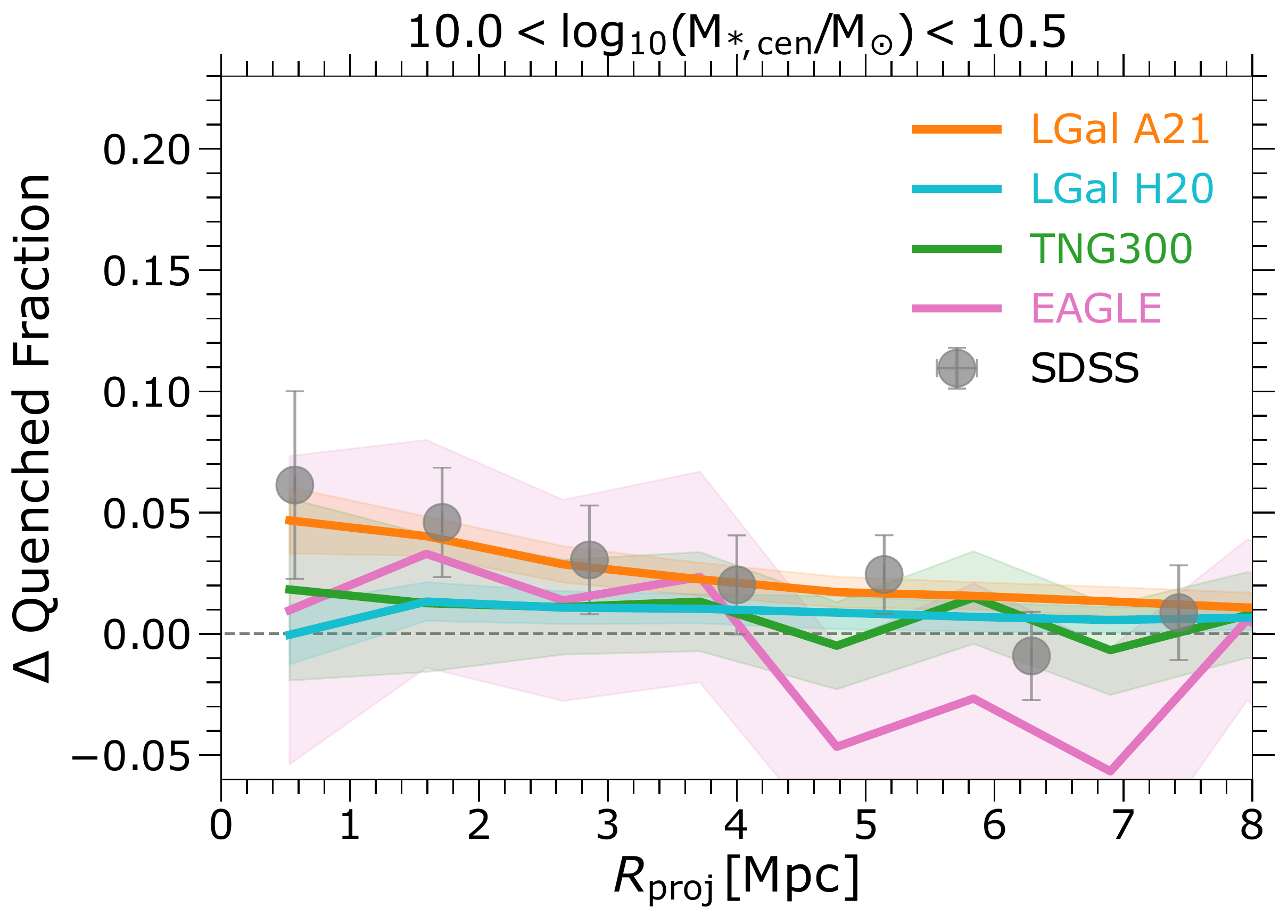}
    \includegraphics[width=0.33\textwidth]{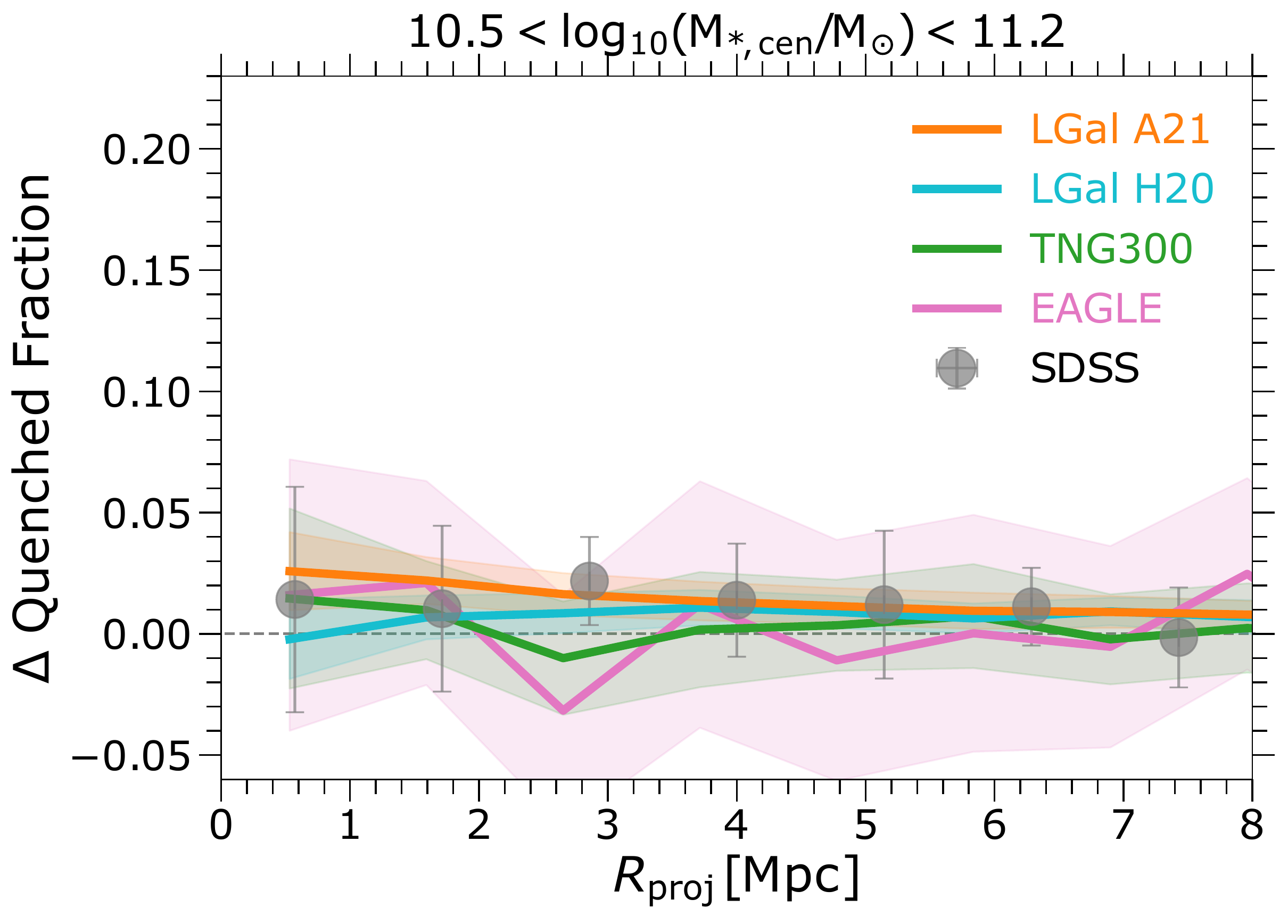}
    \par\medskip
    {\LARGE Only \textbf{satellite} secondary galaxies included}\par\medskip
    \includegraphics[width=0.33\textwidth]{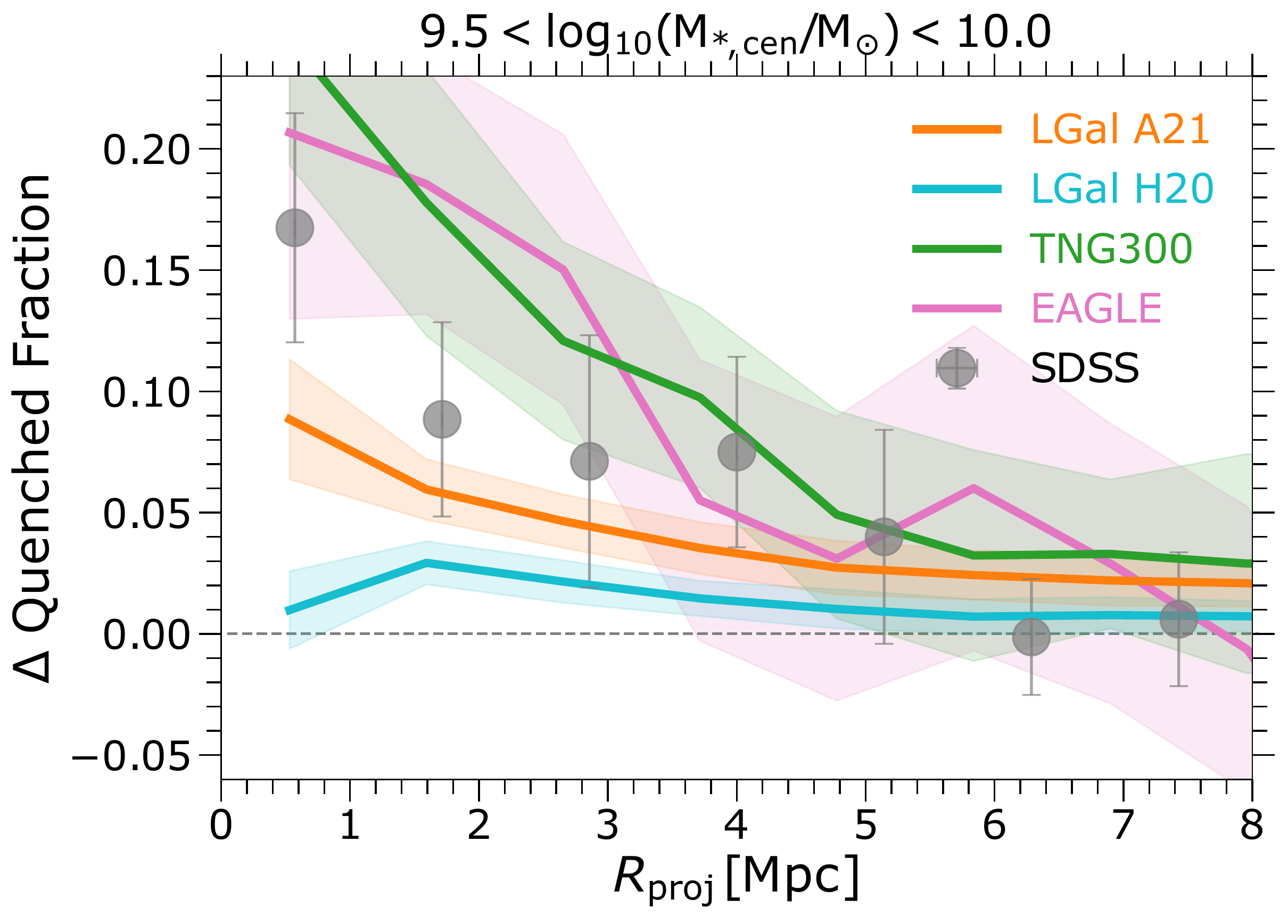}
    \includegraphics[width=0.33\textwidth]{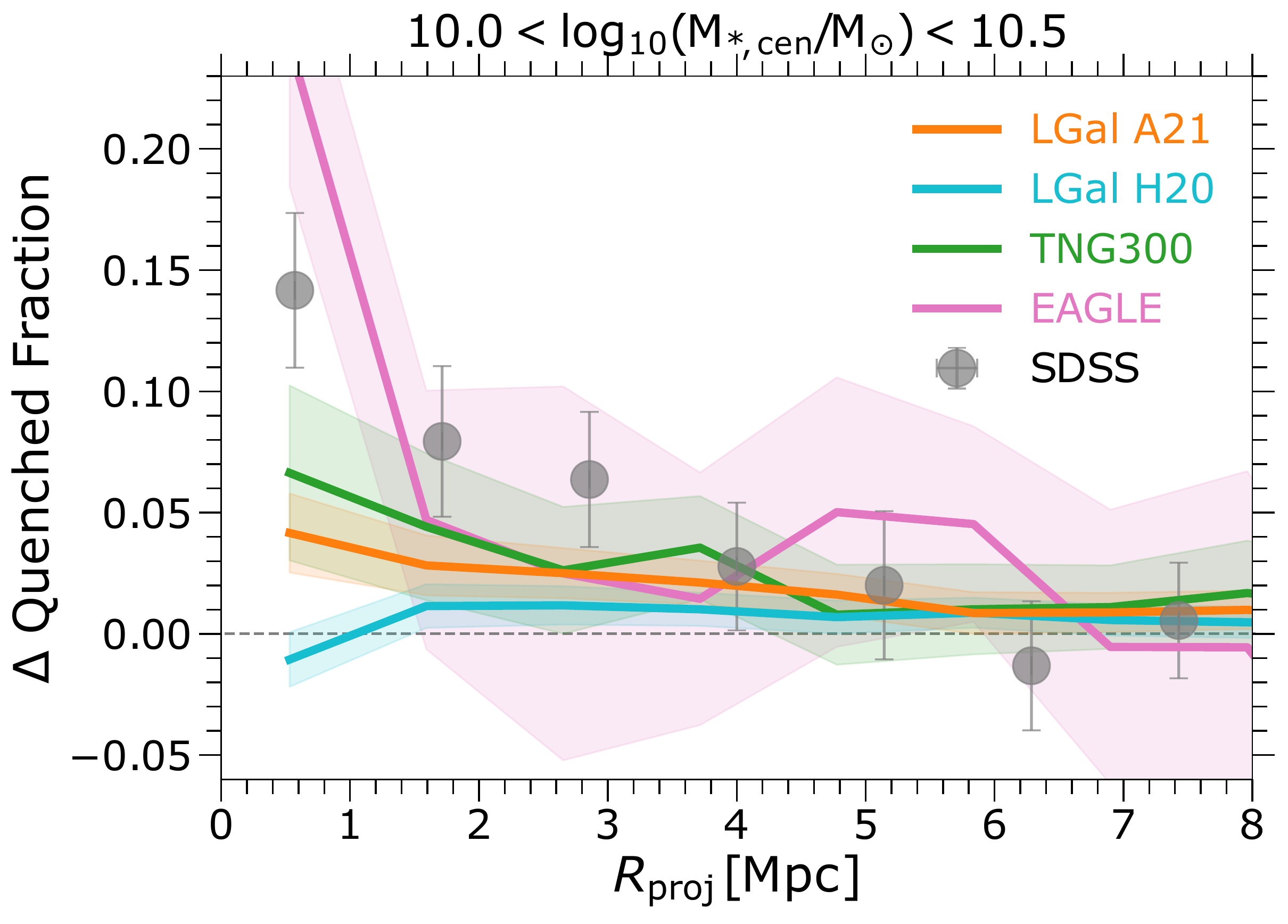}
    \includegraphics[width=0.33\textwidth]{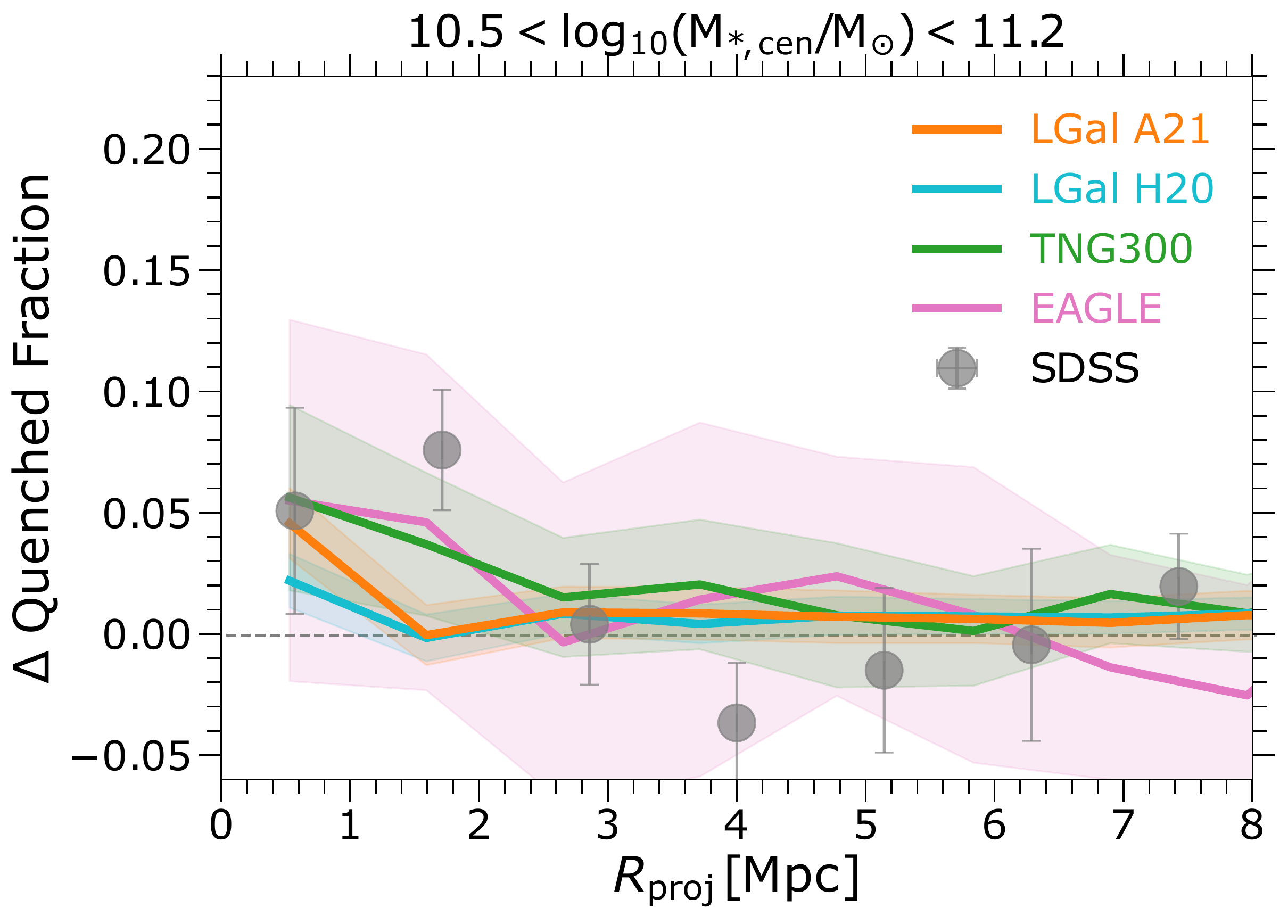}
    {\LARGE The excess of the fraction of satellites near quenched primary galaxies}\par\medskip
    \includegraphics[width=0.33\textwidth]{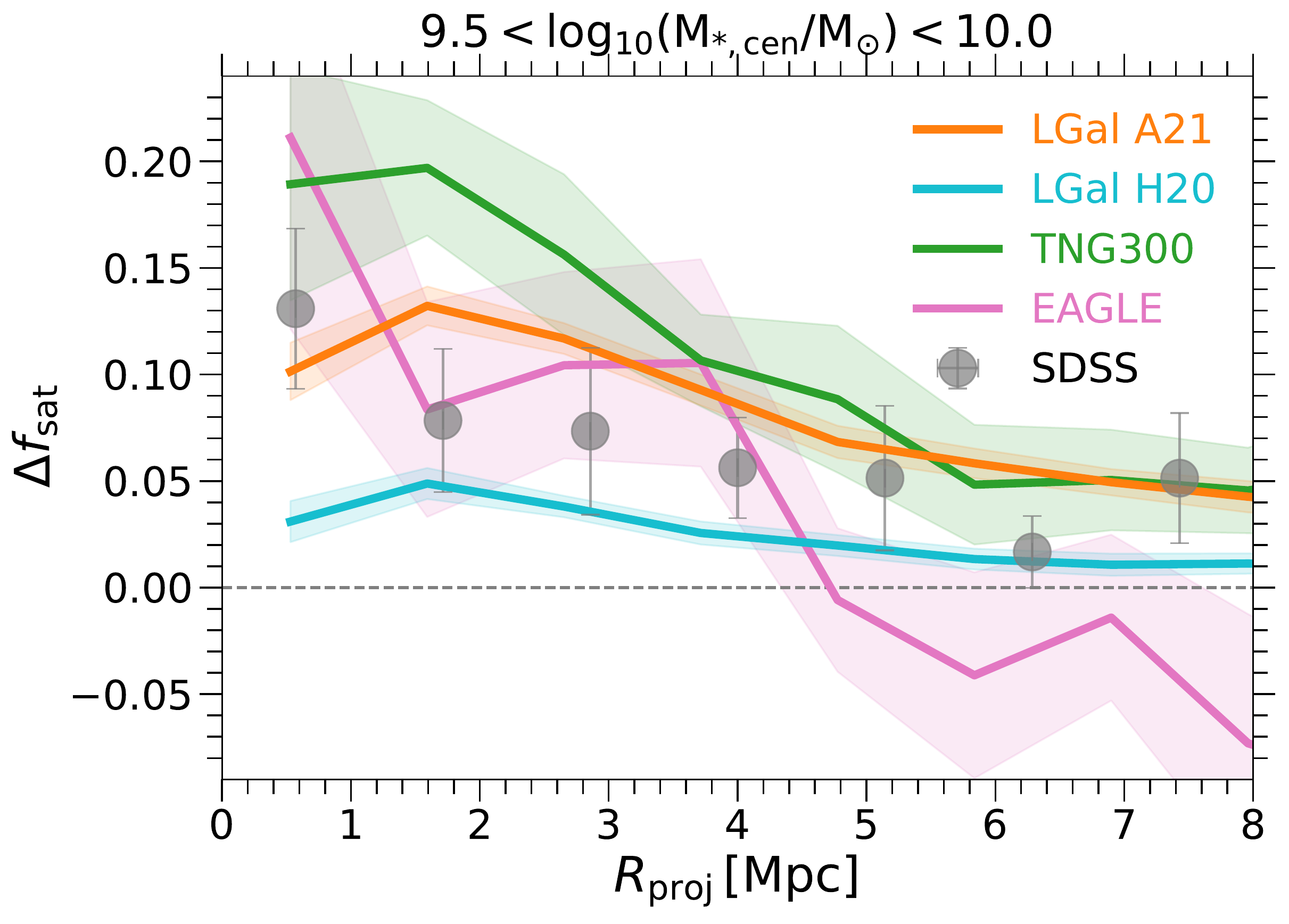}
    \includegraphics[width=0.33\textwidth]{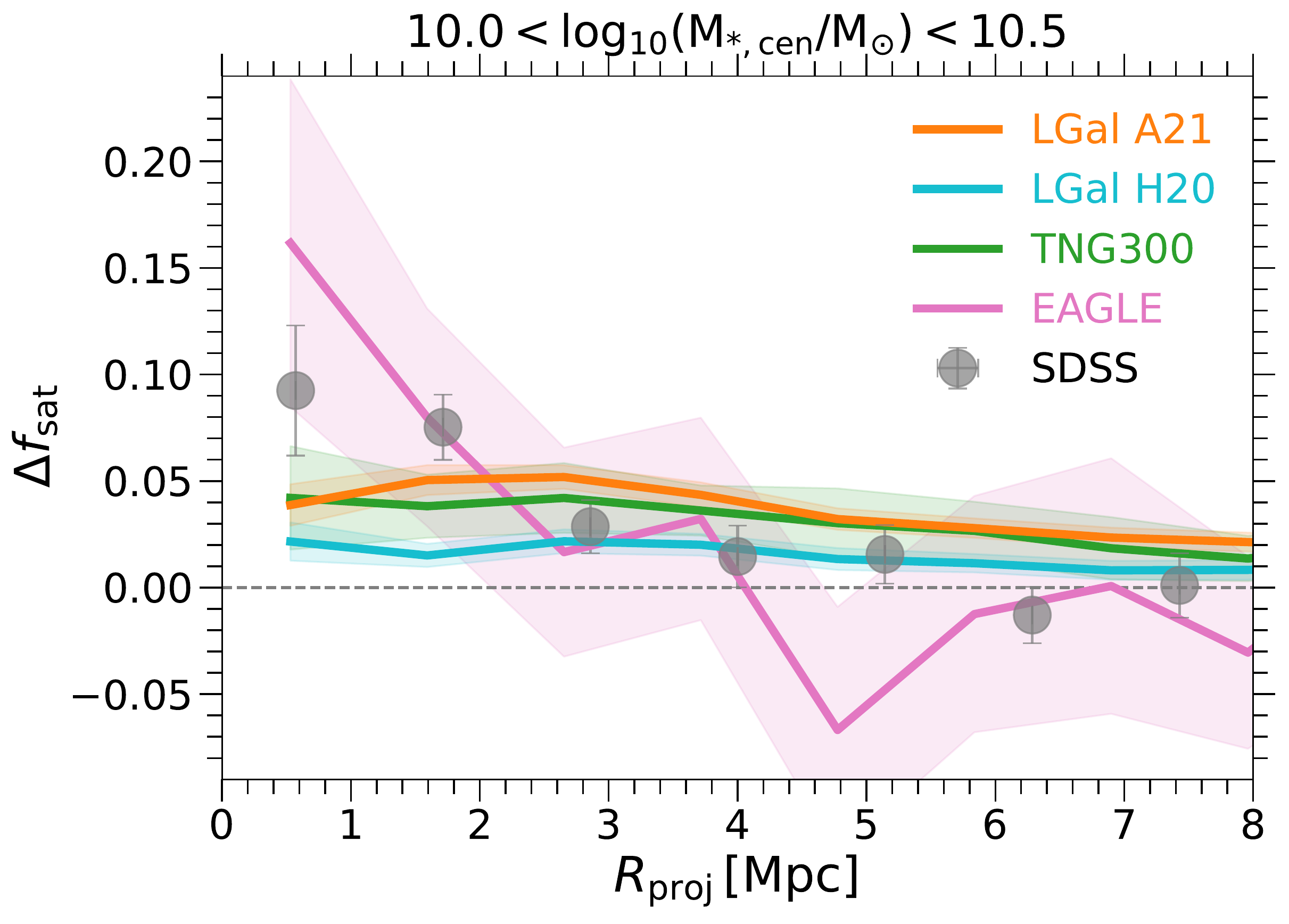}
    \includegraphics[width=0.33\textwidth]{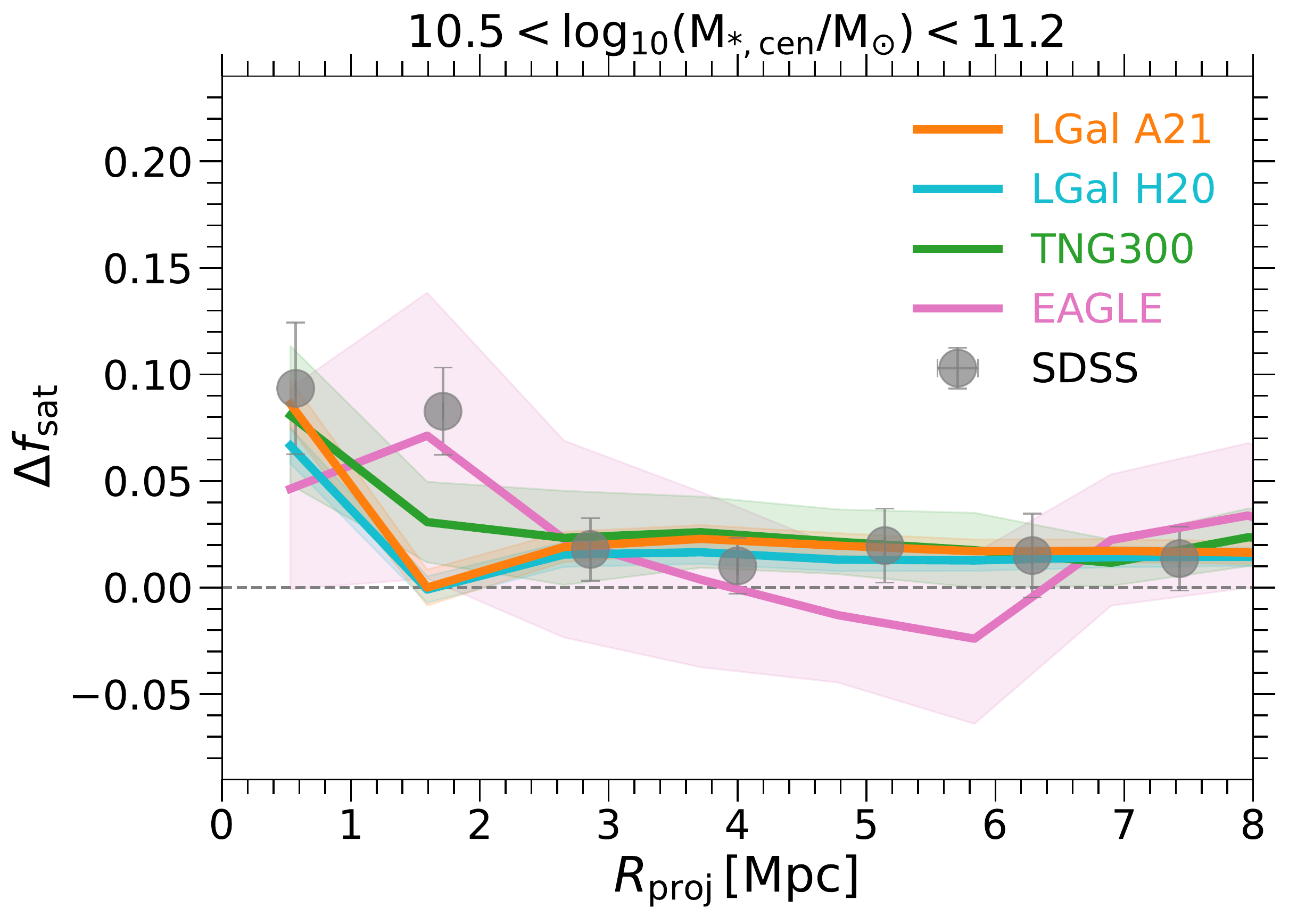}
    \caption{The conformity signal (Eq. \ref{eq: conformity_signal_f_q}) as a function of projected halocentric distance at $z\sim 0$. The top and middle panels show the conformity signal coming from central and satellite secondary galaxies, respectively. The bottom panel shows the difference between the fraction of satellite galaxies in the vicinity of quenched and star-forming primary galaxies. The observational error bars and the shaded regions around the simulation results show the uncertainty of the results, derived using the bootstrap method. The figure is divided into three columns based on the primary galaxies' stellar masses.}
\label{Fig: Conformity_signal_quenching_cen_sat}
\end{figure*}
More massive galaxies are on average less star-forming. As a result, if galaxies in the vicinity of quenched primaries were on average more massive than galaxies in the vicinity of star-forming primaries, this could cause a conformity signal. To test this idea, we examine the conformity signal for the stellar mass of galaxies, as given in Eq. \ref{eq: conformity_signal_mass}. In Fig. \ref{Fig: Conformity_signal_Mstar_SDSS}, we show the difference between the median stellar mass of galaxies in the vicinity of quenched primaries and those in the vicinity of star-forming primaries. No conformity signal can be seen in the observations or in the models. Therefore, we find no correlation between the star formation rate of a central galaxy and the median stellar mass of its neighbours.

\subsection{The contribution of central and satellite secondary galaxies to the conformity signal}
\label{subsec: secondary_cens_sats}
Secondary galaxies, which reside in the vicinity of primary galaxies, can be both central and satellite galaxies. Here, we investigate the contribution of central and satellite secondary galaxies to the conformity signal. Fig. \ref{Fig: Conformity_signal_quenching_cen_sat} shows the conformity signal where a) only central secondary galaxies are included (top panel), and b) only satellite secondary galaxies are included (middle panel). As here we compare with the SDSS observations, all central and satellite galaxies are identified using \textsc{CenSat}. We note that in this particular analysis, we do not use DESI and its mocks, due to a higher fraction of misclassified central and satellite galaxies (see Fig. \ref{Fig: Matched_Fraction2}).

Regardless of the mass of the primary galaxy, the satellite secondary galaxies in their vicinity typically contribute to the conformity signal more than the central secondaries. Nevertheless, in the vicinity of low- and intermediate-mass primaries, the contribution of central secondaries is still significant and far from negligible. We note that in some cases the contribution of the central secondaries to the observed conformity signal is comparable to or even more than satellite secondaries. For example, this can be seen at $R_{\rm proj}\sim 3\, \rm Mpc$ from low-mass primaries (left column of Fig. \ref{Fig: Conformity_signal_quenching_cen_sat}, $9.5<\log_{10}(M_{\star}/M_{\odot})<10$), where the signal from central and satellite secondaries is $\sim 0.08$ and $\sim 0.07$, respectively.

\begin{figure*}
    \centering
    % \textbf{\LARGE $z = 0$}\par\medskip
    \includegraphics[width=0.33\textwidth]{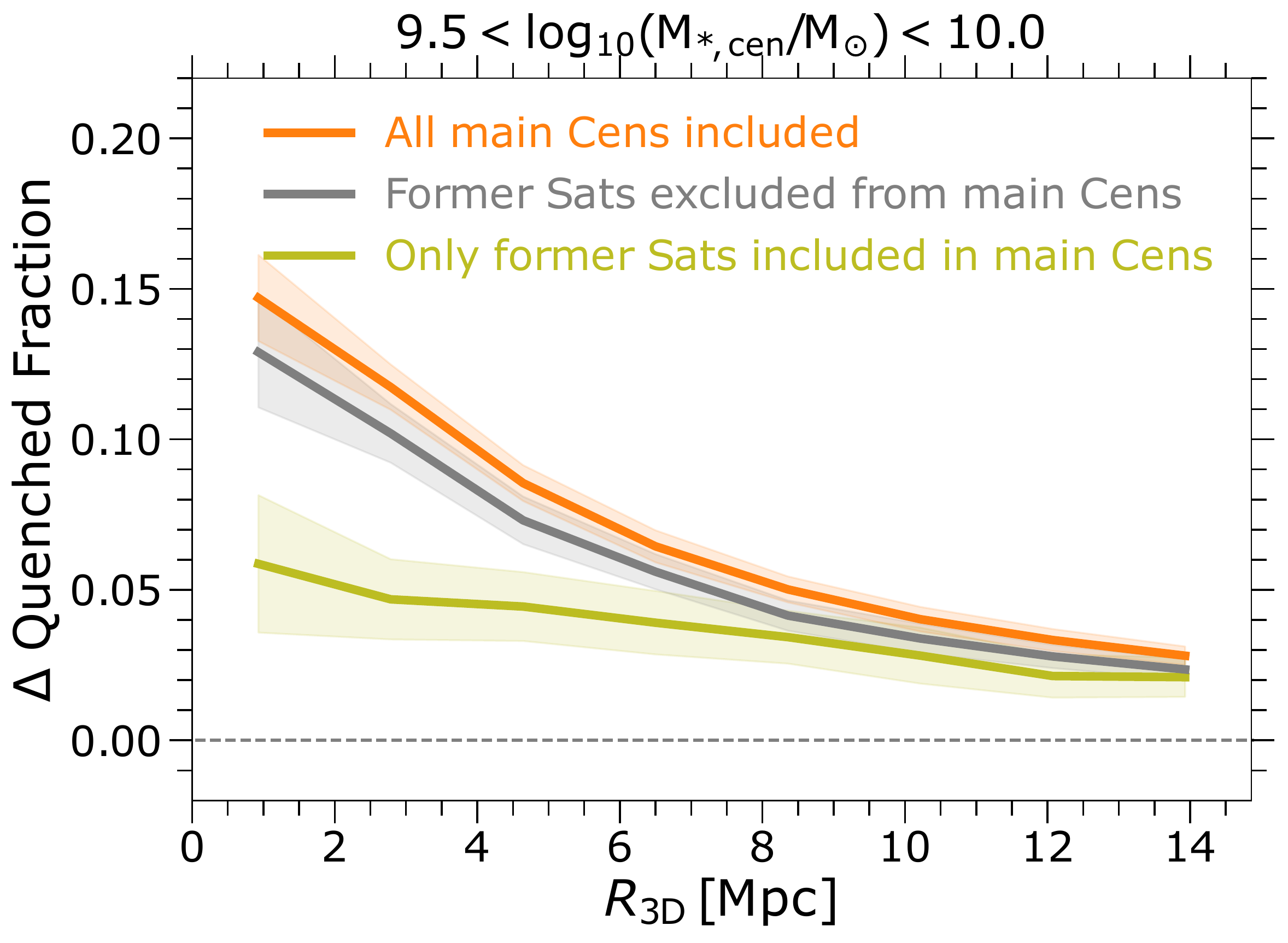}
    \includegraphics[width=0.33\textwidth]{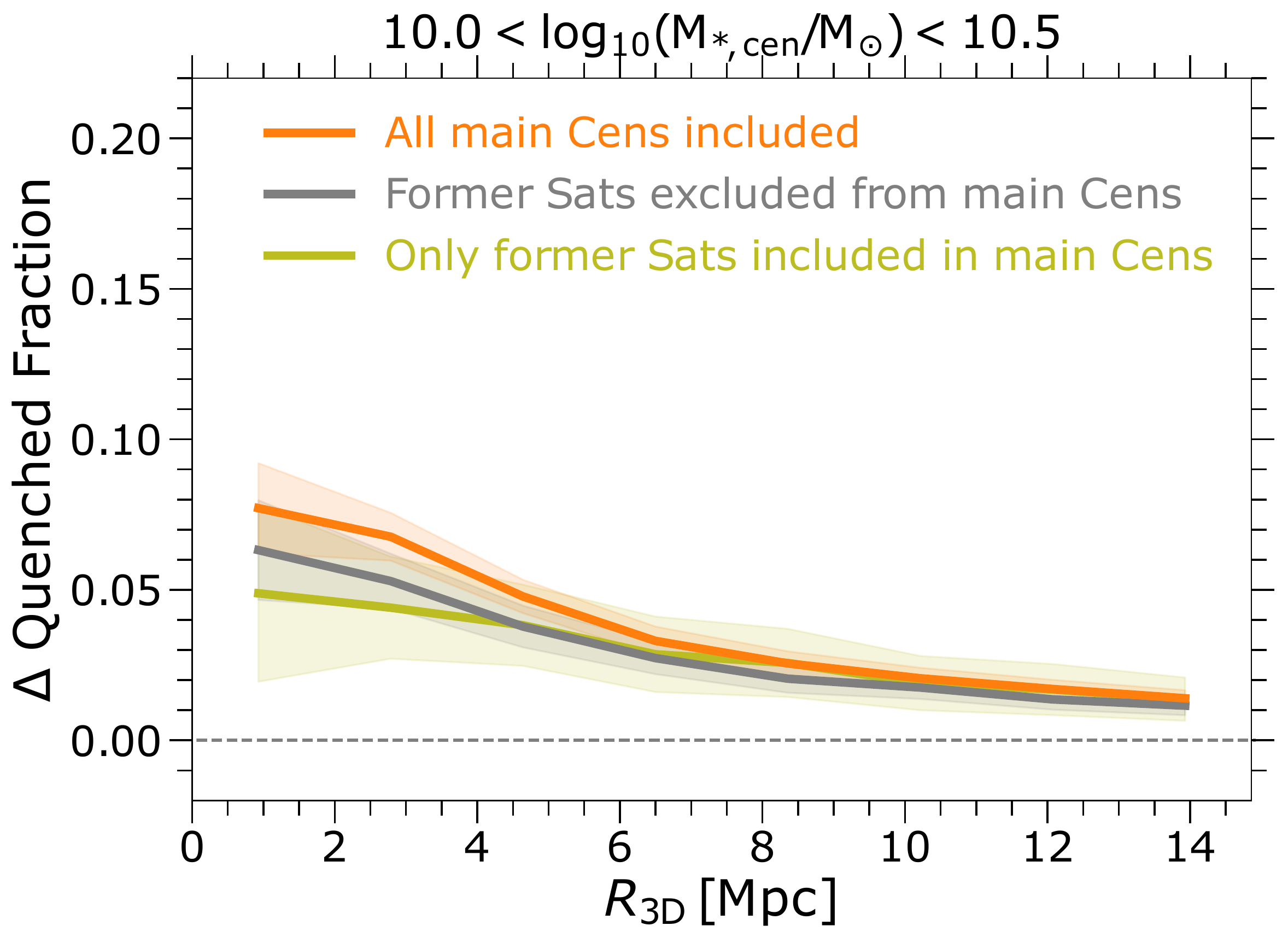}
    \includegraphics[width=0.33\textwidth]{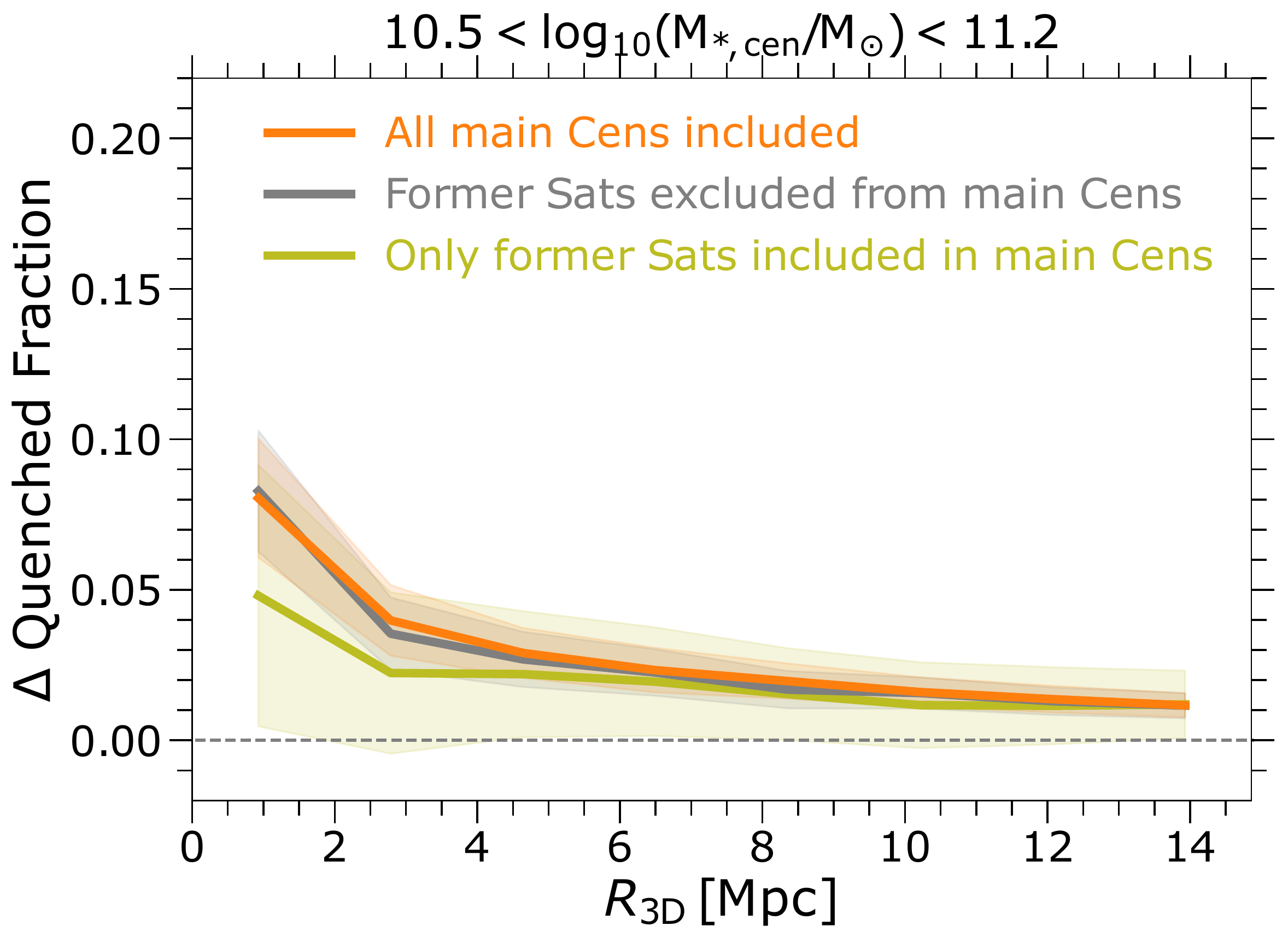}

    \caption{The galactic conformity signal in LGal-A21 at $z=0$. The orange lines are the same as the orange line in the top row Fig. \ref{Fig: Conformity_quenching_Mpc_sim} and show the signal when all primary galaxies are included. The grey lines show the signal when all former satellites are excluded from the "main" central galaxies, and the yellow lines show the signal when only former satellites are included in the main central (primary) galaxies. The shaded regions show the uncertainty of the results, derived using the bootstrap method. For all three cases, all galaxies in the vicinity of the primaries are included.}
\label{Fig: signal_backsplash}
\end{figure*}

Comparing the conformity signal generated from central secondaries in the models and SDSS (top panel of Fig. \ref{Fig: Conformity_signal_quenching_cen_sat}), we find that LGal-A21, TNG, and EAGLE are in relatively good agreement with observations, while LGal-H20 completely fails to produce any conformity signal. Given that LGal-H20 does not implement gas stripping for central galaxies and the other models do, it is very likely that the central secondaries which produce the conformity signal became quenched when they were centrals.

In the middle panel of Fig. \ref{Fig: Conformity_signal_quenching_cen_sat}, the conformity signal coming from satellite secondary galaxies is underestimated in LGal-A21, comparing to the observations. This may be due to the lack of cold star-forming gas stripping in LGal-A21. The signal in both the TNG and EAGLE hydrodynamical simulations is stronger than LGal-A21, which further confirms this hypothesis. Nevertheless, in the vicinity of low-mass primary galaxies ($9.5<\log_{10}(M_{\star}/M_{\odot})<10$), TNG and EAGLE overestimate the signal. We suspect that a combination of gas stripping processes with other intrinsic physical processes such as supernova and black hole feedback may have influenced this. We will investigate this in future work.

In order to further understand the role of central and satellite secondary galaxies in the conformity signal, we show the difference between the fraction of satellites in the vicinity of quenched and star-forming primary galaxies in the bottom panel of Fig. \ref{Fig: Conformity_signal_quenching_cen_sat}. In SDSS and all the models except for LGal-H20, we find a higher fraction of satellite galaxies in the vicinity of quenched primary galaxies. The lack of an excess of satellites near quenched primary galaxies in LGal-H20 is due to the model's lack of environmental quenching for central galaxies.

Finally, we conclude that the conformity signal is originated both from central and from satellite secondary galaxies. This indicates that a fraction, but not all, of the conformity signal originates from primary galaxies near massive systems which host quenched satellites. These satellites reside both within and in the outskirts of galaxy clusters and groups, whose environments are able to stop star formation in their satellites and generate a conformity signal.

\subsection{Today's centrals, former satellites: The case of backsplash and fly-by galaxies}
\label{subsubsec: conformity_backsplash}
Today's central galaxies could have been satellites at some point in their evolutionary history. We call these galaxies "former satellites". Two common examples include backsplash and fly-by galaxies. Backsplash galaxies are former satellites that moved away from their former host halo but will be captured by the halo in the future. On the other hand, fly-by galaxies pass through a halo but are not gravitationally bound to the halo and will not return to the halo in the future. Both backsplash and fly-by galaxies usually lose gas and dark matter due to stripping inside and in the outskirts of their former host haloes. Therefore, they are likely to be quenched and to be found near other haloes. These could impact the conformity signal, given that at $z=0$ they are classified as centrals.

To discover the contribution of former satellites to the conformity signal, we make a "former satellite" catalogue for LGal-A21 using the subhalo merger trees that describe the evolution of each subhalo/galaxy in the model. We then calculate the conformity signal in Fig. \ref{Fig: signal_backsplash} in the vicinity of the following primary galaxies: a) all centrals (orange lines), b) all centrals excluding former satellites (grey lines), c) former satellites (yellow lines). The orange lines are the same as the orange lines in the top row of Fig. \ref{Fig: Conformity_signal_quenching_models_redshift}. Comparing the orange and grey lines, we see that excluding the former satellites would decrease the conformity signal by a few per cent on the largest scales.

On smaller scales $(R_{\rm 3D}<2 \,\rm Mpc)$, in the vicinity of low- and intermediate-mass primaries (left and middle panels of Fig. \ref{Fig: signal_backsplash}), the contribution of "former satellites" to the signal goes up to 20\% (1 - the ratio between the grey and orange lines) and stays significant even up to 5 Mpc from the centres of haloes. Therefore, the predictions of the LGal-A21 model further support the idea that environmental effects are the main physical processes that create the conformity signal on large scales.

Furthermore, the low value of the conformity signal near former satellites (yellow lines) is caused by the fact that former satellites reside near more massive objects which make their neighbours on average more quenched than the average field value. Therefore, galaxies in the vicinity of former satellites have a typically high quenched fraction, regardless of the star formation of the central galaxy (i.e. the former satellite), which does not produce a significant conformity signal. Nevertheless, the non-zero value of the yellow line could come from "former satellites" used to be a part of haloes with different masses and, therefore, different abilities in quenching their satellites.

%%%%%%%%%%%%%%%%%%%% DISCUSSION AND SUMMARY %%%%%%%%%%%%%%%%%%
\section{Summary and Discussion}
\label{sec: summary}
The strength and the origin of the correlation between central (primary) galaxies and their neighbours (secondary galaxies), commonly known as galactic conformity, is relatively well understood on small scales ($R<\rm 1 \,Mpc$) but is a matter of debate on large scales (${\rm 1\, Mpc}<R<{\rm 5\, Mpc}$). In this paper, we employed two observational datasets, SDSS and DESI, and four galaxy formation models, LGal-A21, LGal-H20, TNG, and EAGLE, to find the amplitude and uncover the origin of the conformity signal.

It has been shown that the conformity signal could be susceptible to the method employed to distinguish central galaxies from satellites. In a recent study, \cite{Lacerna2018conformity} showed how different isolation criteria for selecting central galaxies leads to a rather considerable change in the amplitude of the conformity signal \citep[see also][]{Sin2017Conformity,Tinker2018Conformity}. In general, satellites within groups and clusters tend to be more quenched than field galaxies with the same stellar mass. Therefore, if such a satellite galaxy is misclassified as a central, its properties (e.g. star formation) would show strong correlations with galaxies in its vicinity, since all of them are influenced by environmental effects within the cluster or group in which they reside. This could lead to an artificial conformity signal. On the other hand, if a central galaxy is misclassified as a satellite, it would work the opposite way.

To avoid such problems, in section \ref{subsec: conformity_sims} we analysed pure simulation results with well-known halo finders such as FOF and \textsc{Subfind} in 3D space to avoid any projection effects and minimise possible misclassification of satellites as centrals. Moreover, we devised \textsc{CenSat}, a new algorithm to identify central and satellite galaxies in two spatial and one line-of-sight velocity (redshift) dimension. We calibrated \textsc{CenSat} against simulation results and \textsc{Subfind}. As a result of this calibration, all galaxies in the vicinity of centrals out to a projection distance of $R = 1.5R_{\rm 200,halo}$ are considered satellites. Therefore, it is very unlikely that \textsc{CenSat} misclassifies a satellite within $R_{\rm 200,halo}$ as a central.

Considering the aforementioned points, we first analysed the conformity signal in the galaxy formation models by employing three-dimensional distances and using \textsc{Subfind} to identify central and satellite galaxies. Our principal results are as follows:
\begin{itemize}
    \item The conformity signal in the quenched fraction of galaxies is present in LGal-A21, TNG, and EAGLE, out to a 3-dimensional distance of at least $R_{\rm 3D} = 5 \rm \, Mpc$ from $z=0$ out to at least $z=2$. In contrast, the signal is very weak in LGal-H20 (Figs. \ref{Fig: Conformity_quenching_Mpc_sim},\ref{Fig: Conformity_signal_quenching_models_redshift}).
    \item For reference, at $z=0$ in the vicinity of low-mass primary galaxies ($9.5<\log_{10}(M_{\star}/{\rm M_{\odot}})<10$) the amplitude of the signal at $R_{\rm 3D}\sim 3 \rm \,Mpc$ is: $\sim 0.28$ in TNG, $\sim 0.25$ in EAGLE, $\sim 0.12$ in LGal-A21, and $\sim 0.03$ in LGal-H20 (Fig. \ref{Fig: Conformity_signal_quenching_models_redshift}).
\end{itemize}

Next, we investigated galactic conformity in SDSS and DESI observations and compared the results with the galaxy formation models by making careful mock catalogues. We operated \textsc{CenSat} to identify central and satellite galaxies uniformly both in the models and in the observations. Our main findings are:
\begin{itemize}
    \item In the SDSS and DESI observations, the conformity signal is present both in the fraction of quenched galaxies and in their specific star formation rates out to a projected halocentric distance of $R_{\rm proj}=4\, \rm Mpc$, with mean amplitude slightly higher in SDSS than in DESI (Figs. \ref{Fig: Conformity_signal_quenching_SDSS} and \ref{Fig: Conformity_signal_ssfr_SDSS}).
    \item Among the mock galaxy catalogues that we have examined, LGal-A21, TNG, and EAGLE are in relatively good agreement with observations on large scales $(R_{\rm proj}>1 \,\rm Mpc)$, while LGal-H20 shows almost no sign of galactic conformity. For reference, at $R_{\rm proj} \sim 3\rm \, Mpc$ in SDSS and its mock galaxy catalogues, the signal in the vicinity of low-mass galaxies ($9.5<\log_{10}(M_{\star}/{\rm M_{\odot}})<10$) is: $\sim 0.13$ in TNG, $\sim 0.08$ in EAGLE, $\sim 0.07$ in SDSS, $\sim 0.07$ in LGal-A21, and $\sim 0.01$ in LGal-H20 (Figs. \ref{Fig: Conformity_signal_quenching_SDSS}, \ref{Fig: Conformity_signal_ssfr_SDSS}).
    \item The signal is stronger in the vicinity of low-mass primary galaxies ($9.5<\log_{10}(M_{\star}/{\rm M_{\odot}})<10$) and decreases with the stellar mass of the primary galaxy, although it remains non-negligible even in the vicinity of more massive galaxies ($10.5<\log_{10}(M_{\star}/{\rm M_{\odot}})<11.2$).
    \item We find no conformity in the galaxy stellar mass in the observations or the models; i.e. there is no difference between the median stellar mass of galaxies in the vicinity of quenched primary galaxies with those in the vicinity of star-forming primaries (Fig. \ref{Fig: Conformity_signal_Mstar_SDSS}).
\end{itemize}
    
To uncover the origin of the conformity signal, we took advantage of the differences between the models. We make our conclusion with the help of the following:
\begin{itemize}
    \item LGal-H20 is the only model that is similar to traditional Halo Occupation Distribution (HOD) models in that environmental effects outside the halo boundary are not taken into account. LGal-A21 applies environmental processes uniformly to all galaxies, including both centrals and satellites, and TNG and EAGLE resolve most of these effects naturally.
    \item We analysed the contribution of "former satellites" (backsplash and fly-by galaxies) to the conformity signal in LGal-A21. These "currently central galaxies" have been satellites at some point in their evolutionary history. We found that "former satellites" contribute to 0-20\% of the conformity signal, although their contribution becomes smaller at large scales (Fig. \ref{Fig: signal_backsplash})
    \item Moreover, we examined the contribution of the central and satellite secondary galaxies to the conformity signal in SDSS and the models. We found that both groups make a non-negligible impact on the signal, although the conformity signal generated by the satellite secondaries is typically stronger.
\end{itemize}

Based on the above points, we conclude that the conformity signal on large scales is likely produced by environmental effects, possibly with a considerable contribution from gas stripping operating on satellites beyond the halo boundary and on central galaxies in the outskirts of massive haloes. This may occur when moving through the warm-hot intergalactic medium or passing through filaments. This complements the more substantial gas stripping within halo boundaries. The contribution of former satellites (currently primaries) as well as satellite secondary galaxies to the signal support the mentioned scenario, that environmental effects generate the conformity signal. Furthermore, it shows that a fraction of the conformity signal comes from primary centrals near massive systems.

Finally, we check to see whether our results and conclusions are influenced by the misclassification of satellite galaxies as centrals. A comparison between the conformity signal found in 3D in our simulation snapshots (Fig. \ref{Fig: Conformity_signal_quenching_models_redshift}) and that found in our mock catalogues when analysed with \textsc{CenSat} (Fig. \ref{Fig: Conformity_signal_quenching_SDSS}) is shown in in Fig. \ref{Fig: signal_sims_mocks} in Appendix \ref{app: sims_and_mocks}. This demonstrates not only that we do not overestimate the conformity signal, but also that the actual 3D signal could be up to 50\% larger than the signal we get in 2D projected galaxy data.

The remaining tensions between the models and observations include an underestimation of the observed conformity signal at $R<3\rm \, Mpc$ in the vicinity of intermediate-mass primary galaxies ($10<\log_{10}(M_{\star}/{\rm M_{\odot}})<10.5$) in the models. We argue that one possible explanation is stronger gas stripping in the real data due to a combination of ram-pressure stripping and feedback processes, which could occur in multiple ways that we will examine in future work. For instance, the gas ejected from a galaxy or its subhalo due to supernova or AGN feedback is less bound to the galaxy and, therefore, more easily stripped by its surrounding medium. Furthermore, an enhancement in ram-pressure stripping could happen near AGN hosts if infalling galaxies hit ejected gas moving fast towards them in the opposite direction. Evaluating these scenarios requires studying the impact of baryonic feedback processes on gas distribution and kinematics from tens of kiloparsecs to tens of Megaparsecs scales, as well as their connections with the local background environment (LBE) properties of galaxies.

\section*{Data Availability}
We have made the "former satellites" (backsplash and fly-by galaxies) catalogues of LGal-A21 publicly available at \href{https://lgalaxiespublicrelease.github.io/}{https://lgalaxiespublicrelease.github.io/}, where the LGal-A21 and LGal-H20 semi-analytical models are already publicly available.
The TNG (\href{https://www.tng-project.org/}{https://www.tng-project.org/}) and EAGLE (\href{http://icc.dur.ac.uk/Eagle/}{http://icc.dur.ac.uk/Eagle/}) simulations, as well as the SDSS (\href{https://www.mpa-garching.mpg.de/SDSS/DR7/}{https://www.mpa-garching.mpg.de/SDSS/DR7/}) and DESI (\href{http://cdsarc.u-strasbg.fr/viz-bin/cat/J/ApJS/242/8}{http://cdsarc.u-strasbg.fr/viz-bin/cat/J/ApJS/242/8}) observations are also publicly available. The codes performed for analysis and the other datasets/plots generated in this paper are available from the corresponding author upon reasonable request.

\section*{Acknowledgements}
MA would like to thank Dylan Nelson, Annalisa Pillepich, and Volker Springel for fruitful discussions. MA acknowledges funding from the Deutsche Forschungsgemeinschaft (DFG) through an Emmy Noether Research Group (grant number NE 2441/1-1).

\begin{figure*}
\label{Fig: signal_sims_mocks}
    \centering
    \includegraphics[width=0.33\textwidth]{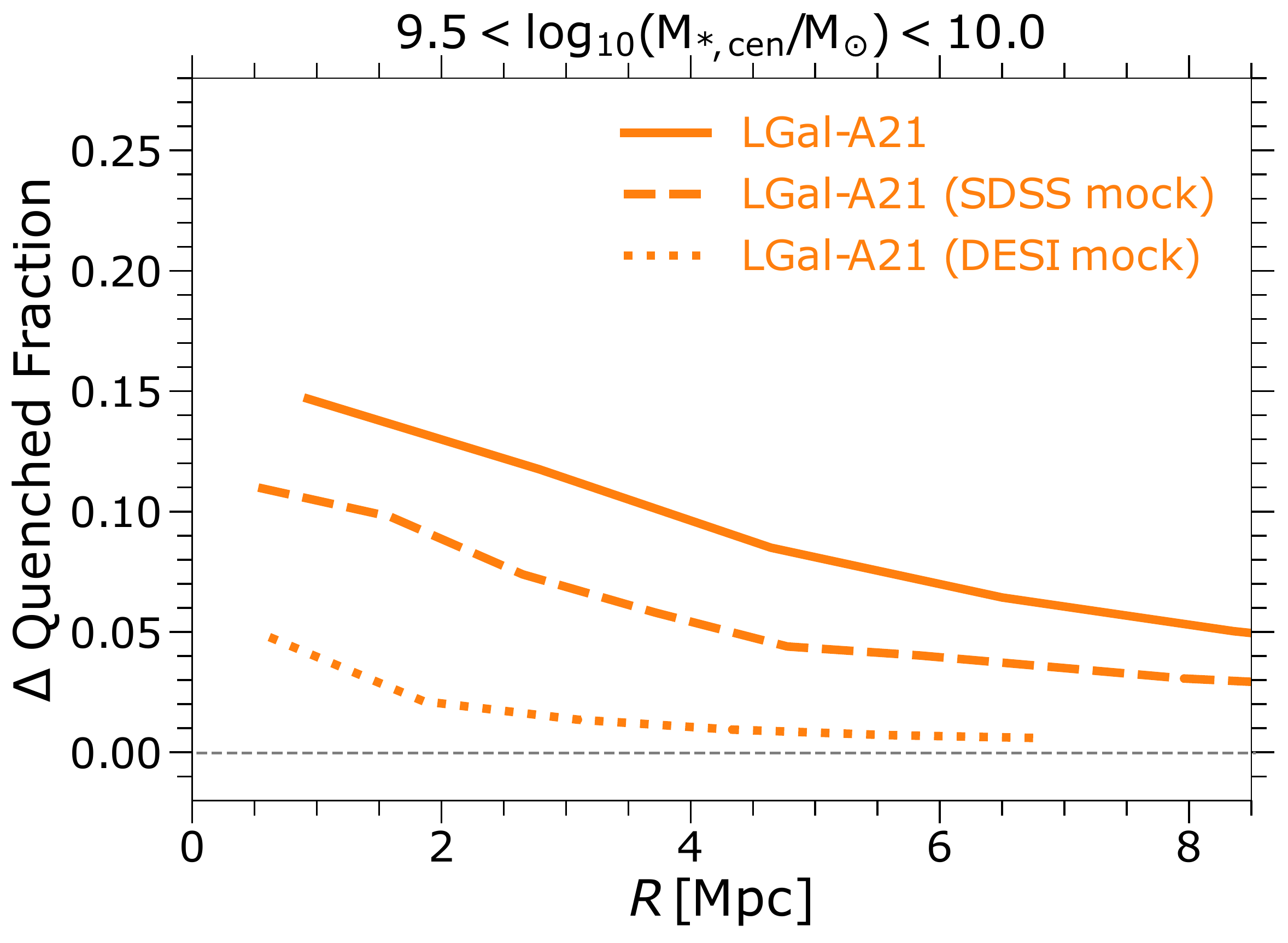}
    \includegraphics[width=0.33\textwidth]{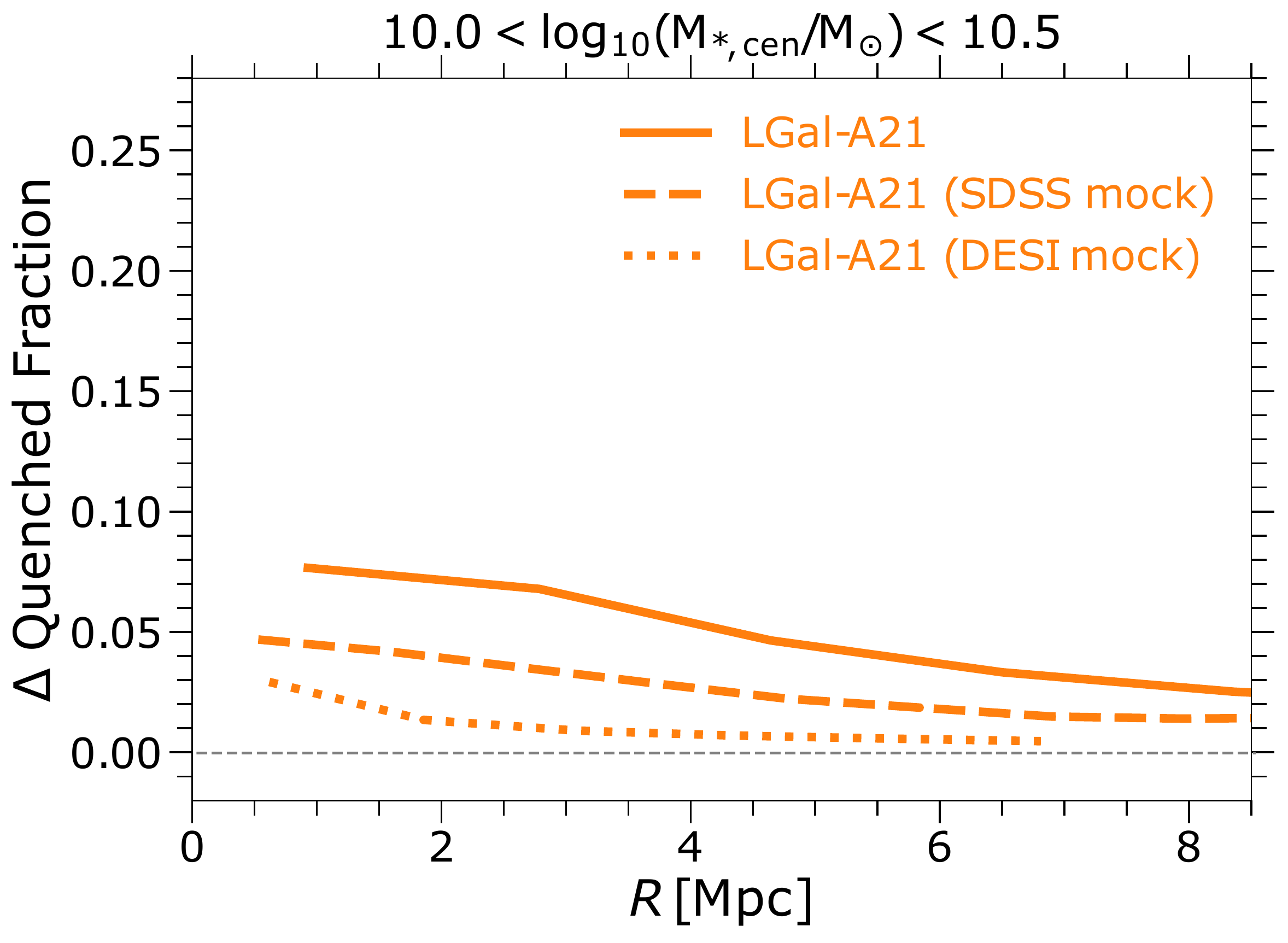}
    \includegraphics[width=0.33\textwidth]{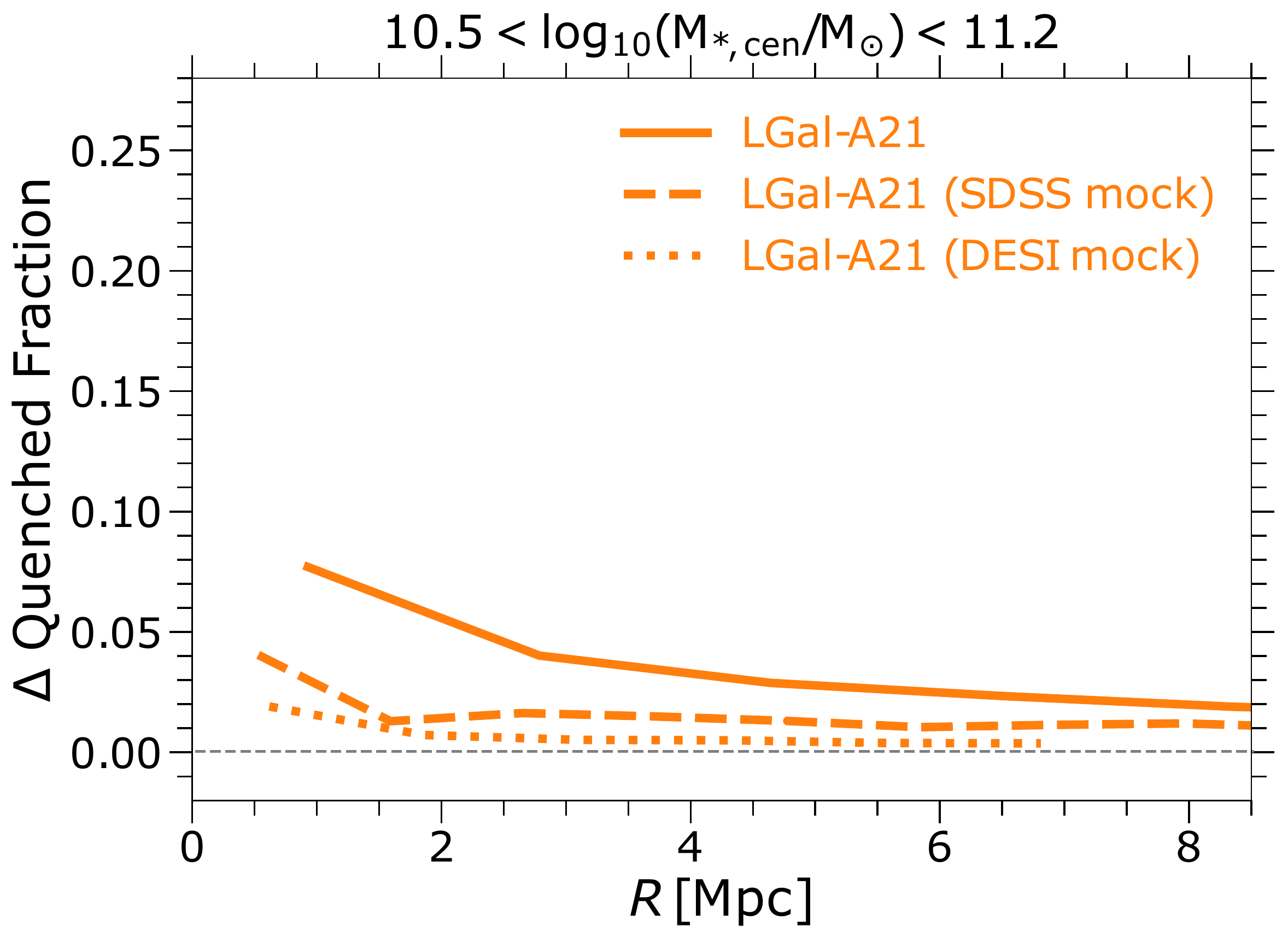}
    \includegraphics[width=0.33\textwidth]{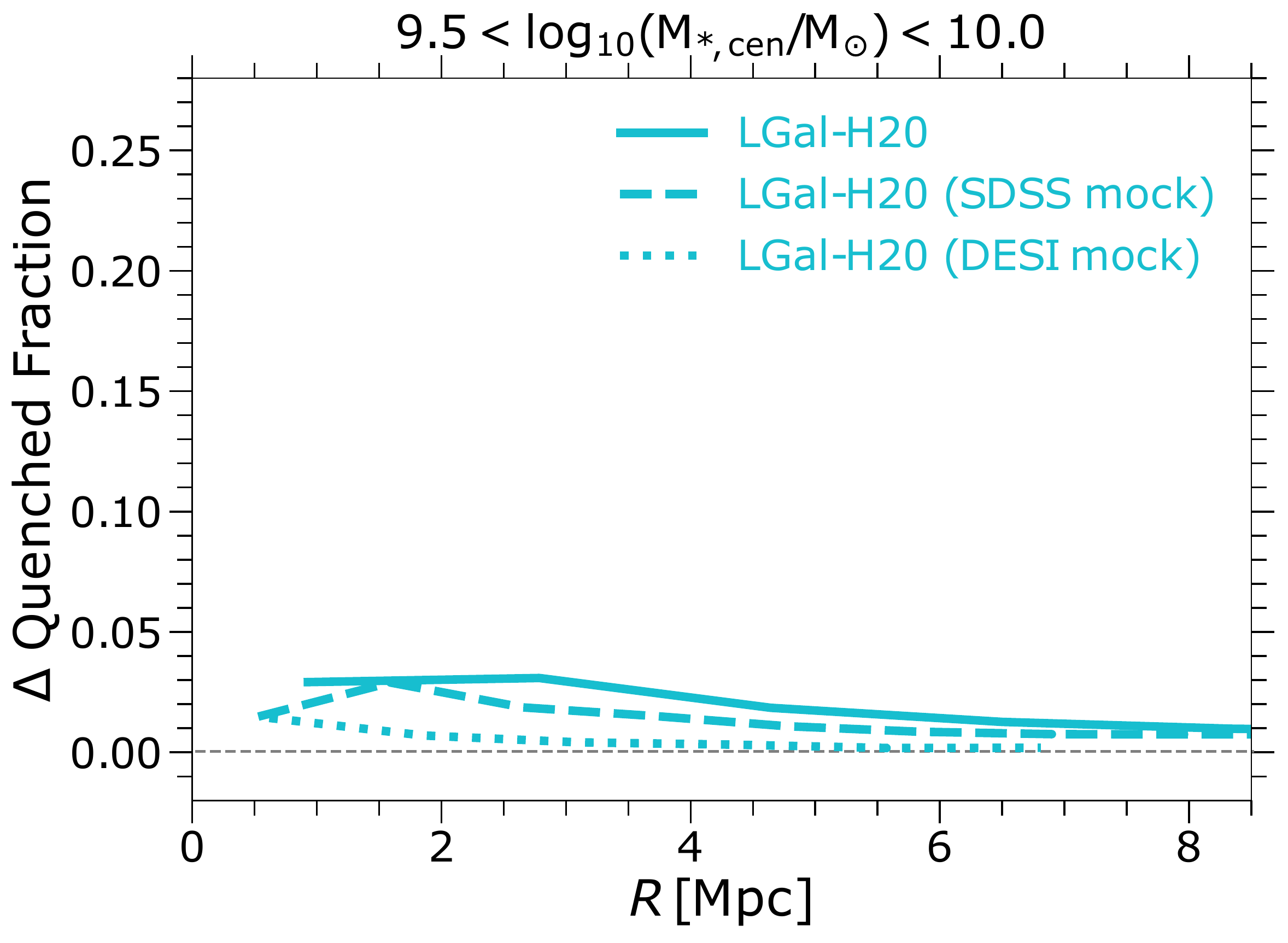}
    \includegraphics[width=0.33\textwidth]{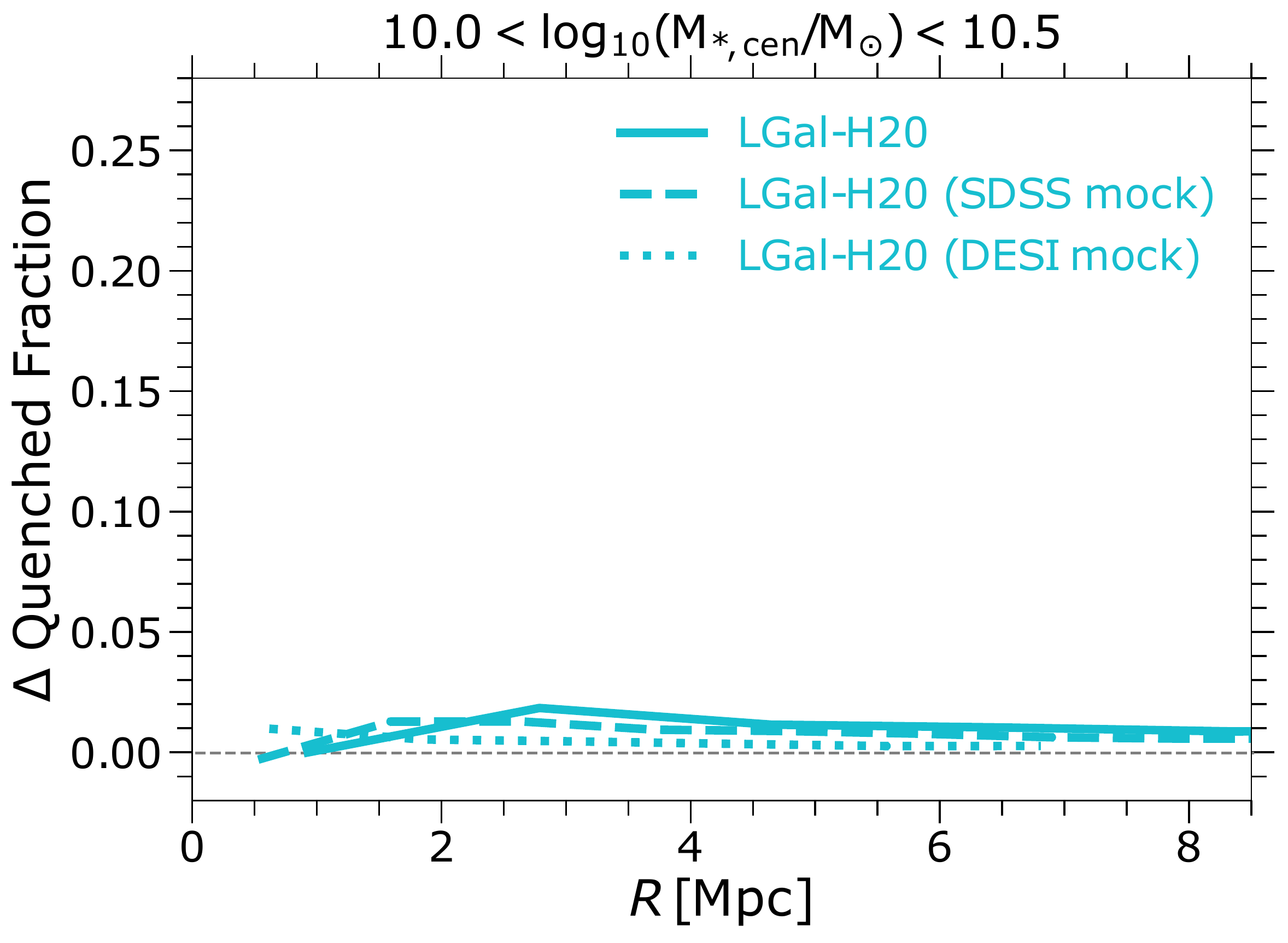}
    \includegraphics[width=0.33\textwidth]{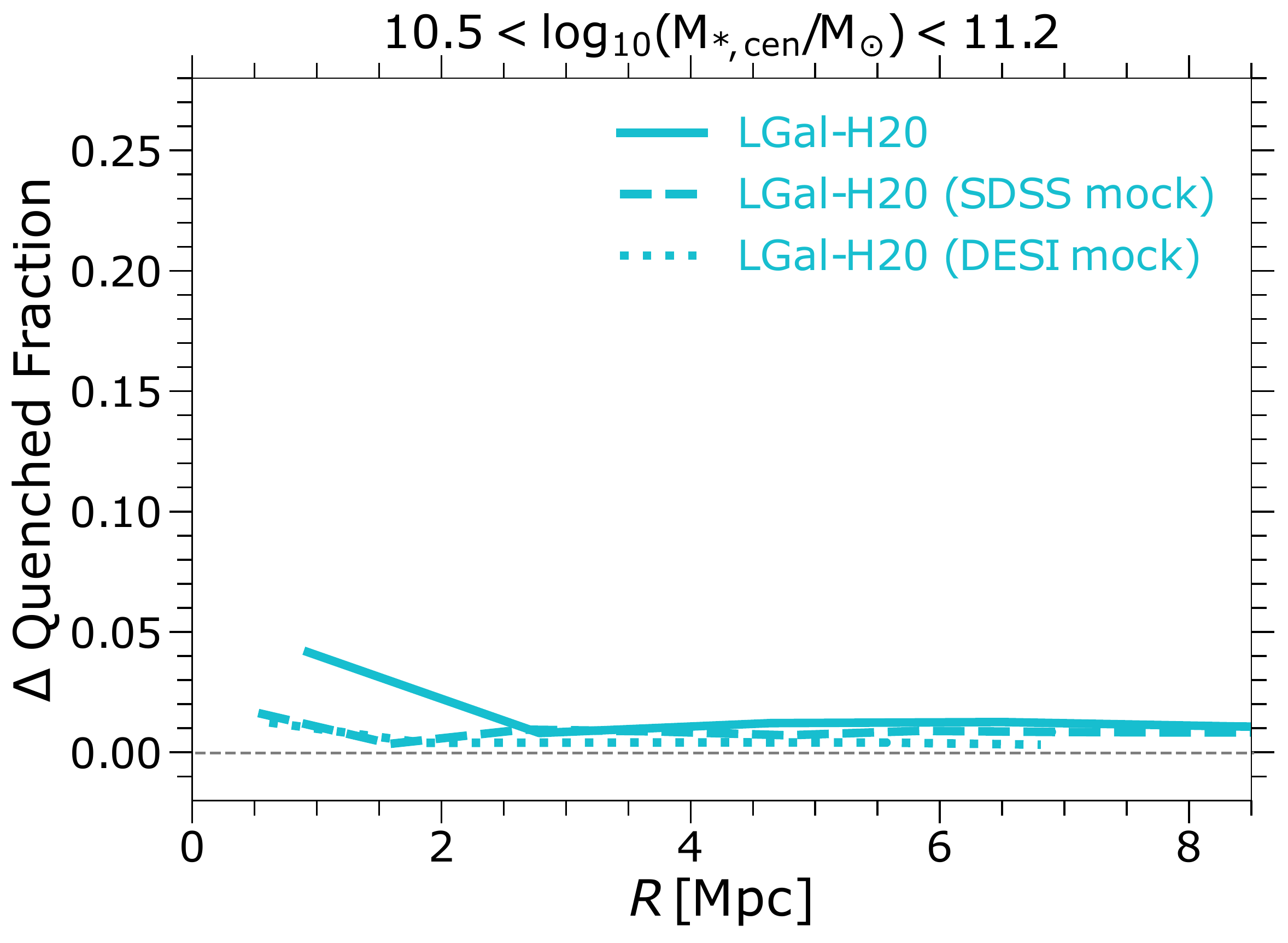}
    \includegraphics[width=0.33\textwidth]{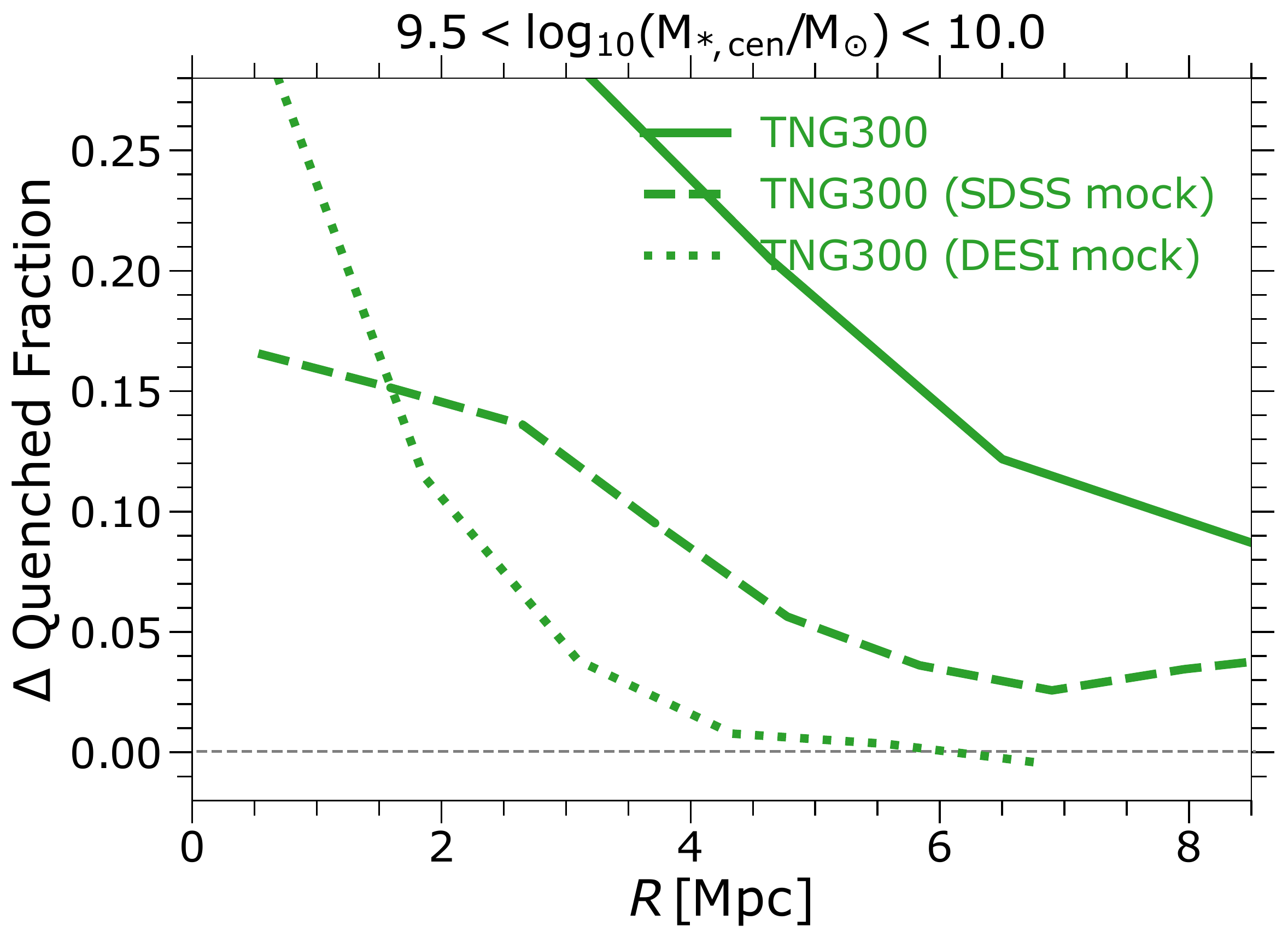}
    \includegraphics[width=0.33\textwidth]{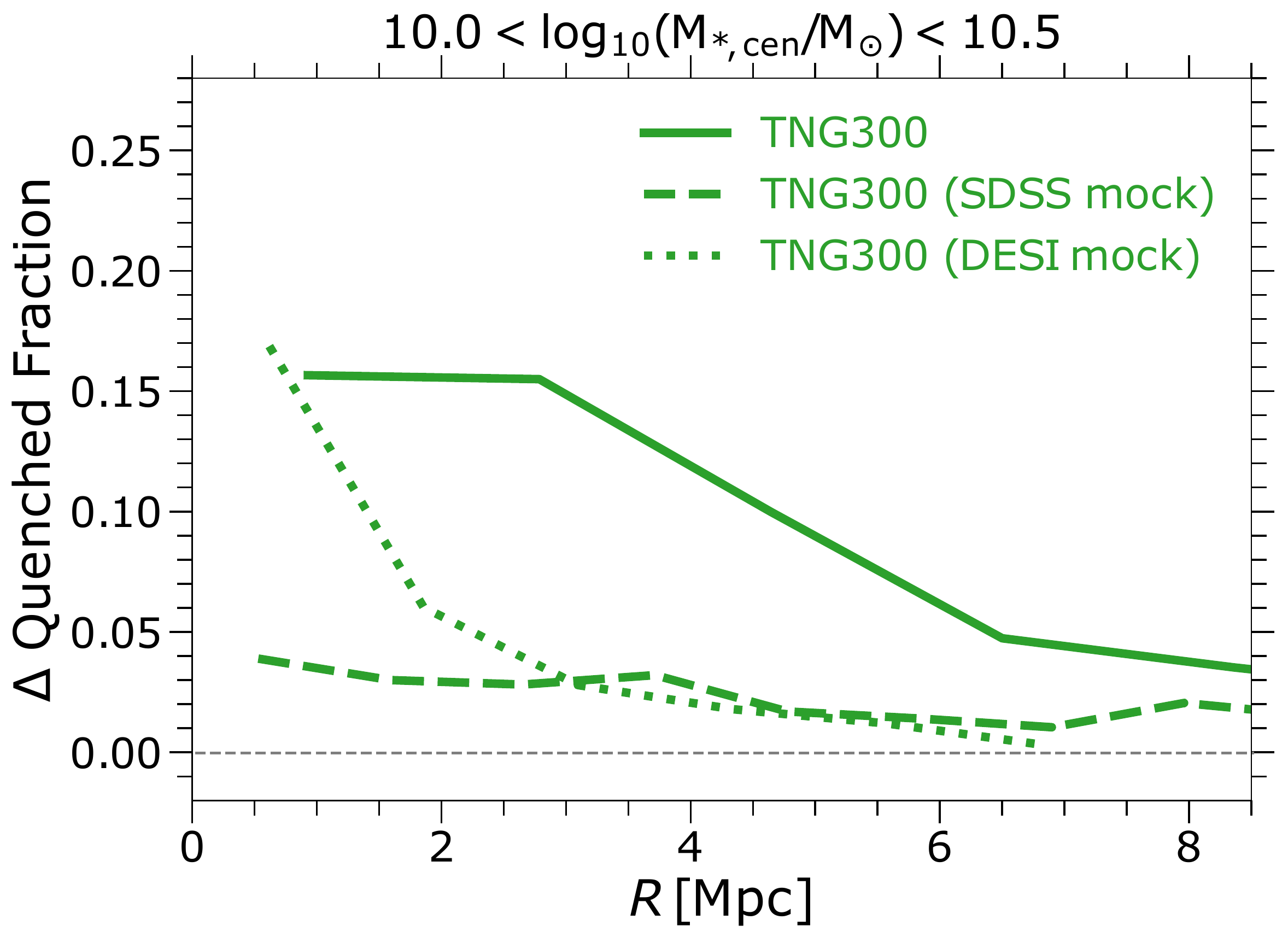}
    \includegraphics[width=0.33\textwidth]{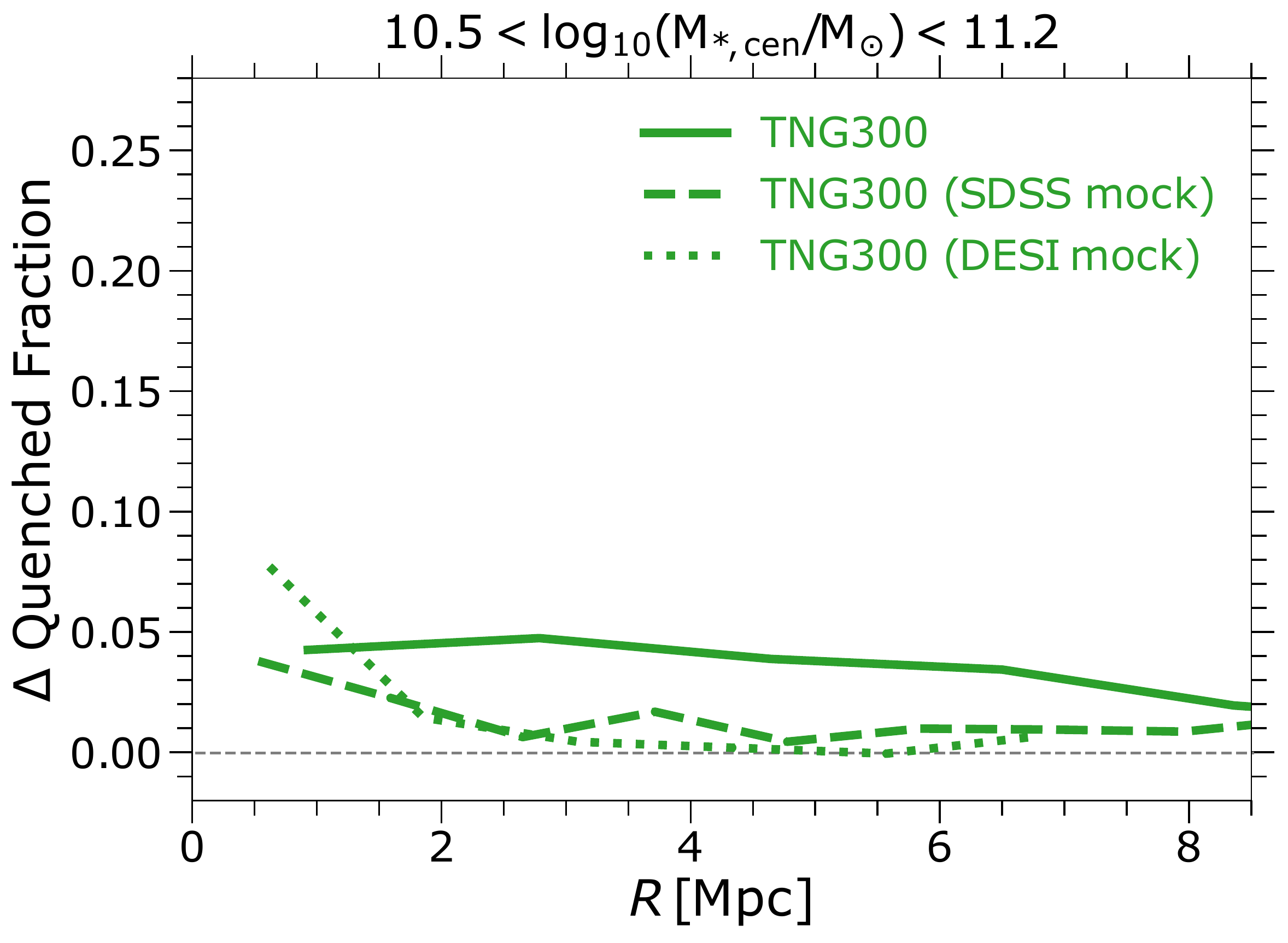}
    \includegraphics[width=0.33\textwidth]{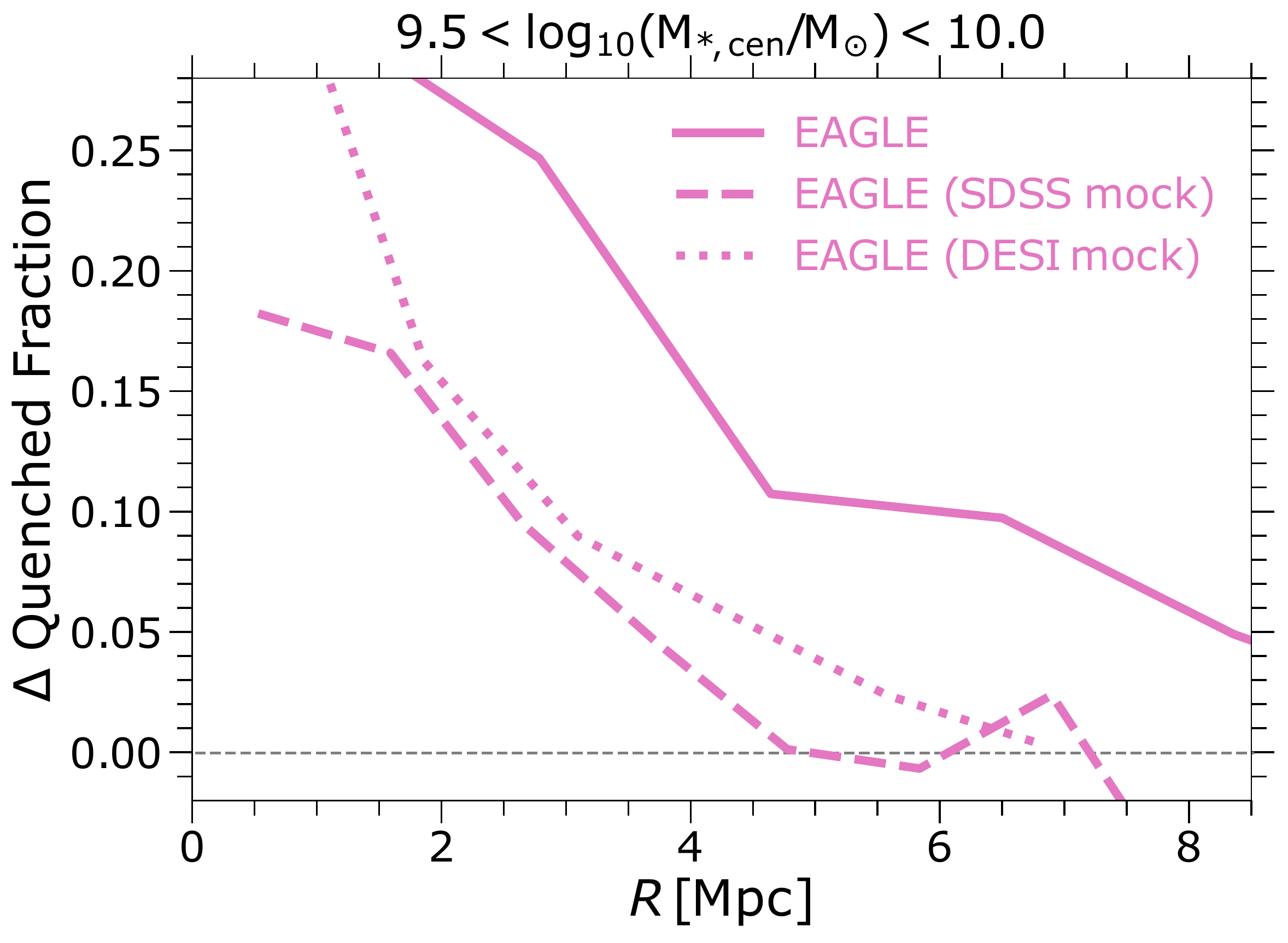}
    \includegraphics[width=0.33\textwidth]{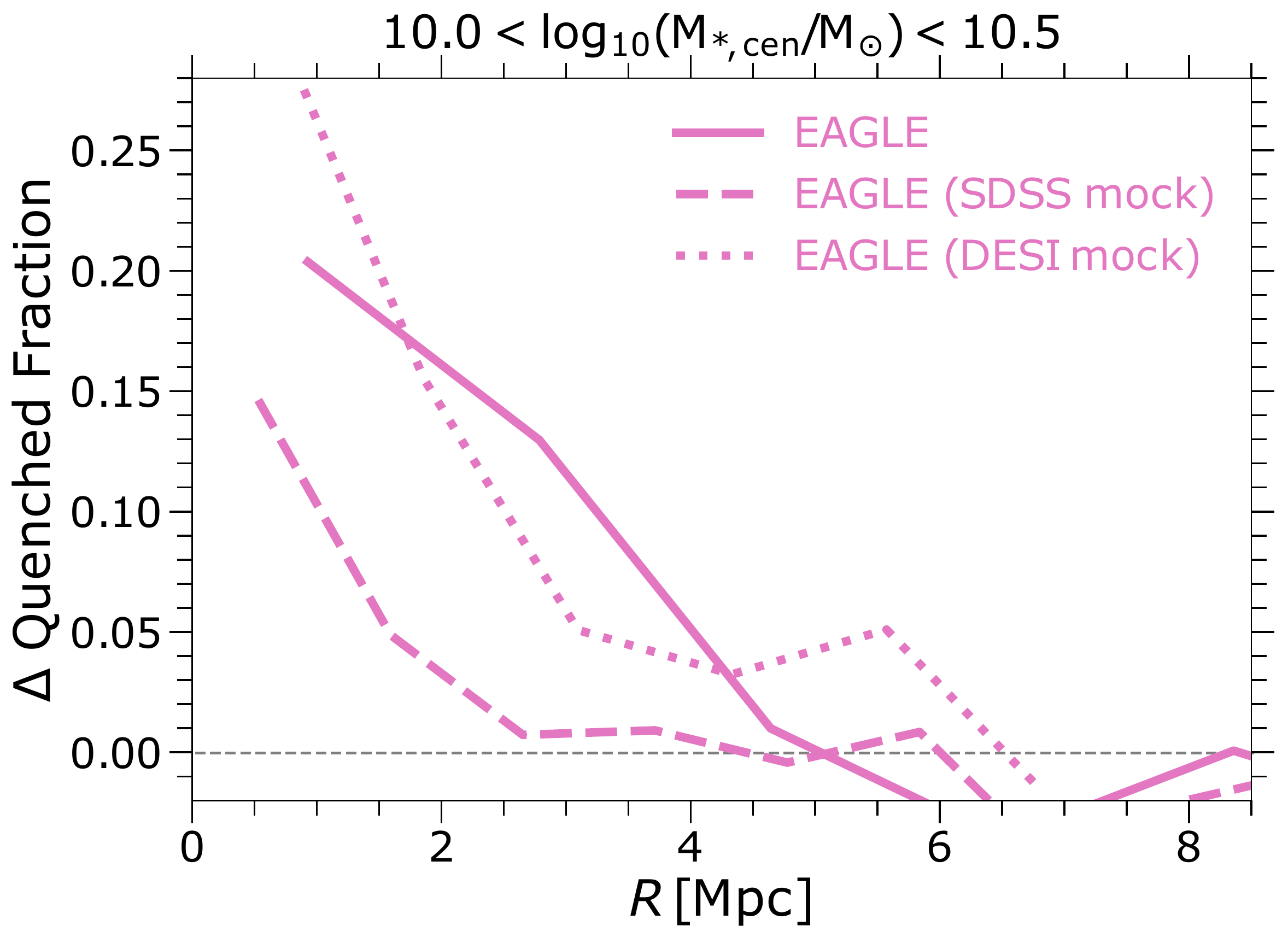}
    \includegraphics[width=0.33\textwidth]{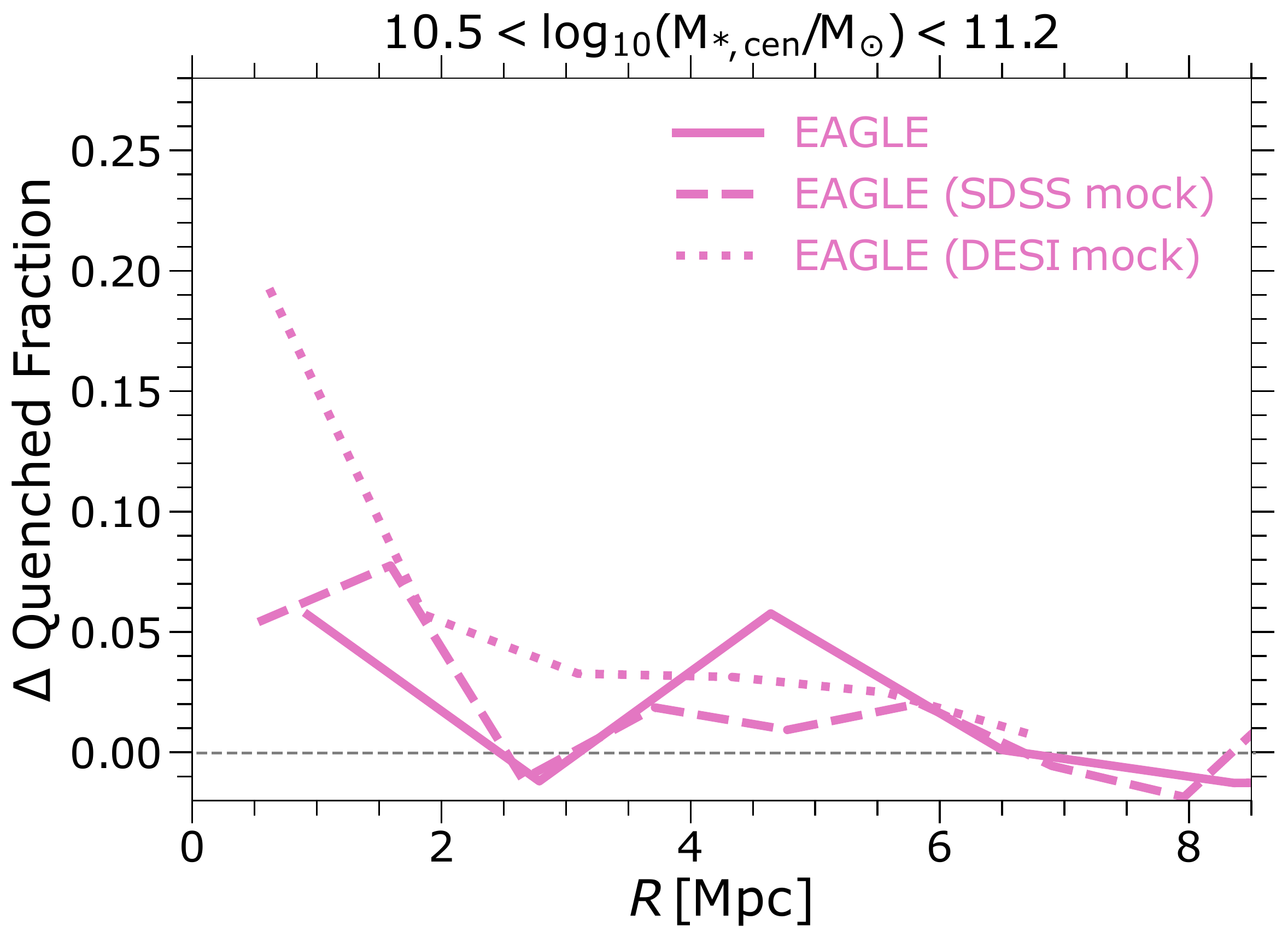}
    \caption{The conformity signal in the LGal-A21, LGal-H20, TNG, and EAGLE models and mock catalogues made for the SDSS and DESI observations. Different columns show the primary galaxies' stellar masses. The linestyles correspond to the following. Solid lines: direct simulation output, dashed lines: mock galaxy catalogues made for SDSS, and dotted lines: mock galaxy catalogues made for DESI. The lines are identical to the lines plotted in Figs. \ref{Fig: Conformity_signal_quenching_models_redshift} and \ref{Fig: Conformity_signal_quenching_SDSS}. Overall, the signal extracted directly from the simulation in 3D (solid lines) is stronger and the amplitude of the signal decreases with projection.}
\end{figure*}

%%%%%%%%%%%%%%%%%%%% REFERENCES %%%%%%%%%%%%%%%%%%

\vspace{-1em}
\bibliographystyle{mnras}
\bibliography{refbibtex}

%%%%%%%%%%%%%%%%% APPENDICES %%%%%%%%%%%%%%%%%%%%%

\appendix
\section{The amplitude of the signal in the simulations and mock catalogues}
\label{app: sims_and_mocks}
\begin{figure*}
\label{Fig: TNG_consistency}
    \centering
    \includegraphics[width=0.33\textwidth]{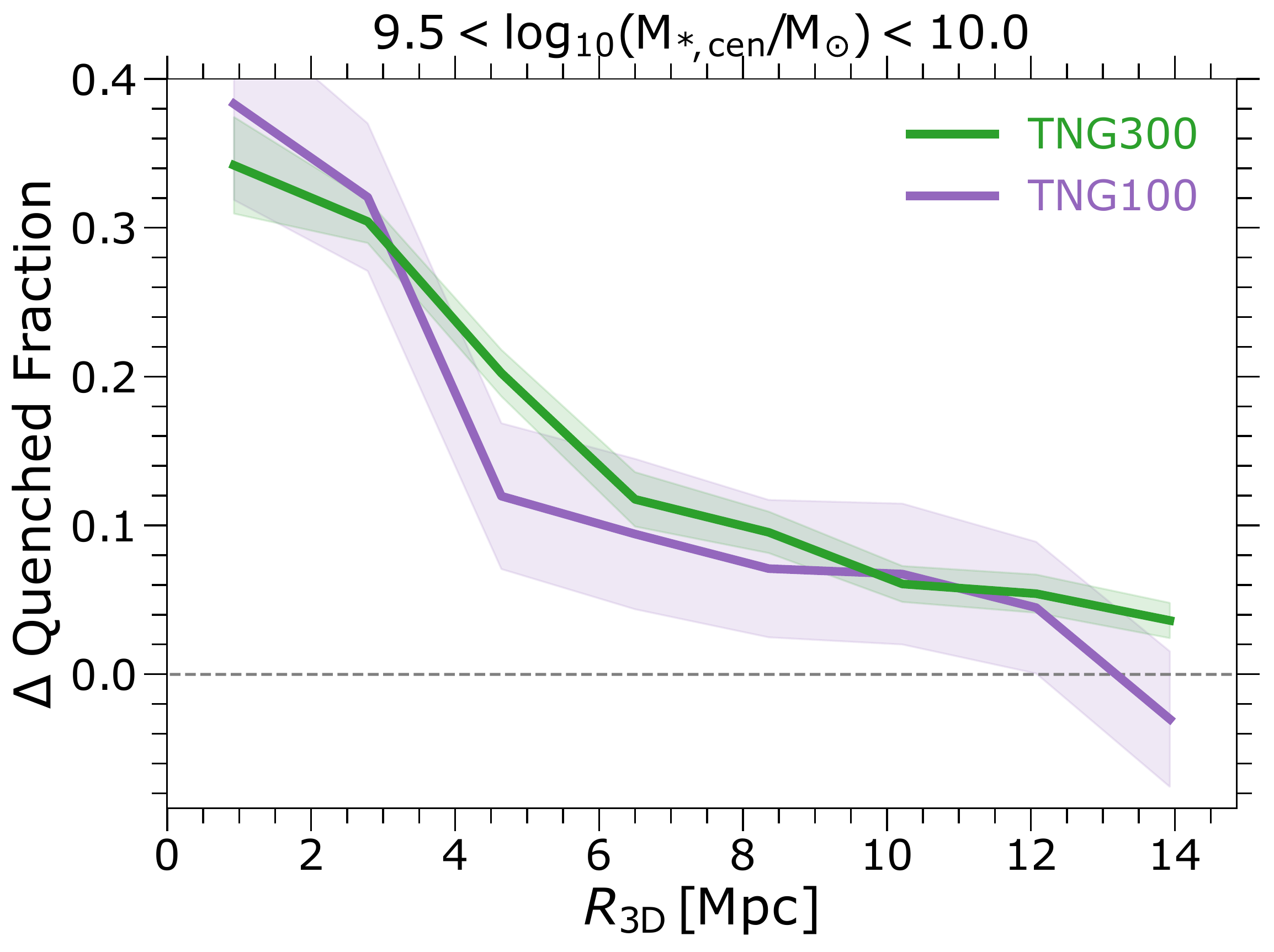}
    \includegraphics[width=0.33\textwidth]{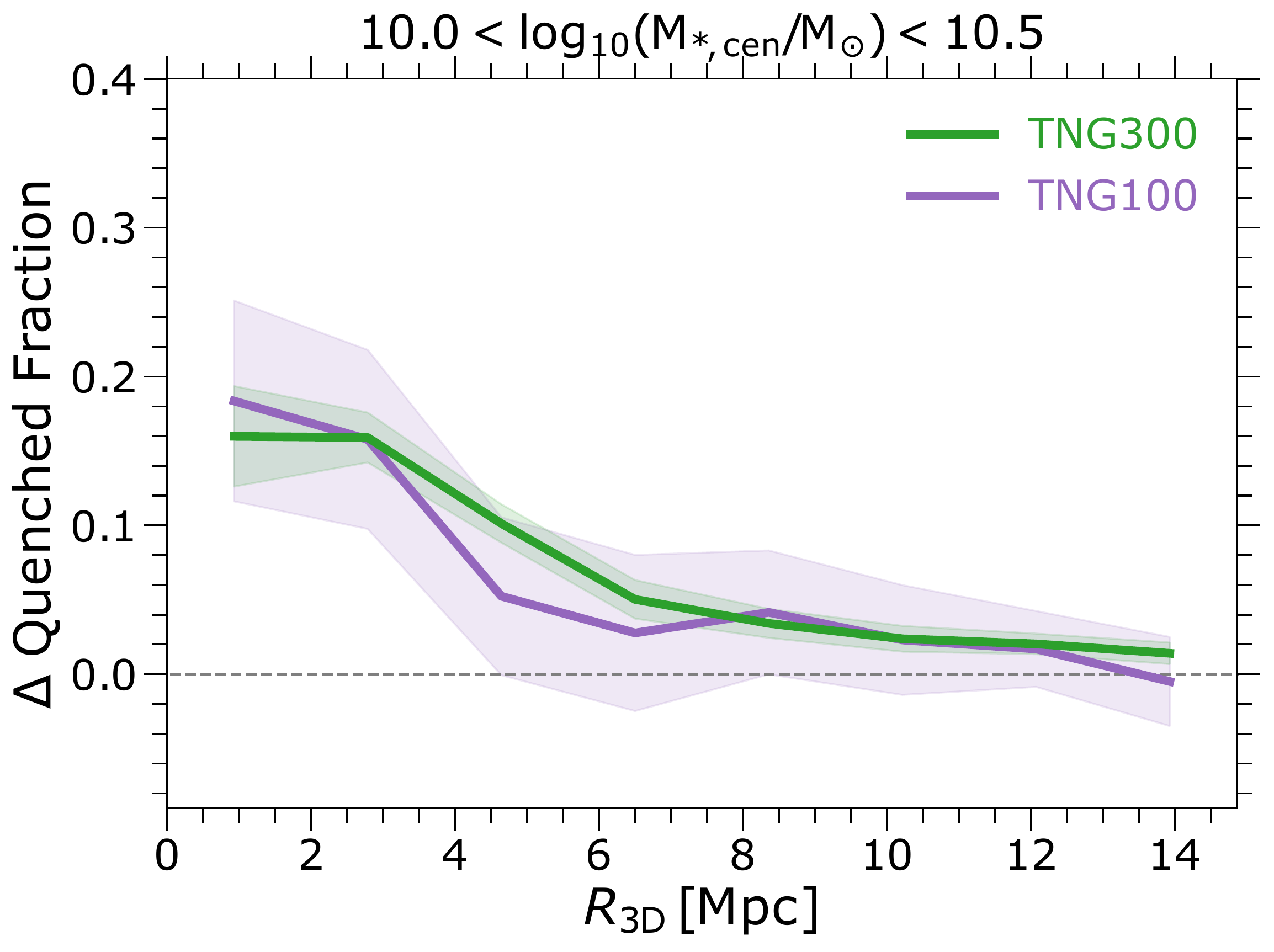}
    \includegraphics[width=0.33\textwidth]{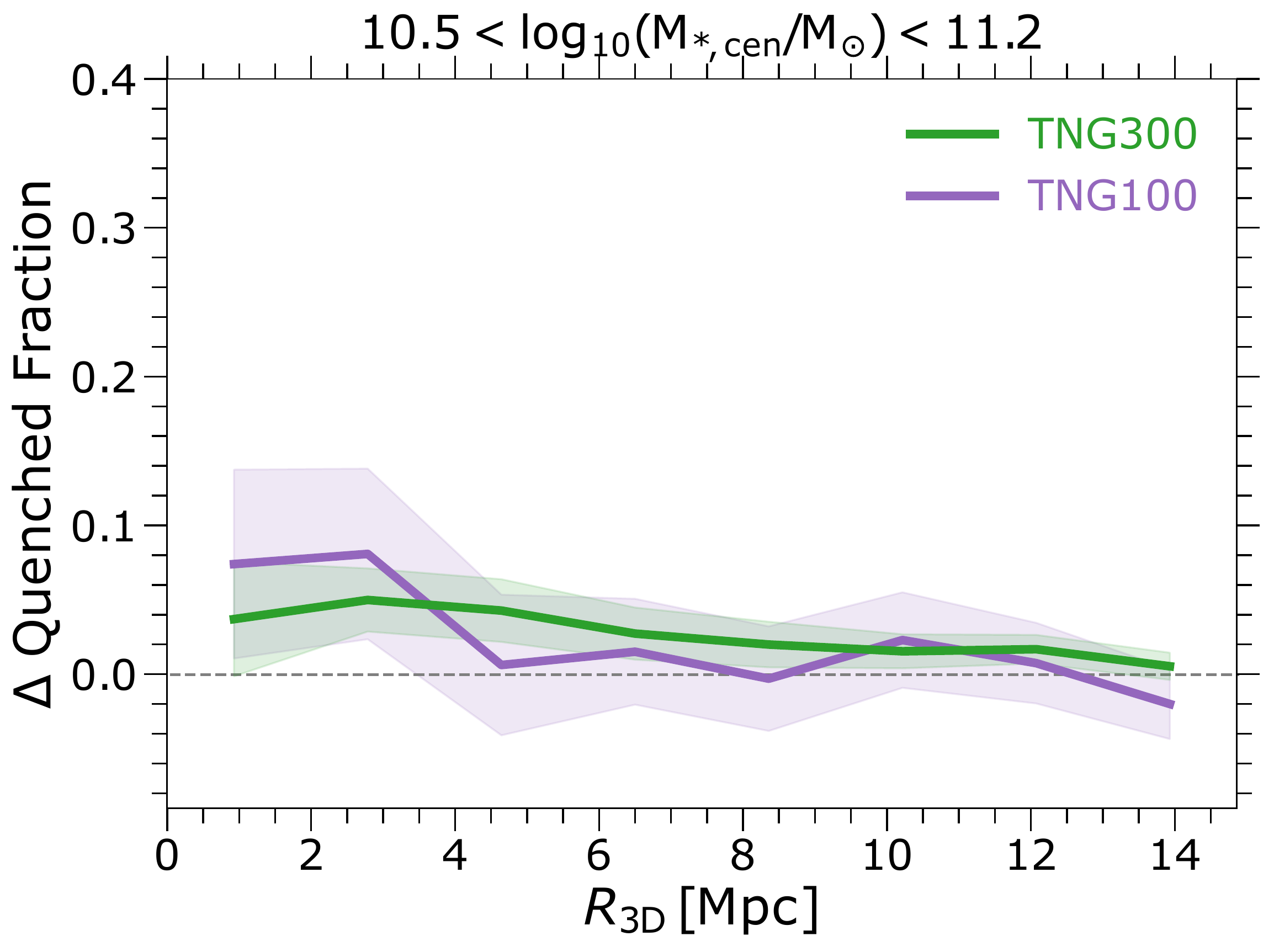}
    \caption{The conformity signal in TNG300 and TNG100 as a function of 3D halocentric distance. The shaded regions show the uncertainty of the results, derived using the bootstrap method. All central and satellite galaxies are identified using \textsc{SUBFIND}. The convergence of the signal between the two TNG models is relatively good.}
\end{figure*}
Fig. \ref{Fig: signal_sims_mocks} shows the conformity signal in the models and their mock galaxy catalogues. Each row corresponds to a model. The results are split into three columns based on the stellar masses of the primary galaxies. In almost all cases, the signal's amplitude is higher when extracted directly from the simulation in three dimensions. The signal's amplitude decreases in the mock catalogues because they contain projections in the direction of line-of-sight velocity to mimic observations. Moreover, the signal's amplitude is lowest in DESI mock catalogues because we operate a deeper projection compared to SDSS mocks to overcome the uncertainties in DESI's photometric redshifts. Therefore, we conclude that not only the "true" conformity signal is not an artefact of our methodology (e.g. the \textsc{CenSat} algorithm), but also the 3D amplitude of the signal is very likely to be even larger than reported in this work.

\section{The consistency of the conformity signal in different TNG runs}
\label{app: TNG_consistency}
The free parameters of the TNG simulations are calibrated at the TNG100 resolution against several observations such as the stellar mass function and the stellar-to-halo mass ratio at z = 0. However, in this work we used the TNG300 version which has a larger box size ($\sim 300 \, \rm Mpc)$ but a lower resolution. The statistical properties of galaxies in TNG300 do not quite match the calibrated TNG100 simulation, because of a lack of resolution convergence in hydrodynamical simulations such as TNG. Although a simple method is suggested to rescale the stellar masses of "central" galaxies to overcome this issue \citep[see][]{pillepich2018Simulating}, we are not aware of any good solution for other properties of galaxies such as star formation rates. Therefore, we simply take the simulation's original outputs, without any rescaling applied.

For our research in this paper, the most important properties of galaxies are the stellar mass and star formation rate. Overall, we checked and found that TNG300 galaxies are both less massive and less star-forming than TNG100 galaxies of the same halo mass. This also cause differences in the quenched fraction of galaxies between the two models. Nevertheless, it does not change the conformity signal, as shown in Fig. \ref{Fig: TNG_consistency}. That is mainly because the quenched fraction of galaxies near quenched and star-forming primary galaxies are shifted by a similar factor in TNG300 comparing to TNG100. As a result, the conformity signal converges between the two models and, therefore, leaves our analysis robust.

\label{lastpage}

\end{document}